\documentclass[12pt]{article}
\usepackage{amsmath}
\usepackage{amssymb}
\usepackage[spanish]{babel}
\usepackage[latin1]{inputenc}

\begin{document}
\vskip 1truein
\begin{center}
{\bf SOBRE EL PROBLEMA DEL ACOPLAMIENTO DE CAMPOS DE ESPINES ALTOS
EN DIMENSI\'ON $2+1$}
 \vskip 5pt
{\bf Rolando Gaitan D.}{\footnote {e-mail: rgaitan@uc.edu.ve}} \\
{\it Tesis Doctoral (Junio 2005), Centro de F\'{\i}sica Te\'orica y
Computacional, Facultad de Ciencias, Universidad Central de
Venezuela, Caracas 1041-A, Venezuela.\\ Tutor: Dr. P\'{\i}o J.
Arias.}\\
\end{center}
\vskip .5truein
\begin{abstract}
Se aborda el problema del acoplamiento de campos de espines altos
con un {\it background} no din\'amico, teniéndose particular
inter\'es en espacios $2+1$ dimensionales. Partiendo de una
formulaci\'on lagrangiana de campos apropiada, se estudian la
causalidad y  la conservaci\'on del n\'umero de grados de libertad
locales en una teor\'{\i}a con interacci\'on con campos
gravitacionales no din\'amicos, verific\'andose que tal clase de
teor\'{\i}a s\'olo es consistente en ciertos tipos de
espacio-tiempo, como los de de (Anti-) Sitter. Por otro lado, se
enfoca el problema del acoplamiento del campo gravitacional (ahora
como objeto din\'amico) con campos materiales como fuentes, desde
el punto de vista de una formulaci\'on de calibre de tipo
Yang-Mills. Allí se encuentran posibles restricciones sobre la
forma en c\'omo se distribuyen los campos materiales, y se muestra
que la introducci\'on de campos auxiliares acoplados con la
conexi\'on de calibre eliminan las restricciones sobre tales
campos materiales. El modelo de la formulaci\'on de calibre de la
gravedad topol\'ogicamente masiva con constante cosmol\'ogica, es
brevemente discutido, mostr\'andose que posee una ecuación de
campo consistente con la esperada en el límite de torsi\'on nula.

\end{abstract}

\newpage
{\Large {\bf \'{\i}ndice}}

\vskip .2truein

{\bf \large{ 1) Introducci\'on}}

\vskip .2truein

{\bf \large{2) Teoría Autodual de espín 2 en un espacio-tiempo
plano}}

{\bf 2.1) La teoría autodual de espín 2 y su análisis Lagrangiano
(p\'ag. 12)}

{\bf 2.2) La acción reducida (p\'ag. 15)}

{\bf 2.2.1) El \'algebra de operadores en la teoría autodual
(p\'ag. 23)}

{\bf 2.2.2) Separación de la parte transversal-sin traza de
${h^{(s)}}_{\mu \nu }$ (p\'ag. 27)}

{\bf 2.3) Generadores del álgebra de Poincaré  (p\'ag. 35)}

\vskip .2truein

{\bf \large{3) Teoría autodual de espín 2 en espacios de curvatura

constante}}

{\bf 3.1) Vínculos Lagrangianos en un espacio curvo  (p\'ag. 41)}

{\bf 3.2) Teoría autodual en un espacio de dS/AdS  (p\'ag. 45)}

{\bf 3.2.1) Acción reducida  (p\'ag. 51)}

\vskip .2truein

{\bf \large{4) Formulación $GL(N,R)$ de calibre de la gravedad}}

{\bf 4.1) Teoría libre  (p\'ag. 56)}

{\bf 4.1.1) Ecuaciones de campo  (p\'ag. 60)}

{\bf 4.2) Acoplamiento con materia  (p\'ag. 64)}

{\bf 4.2.1) Inclusión de campos auxiliares  (p\'ag. 72)}

{\bf 4.3) La gravedad topológicamente masiva  (p\'ag. 77)}

{\bf 4.3.1) Formulación de calibre $GL(3,R)$ topológica masiva
(p\'ag. 78)}

\vskip .2truein

{\bf \large{5) Conclusiones}  (p\'ag. 82)}

\vskip .2truein

{\bf \large{6) Ap\'endices}  (p\'ag. 87)}

\vskip .2truein

{\bf \large{Referencias}  (p\'ag. 98)}

\newpage
{\large {\bf 1) Introducci\'on}} \vskip .5truein

La obtención de una descripci\'on consistente de la interacci\'on
con campos de espines altos posee particular importancia ya que
nos permitir\'{\i}a establecer un puente entre estos campos y el
mundo observable.

El inter\'es por el estudio de campos de espines altos tiene
muchos afluentes. Por ejemplo, podemos notar que la teor\'{\i}a de
cuerdas incluye una cantidad infinita de exitaciones masivas con
todos los espines posibles  y por tanto permitir\'{\i}a  alguna
descripci\'on consistente de la interacci\'on de campos  (por
ejemplo con un campo  electromagn\'etico o gravitacional externos)
con espines arbitrarios.

Existe una evidencia  a favor de la introducci\'on de la
teor\'{\i}a de cuerdas en el problema del acoplamiento \cite{r14}
como es el caso en que las ecuaciones de movimiento consistentes
para el campo masivo de esp\'{\i}n 2 puedan ser construidas a
partir de una serie infinita de t\'erminos en un {\it background}
gravitacional arbitrario. Este tipo de series infinitas aparecen
de manera natural en teor\'{\i}a de cuerdas, raz\'on por la cual
\'esta podr\'{\i}a imponerse como una aproximación consistente
para la descripci\'on de la interacci\'on de espines altos. En
este sentido, existe también un estudio de la propagación
consistente de espín 2 con interacción gravitacional \cite{r3} que
respalda la idea de considerar un conjunto infinito de campos
masivos.

Si consideramos la teor\'{\i}a de campos ordinaria, las
formulaciones lagrangianas cl\'asicas con interacci\'on para
campos de espines altos son conocidas para ciertos espacios. Por
ejemplo, muchos autores \cite{r1}-\cite{r9} han abordado
teor\'{\i}as de campos masivos con espines enteros en un
espacio-tiempo de curvatura constante, y entre otros \cite{r10} se
incorporan, adem\'as,  espacios no Einstenianos.

Entre  las posibles interacciones, la electromagn\'etica ha sido
considerada y ha servido de marco para el estudio de muchos
modelos. Podemos mencionar la teor\'{\i}a de un campo masivo de
esp\'{\i}n 2 en un campo electromagn\'etico homog\'eneo \cite{r11}
en el contexto de la cuerda bos\'onica en dimensi\'on d=26, y un
resultado an\'alogo (pero en teor\'{\i}a ordinaria de campos)
tambi\'en es obtenido \cite{r12}. Hay otro estudio similar para
esp\'{\i}n 3 \cite{r13}. Sin embargo, en este trabajo centraremos
nuestra atención en el problema de la interacción con la
gravitación.

Entonces, en el contexto de la teor\'{\i}a de campos ordinaria
surge la pregunta crucial: {\it ¿qué obstáculos ocurren en la
construcción de una teor\'{\i}a consistente de campos de espines
altos con interacci\'on externa?}.

La raz\'on esencial es que existen por lo menos dos formas
mediante las cuales la interacci\'on destruye la consistencia de
una teor\'{\i}a de espín alto. Primero, la interacci\'on puede
cambiar el n\'umero de grados din\'amicos de libertad. Por
ejemplo, un campo masivo de espín $s$ en un espacio-tiempo
Minkowskiano $3+1$ dimensional, est\'a descrito por un tensor
simétrico, transverso y sin traza de rango $s$ que satisface las
condiciones:
\begin{equation}
(\Box{}-m^2)\phi _{{\mu _1}{\mu _2}...{\mu _s}} =0, \,\,\,
{\partial}^\mu \phi _{\mu {\mu _1}...{\mu _{s-1}}} =0, \,\,\,
{{\phi}^\mu }_{\mu {\mu _1}...{\mu _{s-2}}} =0\, \, .
\label{ecua1}
\end{equation}
En las referencias \cite{r15},\cite{r16} se muestra que para
reproducir estas ecuaciones a partir de un lagrangiano es
necesario introducir $s-1$ campos auxiliares $\chi _{{\mu _1}{\mu
_2}...{\mu _{s-2}}} $, $\chi _{{\mu _1}{\mu _2}...{\mu _{s-3}}}
$,...,$\chi$. Estos campos sim\'etricos y sin traza se anulan en
la capa de masas, pero la presencia de ellos en la teor\'{\i}a
provee una descripci\'on lagrangiana de  las condiciones
(\ref{ecua1}) (en espacios de mayor dimensi\'on aparecen campos
con una estructura tensorial m\'as compleja pero la situaci\'on
general permanece id\'entica, esto es, la descripci\'on
lagrangiana siempre precisa la presencia de grados de libertad
auxiliares no f\'{\i}sicos).

La cuesti\'on es, que estos campos auxiliares crean problemas
cuando uno trata de hacer aparecer la interacci\'on en la
teor\'{\i}a. Una interacci\'on arbitraria hace, en general, que
los campos auxiliares se propaguen, modificando el n\'umero de
grados de libertad locales. Por consiguiente, si se exige la
ausencia de estos grados de libertad, entonces aparecer\'{\i}an
restricciones adicionales sobre la forma posible de la
interacci\'on.

El otro problema que podr\'{\i}a surgir en la construcci\'on de
teor\'{\i}as de interacci\'on con espines altos es el relacionado
con la posible violaci\'on de la causalidad, el cual ha sido
notado por diversos autores \cite{r10},\cite{r17}-\cite{r19}. La
situación consiste en lo siguiente: consideremos un campo de
esp\'{\i}n entero descrito mediante un tensor cuyas componentes
son $\phi ^B$ ($B$ simboliza un conjunto de \'{\i}ndices mixtos)
en un espacio $N$ dimensional. Haciendo uso del sistema de
v\'{\i}nculos lagrangianos \cite{r20} de la teor\'{\i}a con
interacci\'on, podemos reescribir las ecuaciones de movimiento
(obtenidas a partir de las variaciones sobre la acci\'on $S= \int
d^Nx \sqrt {-g}\,L(\phi ,
\partial \phi )$) de la forma
\begin{equation}
{M_{AB}}^{\mu \nu }{\partial}_{\mu }{\partial}_{\nu }{\phi }^B
+... =0, \,\,\, \mu ,\nu  =0,...,N-1 \, \, . \label{ecua2}
\end{equation}
El objeto ${M_{AB}}^{\mu \nu }$ permite definir la {\it matriz
caracter\'{\i}stica}, $M_{AB}(n)\equiv {M_{AB}}^{\mu \nu }n_\mu
n_\nu $, que posee como argumento al multivector $n_\mu$ de $N$
componentes. La {\it ecuaci\'on caracter\'{\i}stica }
correspondiente es $\det M_{AB}(n) =0$. Las soluciones de \'esta
definen una hipersuperficie con $n_\mu$ normal a la misma. De esto
sigue que, si para cualquier $n_i$ ($i=1,2,...,N-1$), las posibles
componentes $n_0 = n_0 (n_i)$ (obtenidas despu\'es de despejar
$n_0$  de la ecuaci\'on caracter\'{\i}stica) son reales, entonces
el sistema de ecuaciones diferenciales (\ref{ecua2}) es llamado
{\it hiperb\'olico} y describe un proceso de propagaci\'on. Un
sistema de ecuaciones hiperb\'olico es llamado {\it causal} si la
ecuaci\'on caracter\'{\i}stica no posee, dentro de las posibles
soluciones, vectores tipo tiempo (de lo contrario implicar\'{\i}a
que la hipersuperficie ortogonal ser\'{\i}a de tipo espacio y por
tanto, los puntos de ella estar\'{\i}an conectados de manera no
causal). Entonces, la cuesti\'on es que si introducimos
interacci\'on en la teor\'{\i}a, la matriz caracter\'{\i}stica
$M_{AB}(n)$ se puede modificar de manera tal que la causalidad
podría ser violada.

Queremos subrayar que la posible inconsistencia mencionada en la
parte anterior, relacionada tanto con la no conservaci\'on de los
grados de libertad como la violaci\'on de la causalidad, y
enmarcada en la teor\'{\i}a de campos presupone un espacio-tiempo
no din\'amico. Desde otro punto de vista podr\'{\i}a ser v\'alido
preguntarse qu\'e suceder\'{\i}a si el {\it background} fuese
din\'amico. En otras palabras, se plantea el problema de la
consistencia,  considerando al campo gravitacional como un objeto
din\'amico.

En t\'erminos de la teor\'{\i}a de campos, se podrían destacar dos
enfoques para explorar el acoplamiento de un campo gravitacional
dinámico con campos materiales. Por un lado podr\'{\i}a pensarse
en agregarle a la densidad lagrangiana de Hilbert-Einstein una
serie de t\'erminos lagrangianos de interacci\'on no minimales,
constru\'{\i}dos mediante contracciones de componentes del tensor
de curvatura de Riemann-Christoffel con los campos materiales, al
estilo de los términos introducidos, por ejemplo en las
referencias \cite{r1},\cite{r10}.

Sin embargo, existe otra corriente, relativamente menos explorada
como la de estudiar el acoplamiento de la gravitaci\'on con la
materia pero desde el punto de vista de una formulaci\'on de
calibre de la gravedad. El deseo natural de que la gravitaci\'on
pueda ser tratada, al menos hasta cierto nivel de analog\'{\i}a
como una teor\'{\i}a de calibre ha sido un tema considerado por
muchos autores \cite{r21}-\cite{r38}. Esta motivaci\'on quiz\'as
se ha mantenido esencialmente debido al \'exito alcanzado por la
teor\'{\i}a de calibre de Yang-Mills \cite{r39} en su aplicaci\'on
al modelo electro-d\'ebil \cite{r40}-\cite{r42}, entre otras, con
la esperanza de que algo similar ocurriese con la gravitaci\'on.

Cuando en general hablamos de formulaciones de calibre de la
gravedad nos referimos heurísticamente a construcciones donde este
campo est\'e descrito mediante un conexi\'on sobre cierto fibrado,
y que la formulación Lagrangiana sea, por ejemplo del tipo
Yang-Mills \cite{r25},\cite{r38}. Dentro del gran conjunto de
propuestas, enfocaremos nuestro interés en la que se fundamenta en
el fibrado de referenciales con grupo de calibre \,$GL(N,R)$
\cite{r25},\cite{r38}, por tener ésta un origen muy intuitivo a
partir de los conceptos geométricos involucrados en la Relatividad
General.

En definitiva, sea cual sea la formulaci\'on de calibre que
adoptemos (asumiendo que la naturaleza refleja parte de su
funcionamiento de esa manera) se hace necesario tener un esquema
consistente para el acoplamiento del campo gravitacional con
fuentes materiales, lo cual, de ser posible constituir\'{\i}a un
aporte significativo en la verificaci\'on y veracidad de tal
modelo.

Este trabajo está organizado como sigue. En el Capítulo 2
revisamos la teoría del modelo autodual de espín 2 en un
espacio-tiempo plano 2+1 dimensional. Allí, siguiendo el
procedimiento para la obtención de la acción reducida, se realiza
la discusión de los conmutadores de los operadores
mecánico-cuánticos de tal teoría, así como la de los generadores
del álgebra de Poincaré. En el Capítulo 3, extendemos la teoría
autodual de espín 2 al caso de un espacio-tiempo curvo, revelando
la situación general en la que la preservación del número de
grados de libertad es rota. Como caso particular, son  estudiados
los espacios de dS y AdS, en donde es respetado el número de
grados de libertad y la causalidad. Un posible procedimiento, como
extensión al usado en el caso plano, para la obtención de la
acción reducida en dS/AdS, es implementado.

En el Capítulo 4, exploramos una posible formulación de calibre
para la gra-vitación, basada en la idea del fibrado de
referenciales. Comenzando con el caso del vacío cosmológico, se
muestra que la consistencia de tal teoría con la formulación de
Hilbert-Einstein demanda que la constante cosmológica debe
contribuir de manera cuadrática. Seguidamente (sec. 4.2), se
discute un posible esquema de acoplamiento no minimal con campos
materiales, considerando un espacio $2+1$ dimensional como un
escenario de prueba sencillo. Allí, mostramos que para evitar las
inconsistencias que ocurren entre la dinámica obtenida de esta
formulación, comparada con la que se obtiene de la teoría de
Hilbert-Einstein, debemos introducir ciertas restricciones sobre
los campos materiales. Con la introducción de campos auxiliares,
acoplados de manera no minimal con la gravitación, se muestra que
es posible eliminar las restricciones sobre los campos materiales.
En la sección final (4.4.1), presentamos el modelo de la
formulación de calibre de la gravedad topológicamente masiva con
constante cosmológica. Allí discutimos su consistencia con el
modelo de Deser \cite{r67}.

\newpage
{\large {\bf 2) Teoría Autodual de espín 2 en un espacio-tiempo
plano}} \vskip .5truein

Los estados de una part\'{\i}cula son especificados por los
casimires del \'algebra de Poincar\'e. En dimensi\'on $2+1$, el
\'algebra resulta ser
\begin{eqnarray}
[ J^\mu , J^\nu ]= i\epsilon ^{\mu\nu\lambda}J_\lambda \,\,\,,
\label{t1}
\end{eqnarray}
\begin{eqnarray}
[ \mathcal{P}^\mu,J^\nu ]= i\epsilon
^{\mu\nu\lambda}\mathcal{P}_\lambda \,\,\,,\label{t2}
\end{eqnarray}
\begin{eqnarray}
[ \mathcal{P}^\mu , \mathcal{P}^\nu ]= 0 \, \,\,,\label{t3}
\end{eqnarray}
donde $\mathcal{P}^\mu$ y $J^\mu = \frac12 \,\epsilon
^{\mu\nu\lambda}\mathcal{J}_{\nu\lambda}$ son los generadores
hermíticos de translaciones y rotaciones de Lorentz,
respectivamente. Los casimires correspondientes son
$\mathcal{P}^2$ y $\mathcal{P}^\mu J_\mu $, los cuales act\'uan
sobre estados de una part\'{\i}cula de la forma
\begin{eqnarray}
(\mathcal{P}^2 + m^2)\Psi= 0\,\, \,,\label{t4}
\end{eqnarray}
\begin{eqnarray}
(\mathcal{P}^\mu J_\mu+ sm)\Psi= 0 \,\, \,  . \label{t5}
\end{eqnarray}
La primer ecuaci\'on es la bien conocida condici\'on de capa de
masas, mientras que la segunda especifica la helicidad, con $s$ el
esp\'{\i}n o helicidad de la part\'{\i}cula.

Es conocido que para una part\'{\i}cula masiva con esp\'{\i}n $1$
se requiere un campo de una componente. Sin embargo, usaremos un
objeto (tensor) de tres componentes con la finalidad de abordar
una representaci\'on lineal del grupo de Lorentz. Este campo debe
satisfacer ciertas condiciones subsidiarias con las cuales se
eliminen las componentes no requeridas. Si realizamos
$\mathcal{P}_\mu $ en unidades naturales como $i\,\partial_\mu$, y
${({J_{1}}^\mu)_\lambda }^\rho \doteq i\,{{\epsilon ^\mu}_\lambda
}^\rho $ obtendremos una representaci\'on del \'algebra de Lie
(2.1-2-3) sobre vectores (en general es sabido \cite{r43} que la
escogencia hecha para ${({J_{1}}^\mu)_\lambda }^\rho $ no es
única, pues es posible agregar una parte orbital sin afectar el
álgebra de Poincaré). La condici\'on de Pauli-Lubanski ser\'a
\begin{eqnarray}
{(\mathcal{P}.J + sm )^\mu}_\lambda\,V^\lambda=-\epsilon
^{\mu\nu\lambda}\,\partial_\nu V_\lambda + sm
V^\mu=0\,\,.\label{t6}
\end{eqnarray}
La condici\'on subsidiaria aparece como una restricci\'on de
transversalidad presente en (\ref{t6})
\begin{eqnarray}
\mathcal{P}_\mu\,V^\mu=0\,\,.\label{t7}
\end{eqnarray}
La otra condici\'on subsidiaria ocurre de la componente temporal
de (\ref{t6}) que corresponde a un vínculo.

Pensamos ahora en (\ref{t6}) como la ecuación de movimiento de una
teoría de campos. La manera covariante de ver que solamente
aparecen las componentes f\'{\i}sicas consiste en el formalismo de
Proyectores (Ap\'endice A). En este sentido, la relaci\'on
(\ref{t6}) es reescrita como
\begin{eqnarray}
[-\Box{}^{\frac12} (\mathbf{P}^+ -\mathbf{P}^- )+sm(\mathbf{P}^+
+\mathbf{P}^- )]\mathbf{V}^T=0\,\,,\label{t8}
\end{eqnarray}
donde consideraremos que $V^\mu$ puede ser descompuesto como
$\mathbf{V}=\mathbf{V}^T+\mathbf{V}^L$, con
$\mathbf{P}\mathbf{V}^T=\mathbf{V}^T$, $\mathbf{P}\mathbf{V}^L=0$.
Si miramos a esta ecuaci\'on como la proveniente de una
teor\'{\i}a con una fuente externa, es inmediato ver que para
esp\'{\i}n $s=+1$ ($-1$) habr\'a un polo masivo para
$\mathbf{V}^{T+}$ ($\mathbf{V}^{T-}$ ), confirmando el hecho de
que solamente hay un grado de libertad con masa $m$. Si ''elevamos
al cuadrado'' la ecuaci\'on (\ref{t8}), la condici\'on de capa de
masas es obtenida.

La acci\'on que tiene a (\ref{t6}) como su ecuaci\'on de
movimiento, es la acci\'on autodual \cite{r44}
\begin{eqnarray}
{S_{ad}}^{(1)}=-\frac{ms}{2}\int d^{3}x(\epsilon
^{\mu\nu\lambda}\,a_\mu\partial_\nu a_\lambda +sm\,a_\mu a^\mu
)\,\,,\label{t9}
\end{eqnarray}
con $s=\pm 1$. Las teor\'{\i}as autoduales en espacios de
dimensi\'on impar han recibido una considerable atenci\'on y
particularmente, en dimensi\'on $2+1$ son estudiadas debido a su
conexi\'on con la f\'{\i}sica de altas temperaturas en dimensi\'on
$3+1$ \cite{r45} y con la f\'{\i}sica de la materia condensada
\cite{r46},\cite{r47}.

Otra realizaci\'on para una part\'icula masiva de esp\'in $1$
masivo es el modelo topol\'ogico masivo abeliano
\begin{eqnarray}
S_{TM}=-\frac14 \int d^{3}x (F_{\mu\nu}F^{\mu\nu}-sm\epsilon
^{\mu\nu\lambda}\,a_\mu F_{\nu\lambda})\,\,,\label{t10}
\end{eqnarray}
con $F_{\mu\nu}=\partial_\mu a_\nu - \partial_\nu a_\mu$ y otra
vez $s=\pm 1$. La ecuaci\'on de movimiento es ahora
\begin{eqnarray}
\epsilon ^{\mu\nu\lambda}\,\partial_\nu {^{*}F}_\lambda
+sm{^{*}F}^\mu =0\,\,,\label{t11}
\end{eqnarray}
donde ${^{*}F}^\mu$ es el dual de Poincar\'e de $F^{\mu\nu}$, y la
condici\'on subsidiaria $\partial_\mu {^{*}F}^\mu =0$ es ahora la
identidad de Bianchi asociada con la invariancia de calibre
$\delta a_\mu =\partial_\mu \Lambda$. Este hecho (cambiar una
ecuaci\'on de movimiento por la identidad de Bianchi) es una señal
de la dualidad entre ambos modelos, como de hecho lo son
\cite{r47a}-\cite{r47c}.

Si buscamos realizar una teor\'{\i}a de esp\'{\i}n 2 masivo,
primero tomamos una re-presentaci\'on del \'algebra de Lie
actuando sobre 2-tensores sim\'etricos. Para esto, escogemos
\begin{eqnarray}
{({J_{2}}^\mu)^{\alpha\beta}}_{\rho\sigma}  \doteq
\frac{i}{2}\,({\delta ^\alpha}_\rho\,{\epsilon
^{\beta\mu}}_\sigma+ {\delta ^\beta}_\rho\,{\epsilon
^{\alpha\mu}}_\sigma+{\delta ^\alpha}_\sigma\,{\epsilon
^{\beta\mu}}_\rho+{\delta ^\beta}_\sigma\,{\epsilon
^{\alpha\mu}}_\rho)\,\,,\label{t12}
\end{eqnarray}
el cual satisface (\ref{t1}). Actuando sobre 2-tensores
sim\'etricos, transversos y sin traza (p.ej., ${h^{(s)Tt}}_{\alpha
\beta}$), la condici\'on de Pauli-Lubanski es establecida como
\begin{eqnarray}
{(\mathcal{P}.J_{2} + sm
)^{\alpha\beta}}_{\rho\sigma}\,h^{(s)Tt\rho\sigma}=0\,\,.\label{t13}
\end{eqnarray}
con $s=\pm 2$ y ${(sm)^{\alpha\beta}}_{\rho\sigma}=sm\,{\delta
^{\alpha\beta}}_{\rho\sigma}=\frac{sm}{2}({\delta
^\alpha}_\rho{\delta ^\beta}_\sigma +{\delta ^\beta}_\rho{\delta
^\alpha}_\sigma )$. Expl\'{\i}citamente, la ecuaci\'on (\ref{t13})
es
\begin{eqnarray}
-{\epsilon ^{\alpha\lambda}}_{\rho}\,\partial_\lambda
h^{(s)Tt\rho\beta}+\frac{sm}{2}
h^{(s)Tt\alpha\beta}=0\,\,,\label{t14}
\end{eqnarray}
y puede verse que solo se propaga un modo masivo de los dos
presentes en  ${h^{(s)Tt}}_{\alpha \beta}$, cuando es considerada
como la ecuación de movimiento de una teoría de campos. Para esto,
escribimos (\ref{t13}) en el lenguaje de proyectores usando los
proyectores de 2-tensores generales $h_{\alpha \beta}$ como sigue
\begin{eqnarray}
[-\Box{}^{\frac12} ({{\mathbf{P}}_+}^{(2)} -{{\mathbf{P}}_-}^{(2)}
)+\frac{sm}{2}({{\mathbf{P}}_+}^{(2)} +{{\mathbf{P}}_-}^{(2)}
)]\mathbf{h}^{Tt}=0\,\,,\label{t15}
\end{eqnarray}
y otra vez, como en el caso de esp\'{\i}n $1$ la propagaci\'on
solo est\'a asociada a la componente $\mathbf{h}^{Tt+}$
($\mathbf{h}^{Tt-}$), si el esp\'{\i}n es $s=+2$ ($s=-2$).

Para un tensor general $h_{\mu\nu}$, las ecuaciones que nos
conduce a  (\ref{t14}) son las de la acci\'on de espín 2 autodual
\cite{r48}
\begin{eqnarray}
{S_{ad}}^{(2)}=\frac m2 \int d^3 x (\epsilon ^{\mu \nu \lambda
}{h_\mu }^\alpha
\partial _\nu
h_{\lambda \alpha }-m(h_{\mu \nu }h^{\nu \mu }-h^2 )) \, \,
,\label{t16}
\end{eqnarray}
donde $h$ es la traza del campo.

Para explorar la posibilidad de una formulaci\'on m\'etrica de
${S_{ad}}^{(2)}$ consideremos la descomposici\'on
$h_{\mu\nu}={h^{(s)}}_{\mu\nu}+\epsilon _{\mu \nu \lambda
}V^\lambda$, con ${h^{(s)}}_{\mu\nu}={h^{(s)}}_{\nu\mu}$. En este
caso (\ref{t16}) es
\begin{eqnarray}
{S_{ad}}^{(2)}=\frac m2 \int d^3 x \big(\,\epsilon ^{\mu \nu
\lambda }{{h^{(s)}}_\mu }^\alpha
\partial _\nu
{h^{(s)}}_{\lambda \alpha }-m({h^{(s)}}_{\mu \nu }h^{(s){\mu \nu
}}-{h^{(s)}}^2 )\nonumber \\ +\,2V^\mu (\partial _\mu {h^{(s)}}
-\partial _\nu{{h^{(s)}}_\mu}^\nu)-\epsilon ^{\mu \nu \lambda
}V_\mu
\partial _\nu
V_\lambda -2mV_\mu V^\mu \big) \, \, . \label{t17}
\end{eqnarray}

Desde un punto de vista dinámico, es posible entender el papel de
''eliminador'' de espines bajos que juega la parte antisimétrica
de $h_{\mu\nu}$, si examinamos las ecuaciones de movimiento que se
derivan de la acción (\ref{t17}), es decir
\begin{eqnarray}
\epsilon ^{\mu \nu \lambda }
\partial _\mu{{h^{(s)}}_\nu }^\alpha +\epsilon ^{\mu \nu \alpha
}\partial _\mu{{h^{(s)}}_\nu }^\lambda -2m h^{(s)\lambda
\alpha }+  2m \eta^{\lambda \alpha }h^{(s)}+\nonumber \\
-2\partial_\mu V^\mu \eta^{\lambda \alpha }+
\partial^\lambda V^\alpha+
\partial^\alpha V^\lambda=0
 \, \, , \label{t18}
\end{eqnarray}
\begin{equation}
\epsilon ^{\mu \nu \lambda }\partial _\mu V_\nu +2m V^\lambda
-\partial^\lambda h^{(s)}+ \partial_\mu h^{(s) \mu\lambda }=0
 \, \, . \label{t19}
\end{equation}
Inmediatamente, podemos ver que la traza y la divergencia  de la
ecuación (\ref{t18}) proporcionan, respectivamente
\begin{eqnarray}
 mh^{(s)}-\partial_\mu V^\mu =0
 \, \, , \label{t20}
\end{eqnarray}
\begin{eqnarray}
\epsilon ^{\mu \nu \lambda }
\partial _\nu \mathcal{H}_\lambda -2m\mathcal{H}^\mu+  2m \partial^\mu h^{(s)} +\Box{}V^\mu
-\partial^\mu
\partial_\alpha V^\alpha=0
 \, \, , \label{t21}
\end{eqnarray}
donde $\mathcal{H}_\lambda \equiv
\partial_\alpha{{h^{(s)}}_\lambda }^\alpha$, es el objeto que
representa la propagación de espín 1 del campo autodual simétrico.
La relación (\ref{t20}) podría interpretarse, a primera vista como
que la propagación de espín cero del campo simétrico es eliminada
por la correspondiente al campo antisimétrico, $\partial_\mu V^\mu
$.

Por otro lado, con la ayuda de (\ref{t21}), el ''rotacional'' de
(\ref{t19}) conduce a
\begin{equation}
V_\sigma =0
 \, \, , \label{t22}
\end{equation}
lo cual, junto con (\ref{t20}) nos proporciona la relación
suplementaria $h^{(s)}=0$. Ahora, la ecuación (\ref{t19}) asegura
inmediatamente que
$\mathcal{H}_\lambda\equiv\partial_\alpha{{h^{(s)}}_\lambda
}^\alpha =0$, teniéndose una descripción completa y consistente de
una propagación de espín 2.

Lo anterior nos dice que la parte antisimétrica del campo autodual
juega el rol de eliminar la propagaci\'on de la parte de
esp\'{\i}n $1$, $\mathcal{H}_\lambda  $ de ${h^{(s)}}_{\mu\nu}$.
Si no hubiésemos considerado la existencia de $V_\mu$ desde el
principio, es decir que hubieramos partido con la acción
${S_{ad}}^{(2)}_{(V_\mu =0)}$, la ecuación de movimiento sería
\begin{eqnarray}
\epsilon ^{\mu \nu \lambda }
\partial _\mu{{h^{(s)}}_\nu }^\alpha +\epsilon ^{\mu \nu \alpha
}\partial _\mu{{h^{(s)}}_\nu }^\lambda -2m h^{(s)\lambda \alpha }+
2m \eta^{\lambda \alpha }h^{(s)}=0
 \, \, , \label{t23}
\end{eqnarray}
cuya traza nos indica correctamente que $h^{(s)}=0$ (no hay
propagación de espín cero), pero la divergencia de ésta toma la
forma
\begin{eqnarray}
\epsilon ^{\mu \nu \lambda }
\partial _\nu \mathcal{H}_\lambda -2m\mathcal{H}^\mu=0
 \, \, , \label{t24}
\end{eqnarray}
indicando que propaga espín 1 con masa $2|m|$, y por tanto no
habría una interpretación consistente para la propagación de un
espín 2 puro.

Así, si pensamos en una formulaci\'on m\'etrica para el esp\'{\i}n
$2$ autodual, necesitaremos la presencia de un campo auxiliar que
asegure la no propagaci\'on de la parte de esp\'{\i}n bajo
contenida en  ${h^{(s)}}_{\mu\nu}$. Este campo, en este caso puede
ser tomado como la parte antisim\'etrica de  $h_{\mu\nu}$ y, por
tanto la acci\'on puede ser escrita en la forma compacta
(\ref{t16}).

\vspace{1cm}

{\bf 2.1) La teoría autodual de espín 2 y su análisis
Lagrangiano} \vskip .1truein

Seguidamente revisaremos el an\'alisis de los v\'{\i}nculos
Lagrangianos de la acci\'on autodual de espín 2 \cite{r49},  cuya
acción es (\ref{t16}), con la finalidad de reafirmar que tiene el
espectro esperado. Debido a que esta teor\'{\i}a es de primer
orden, las ecuaciones de movimiento que surgen de la extremal de
${S_{ad}}^{(2)}$ constituyen los nueve v\'{\i}nculos Lagrangianos
primarios siguientes
\begin{eqnarray}
E^{\mu \rho }\equiv m\,\epsilon ^{\mu \nu \lambda }\partial _\nu
{h_\lambda}^\rho + m^2\,(\eta ^{\mu \rho }h - h^{\rho \mu })
\approx 0 \, \, . \label{tl18}
\end{eqnarray}
Obsérvese que  $E^{0 \rho }$ no posee derivadas temporales,
as\'{\i} que su preservaci\'on nos da tres v\'{\i}nculos
secundarios. La preservaci\'on de $E^{i \rho }$ proporciona las
aceleraciones $ {\ddot h}_{k \rho }$
\begin{eqnarray}
{\ddot h}_{k \rho } = \partial _k {\dot h}_{0 \rho } + m\epsilon
_{i k }({\delta ^i }_ \rho  {\dot h}- {{\dot h}_\rho }\,^i ) \, \,
, \label{tl19}
\end{eqnarray}
donde $\epsilon ^{0ij} \equiv \epsilon ^{ij} =\epsilon _{ij}$. La
preservaci\'on de $E^{0 \rho }$  puede verse en la capa de masas
como
\begin{eqnarray}
{\dot{E}}^{0 \rho }\approx \partial _\mu E^{\mu \rho } - m\epsilon
^{\rho \mu \alpha } E_{\mu \alpha  } = -m^3 \epsilon ^{\rho \mu
\alpha } h_{\mu \alpha } \equiv -m^2\,E^\rho \approx 0 \, \, ,
\label{tl20}
\end{eqnarray}
que expresa la propiedad de simetr\'{\i}a de $h_{\mu\nu}$.
Continuando con el procedimiento, ${\dot{E}}^\rho \approx 0$
proporciona tres nuevos v\'{\i}nculos
\begin{eqnarray}
{\dot{E}}^\rho=-m^3 \epsilon ^{\rho \mu \alpha } {\dot h}_{\mu
\alpha } \approx 0 \, \, , \label{tl21}
\end{eqnarray}
de \'estos, ${\dot{E}}^i$ relaciona ${\dot{h}}_{0i}$ con
${\dot{h}}_{i0}$. Si preservamos esta relaci\'on obtenemos la
ace-leraci\'on
\begin{eqnarray}
{\ddot h}_{0i} = {\ddot h}_{i0} =\partial _i {\dot h}_{00} +
m\epsilon _{i j }{\dot h}_{0j} \, \, . \label{tl22}
\end{eqnarray}
Para ${\dot{E}}^0$ podemos ver que en la capa de masas nos conduce
al v\'{\i}nculo
\begin{eqnarray}
{\dot{E}}^0\approx \partial _\mu E^\mu -{E_\mu}^\mu = -2m^4 h
\equiv -2m^4 h^{(s)} \approx 0 \, \, . \label{tl23}
\end{eqnarray}
Preservando \'este, obtenemos un nuevo v\'{\i}nculo que relaciona
${\dot{h}}_{00}$ con ${\dot{h}}_{ii}$. Su posterior
preservaci\'on, nos permite obtener la aceleraci\'on faltante
${\ddot{h}}_{00}$, o sea
\begin{eqnarray}
{\ddot h}_{00} = {\ddot h}_{ii} =\partial _i {\dot h}_{0 i } +
m\epsilon _{i j }{\dot h}_{ij} \, \, . \label{tl24}
\end{eqnarray}
As\'{\i} culmina el procedimiento del análisis de los vínculos
Lagrangianos. Se tienen entonces 16 v\'{\i}nculos Lagrangianos
representados por $E^{\mu \rho }$, $E^\rho$, ${\dot{E}}^\rho$ y
${\dot{h}}^{(s)}$, indicando la existencia de solamente una
excitaci\'on en esta teor\'{\i}a. Es f\'acil ver que el sistema de
v\'{\i}nculos Lagrangianos describe un campos sim\'etrico,
transverso y sin traza que satisface la ecuaci\'on $(\Box{} -
m^2){h^{(s)Tt}}_{\mu \nu}=0$, donde $\Box{} \equiv
\partial _{\mu}\partial ^{\mu}=-\partial _0\partial _0 + \Delta$.

Si el campo autodual es descompuesto como
\begin{eqnarray}
h_{\mu \nu }\equiv { h^{(s)}}_{\mu \nu }+\epsilon _{\mu \nu
\lambda }V^{\lambda} \, \, , \label{tl25}
\end{eqnarray}
podemos tomar $V^{\lambda}=0$, ${\dot V}^{\lambda}=0$ dej\'andonos
diez v\'{\i}nculos
\begin{eqnarray}
h^{(s)}=0\, \, , \label{tl26}
\end{eqnarray}
\begin{eqnarray}
{\dot h}^{(s)}=0\, \, , \label{tl27}
\end{eqnarray}
\begin{eqnarray}
\partial _\mu {{h^{(s)}}^\mu}_\nu=0\, \, ,
\label{tl28}
\end{eqnarray}
\begin{eqnarray}
m{h^{(s)}}_{ii} + \epsilon _{i j }\partial _i {h^{(s)}}_{0 j }
=0\, \, , \label{tl29}
\end{eqnarray}
\begin{eqnarray}
m{h^{(s)}}_{0i} + \epsilon _{kl }\partial _k {h^{(s)}}_{li } =0\,
\, , \label{tl30}
\end{eqnarray}
\begin{eqnarray}
{{\dot h}^{(s)}}\,_{i j }=\frac12 \,\delta _{i j } \,{\dot
h^{(s)}}\,_{k k }= \frac{m}2\,(\epsilon _{ik}{h^{(s)}}_{kj}
+\epsilon _{jk}{h^{(s)}}_{ki})+\frac12\,(\partial _i
{h^{(s)}}_{0j}+\nonumber \\+\,\partial _j {h^{(s)}}_{0i}- \delta
_{i j }\partial _k {h^{(s)}}_{0k})\, \, , \label{tl31}
\end{eqnarray}
que pueden ser resueltos dando
\begin{eqnarray}
{h^{(s)}}_{00}=\frac{(-\Delta)^2}{m} \,h^{TT}\, \, , \label{tl32}
\end{eqnarray}
\begin{eqnarray}
{h^{(s)}}_{0i}=-\frac{(-\Delta)}{m} \,(\epsilon _{ij}\partial _j
h^{TT}+\frac{\partial _i}{m}\,{\dot h}^{TT})\, \, , \label{tl33}
\end{eqnarray}
\begin{eqnarray}
{h^{(s)}}_{ij}= \epsilon _{ik}\epsilon _{jl}\partial _k\partial
_lh^{TT} -\frac{(-\Delta +m^2)}{m^2} \,\partial _i\partial _j
h^{TT}+\frac1m\,(\epsilon _{ik}\partial _k\partial _j+\epsilon
_{jk}\partial _k\partial _i){\dot h}^{TT}\, \, , \label{tl34}
\end{eqnarray}
en t\'erminos de \'unicamente el campo $h^{TT}$ y su velocidad
${\dot h}^{TT}$. La aceleraci\'on de este campo muestra el
car\'acter masivo de la excitaci\'on
\begin{eqnarray}
{\ddot h}^{TT}=-(-\Delta +m^2)h^{TT}\, \, . \label{tl35}
\end{eqnarray}
Si extraemos las partes  ''+'' y ''$-$'' de este tensor
sim\'etrico, transverso y sin traza mediante
\begin{eqnarray}
{h^{(s)Tt\pm }}_{\mu\nu }=\frac12\,\big(\frac{{\Box{}}^{\frac12}
\pm m}{{\Box{}}^{\frac12}} \big){h^{(s)}}_{\mu\nu }\, \, ,
\label{tl36}
\end{eqnarray}
entonces, en la capa de masas (${\Box{}}^{\frac12} \sim m$)
obtenemos
\begin{eqnarray}
{h^{(s)Tt+ }}_{\mu\nu }={h^{(s)}}_{\mu\nu }\,,\,\,\,\,{h^{(s)Tt-
}}_{\mu\nu }=0 \, \, . \label{tl37}
\end{eqnarray}
Si cambiamos $m$ por $-m$ en la acci\'on original, los roles de
${h^{(s)Tt+}}_{\mu\nu }$ and ${h^{(s)Tt-}}_{\mu\nu }$ se
intercambian. De hecho el esp\'{\i}n de la excitaci\'on (como
 mostraremos en la sección $\S$ 2.3), es $s=2\frac{m}{|m|}$.

\vspace{1cm}

{\bf 2.2) La acción reducida } \vskip .1truein

Es bien sabido, que la formulación Hamiltoniana que da pie al
procedimiento canónico de cuantización de Dirac pasa por la
introducción de la acción de la teoría, que escribimos de una
manera genérica como
\begin{eqnarray}
S= \int dt \,L(\phi^n,\dot{\phi}^n)
 \, \,\,\, ,
\label{are1}
\end{eqnarray}
donde $L(\phi^n,\dot{\phi}^n)$ es el lagrangiano que toma valores
sobre las coordenadas gene-ralizadas $\phi^n$, $n=1,...,N$, las
cuales eventualmente representarán campos si pensamos en el índice
$n$ como un índice compuesto de una parte discreta y otra
contínua. En (\ref{are1}) convenimos $\dot{\phi}^n \equiv
\frac{d\phi^n }{dt}$.

Las ecuaciones de movimiento que se derivan de la extremal de
(\ref{are1}) son
\begin{eqnarray}
\frac{\partial L}{\partial \phi^n} -\frac{dP_n }{dt}=0
 \, \,\,\, ,
\label{are2}
\end{eqnarray}
donde $P_n \equiv \frac{\partial L}{\partial \dot{\phi}^n}$ es el
momento canónico conjugado.

El espacio de fase $2N$ dimensional tiene coordenadas generales
$x^A$, $A=1,...,2N$, y canónicas
\begin{eqnarray}
(..., {x_{(c)}}^A,...)=(\phi^1,...,\phi^m,...,\phi^N,
P_1,...,P_m,...,P_N)
 \, \,\,\, ,
\label{are3}
\end{eqnarray}
con las cuales se define la estructura de corchetes de Poisson
\begin{eqnarray}
\{ F,G\}=\frac{\partial F}{\partial \phi^n}\frac{\partial
G}{\partial P_n}-\frac{\partial G}{\partial \phi^n}\frac{\partial
F}{\partial P_n}
 \, \,\,\, ,
\label{are4}
\end{eqnarray}
donde $F$ y $G$ son objetos físicos que toman valores sobre las
coordenadas de la variedad de fase. Los corchetes de Poisson
satisfacen las conocidas relaciones algebraicas
\begin{eqnarray}
\{ F,G\}=-\{ G,F\}
 \, \,\,\, ,
\label{are5}
\end{eqnarray}
\begin{eqnarray}
\{ F,GB\}=\{ F,G\}B+G\{ F,B\}
 \, \,\,\, ,
\label{are6}
\end{eqnarray}
\begin{eqnarray}
\{ F,\{ G,B\}\}+\{ G,\{ B,F\}\}+\{ B,\{ F,G\}\}=0
 \, \,\,\, .
\label{are7}
\end{eqnarray}

Con esta estructura es posible establecer la dinámica de un objeto
físico, entendiendo al Hamiltoniano canónico, ${H^{(c)}}_{(\phi, P
)}=P_n \,\dot{\phi}^n -L(\phi^n,\dot{\phi}^n)$ como el generador
de translaciones temporales, es decir $\dot{A}=\{A,H^{(c)}\}$.
Esto es realizable de manera consistente con las ecuaciones de
Hamilton
\begin{eqnarray}
\dot{\phi}^n=\{\phi^n,H^{(c)}\}=\frac{\partial H^{(c)}}{\partial
P_n}
 \, \,\,\, ,
\label{are8}
\end{eqnarray}
\begin{eqnarray}
\dot{P}_n=\{P_n,H^{(c)}\}=-\frac{\partial H^{(c)}}{\partial
\phi^n}
 \, \,\,\, ,
\label{are9}
\end{eqnarray}
si implícitamente estamos pensando en un Lagrangiano no singular,
donde sea posible despejar todas las velocidades $\dot{\phi}^n$ en
términos de las coordenadas y sus momentos conjugados, a partir de
la relación $P_n = \frac{\partial L}{\partial \dot{\phi}_n}$. Aquí
surge el obstáculo principal de la formulación Hamiltoniana: en
general no es posible realizar esto último, razón por la cual $P_n
= \frac{\partial L}{\partial \dot{\phi}^n}$ puede conducir a
relaciones entre coordenadas y momentos, las cuales reciben el
nombre de {\it vínculos primarios}.

De manera general, un Lagrangiano es identificado como {\it
singular}, verificando la nulidad del determinante de la matriz
$\frac{\partial^2 L}{\partial \dot{\phi}^n \partial
\dot{\phi}^m}$. Esto permite escribir los vínculos primarios de la
forma ${\gamma ^{(1)}}_k(\phi,P)\approx 0$, $k=1,...,K$, y a
partir de éstos extendemos el Hamiltoniano como
\begin{eqnarray}
\widetilde{H}=H^{(c)}+u_k{\gamma ^{(1)}}_k
 \, \,\,\, ,
\label{are10}
\end{eqnarray}
con el cual reestablecemos la dinámica en términos de la
estructura de Poisson tomando en cuenta los vínculos, es decir
\begin{eqnarray}
\dot{A}=\{A,\widetilde{H}\}|_{{\gamma ^{(1)}}_k\approx 0}
 \, \,\,\, .
\label{are11}
\end{eqnarray}
Pero los vínculos primarios, como objetos físicos conservados
deben ser consistentes con la dinámica establecida, lo que puede
conducir a la aparición de los {\it vínculos secundarios},
${\gamma ^{(2)}}_l(\phi,P)\approx 0$, $l=K+1,...,M$. Éstos, como
los primarios son sujetos a la consistencia con la dinámica,
dándose a lugar todo el proceso de la formulación canónica, muy
bien conocido.

Para el procedimiento de Dirac, la distinción fundamental entre
los vínculos no es la de primarios o secundarios, sino cuando
estos se agrupan en vínculos de {\it primera clase} y de {\it
segunda clase}. Esto es, los vínculos que ''conmutan'' y los que
''no conmutan'', según la estructura de Poisson.

De existir únicamente vínculos de primera clase es posible escoger
el camino de la cuantización con estados etiquetados por
representantes de las clases de equivalencia (o módulo
transformaciones de calibre) de los campos. Pero, eventualmente el
análisis de vínculos puede ser cambiado al de un conjunto puro de
 segunda clase, los cuales denotaremos como
$\chi_{\alpha}\approx0$, mediante fijaciones de calibre que
corresponden a la introducción de nuevos vínculos. Es de
observarse que con los vínculos de segunda clase es posible
construir una matriz no singular, $C_{\alpha\beta}\equiv
\{\chi_{\alpha},\chi_{\beta}\}$,  a partir de la cual son
introducidos los corchetes de Dirac de dos objetos físicos
\begin{eqnarray}
\{A,B\}_D=\{A,B\}-\{A,\chi_{\alpha}\}{C^{-1}}_{\alpha\beta}\{\chi_{\beta},B\}
 \, \,\,\, .
\label{are12}
\end{eqnarray}
Estos corchetes permiten obtener finalmente las reglas de
conmutación correctas para los vínculos de la teoría, es decir
$\{A,\chi_{\alpha}\}_D=0$, preparando el paso a una teoría
cuántica consistente.

Son numerosos los ejemplos en los que el procedimiento de
cuantización canónica de Dirac conlleva a un proceso tedioso,
dentro de los cuales, la teoría autodual de espín 2, no es la
excepción \cite{r50}. Sin embargo, existe una aproximación al
pro-blema de la obtención del álgebra de corchetes de Dirac (e
inmediatamente de los conmutadores de los operadores
mecánico-cuánticos), que, hablando superficialmente parte de la
obtención de la acción reducida de la teoría. En nuestro problema
de interés (la teoría autodual), ésto significa ''reducir'' la
teoría a una acción correspondiente a un solo grado de libertad.

La posibilidad de realizar este tipo de procedimiento está
garantizada por el teorema de Maskawa y Nakajima \cite{r51} que
asegura la existencia de una transformación canónica de las
coordenadas $x^A$, de tal manera que los corchetes de Dirac
definidos en la variedad de los puntos del espacio de fase son
iguales a los corchetes de Poisson ({\it inducidos}) de la
subvariedad de las variables reducidas sin vínculos.

Ahora bien, la posibilidad de reducir el problema al de la
subvariedad del espacio de fase o, como llamaremos {\it superficie
de vínculos}, es un hecho intrínsecamente asociado a la geometría
simpléctica del espacio de fase. Seguidamente, revisaremos a
rasgos generales algunos de los elementos de esta geometría
\cite{r52} que justifican la afirmación anterior, así como la
introducción del concepto de acción reducida.

Además de considerarse que el espacio de fase está dotado con
coordenadas $x^A$ y  una estructura de Poisson
(\ref{are4})-(\ref{are7}), se introduce un tensor antisimétrico no
singular de rango 2, llamado {\it 2-forma simpléctica}
\begin{eqnarray}
\sigma^{AB}\equiv \{ x^A,x^B\}
 \, \,\,\, ,
\label{are13}
\end{eqnarray}
con las siguientes propiedades
\begin{eqnarray}
\det (\sigma^{AB})\neq 0
 \, \,\,\, \, \,\,\, \, \,\,\, (invertible),
\label{are14}
\end{eqnarray}
\begin{eqnarray}
\partial_C\sigma_{AB}+\partial_A\sigma_{BC}+\partial_B\sigma_{CA}= 0
 \, \,\,\, \, \,\,\, \, \,\,\, (forma\,\,cerrada),
\label{are15}
\end{eqnarray}
con $\sigma^{AB}\sigma_{BC}={\delta^A}_C$. Si $F$ y $G$ toman
valores en las coordenadas del espacio de fase, podemos reescribir
(\ref{are4}) de la forma
\begin{eqnarray}
\{ F,G\}=\sigma^{AB}\frac{\partial F}{\partial x^A}\frac{\partial
G}{\partial x^B}
 \, \,\,\, .
\label{are16}
\end{eqnarray}

Seguidamente, consideramos secciones del espacio de fase,
aludiendo a las superficies de vínculos (de primera o segunda
clase) que ocurren tras imponer los vínculos. A estas
subvariedades se les puede proporcionar una {\it 2-forma
inducida}, $\sigma_{ij}$ debido a $\sigma_{AB}$.
 Considerando que tales superficies de vínculos poseen ecuaciones
paramétricas denotadas por $x^A=x^A(y^i)$, con $y^i$
($i=1,...,2N-M$) las coordenadas de la superficie de vínculos, la
2-forma inducida es
\begin{eqnarray}
\sigma_{ij}=\sigma_{AB}\frac{\partial x^A}{\partial
y^i}\frac{\partial x^B}{\partial y^j}
 \, \,\,\, ,
\label{are17}
\end{eqnarray}
la cual hereda la propiedad de antisimetría y la identidad de
Bianchi (p.ej.: forma cerrada).

A diferencia de $\sigma^{AB}$, la definición (\ref{are17}) no
garantiza que $\det (\sigma_{ij})\neq 0$, razón por la cual en
general la 2-forma inducida no sea invertible. Esto representa un
serio obstáculo para poder definir la {\it estructura de corchetes
de Poisson inducida}, que tiene la forma
\begin{eqnarray}
\{f,g\}^*=\sigma^{ij}\frac{\partial f}{\partial y^i}\frac{\partial
g}{\partial y^j}
 \, \,\,\, ,
\label{are18}
\end{eqnarray}
donde $f$ y $g$ toman valores sobre los puntos de la superficie de
vínculos.

Esta situación ocurre  (como ejemplo extremo) en el caso de una
superficie puramente de vínculos de primera clase, $\gamma _a
\approx 0$ (con, $\{\gamma_a,\gamma_b\}\approx 0$, $a,b=1,...,M$),
y que pensaremos por simplicidad que éstos son independientes
(caso irreducible). Entendiendo que los vínculos de primera clase
revelan el carácter invariante de calibre de la teoría, es
ampliamente demostrado \cite{r52} que la 2-forma inducida es
singular (de hecho maximalmente degenerada) ya que existen $M$
vectores linealmente independientes definidos con las funciones
vínculos de la forma
\begin{eqnarray}
{X^A}_a\equiv \sigma^{AB}\partial_B\gamma_a
 \, \,\,\, ,
\label{are19}
\end{eqnarray}
que son vectores nulos de la 2-forma simpléctica, pues
\begin{eqnarray}
0\approx \{\gamma_a,\gamma_b\} =
\sigma^{AB}\partial_A\gamma_a\partial_B\gamma_b=\sigma_{AB}{X^A}_a{X^B}_b
 \, \,\,\, .
\label{are20}
\end{eqnarray}
Los vectores ${X^A}_a$ también reciben el nombre de {\it vectores
Hamiltonianos}.

Considerando (\ref{are16}) y (\ref{are19}), y sea $F$ una función
de las coordenadas del espacio de fase, se tiene
\begin{eqnarray}
\partial_aF \equiv {X^A}_a\partial_AF=\{F,\gamma_a\}
 \, \,\,\, ,
\label{are21}
\end{eqnarray}
indicando que los $M$ vectores ${X^A}_a$ generan la
transformaciones infinitesimales de calibre. Esto ayuda a mostrar
que estos vectores son tangentes a la subvariedad de la superficie
de vínculos primarios:
\begin{eqnarray}
{X^A}_a\partial_A\gamma_b=\{\gamma_b,\gamma_a\}\approx0
 \, \,\,\, ,
\label{are22}
\end{eqnarray}
y más aún, con la condición de integrabilidad de Frobenius (lema
del teorema de Frobenius), esto es ''$\{{X^A}_a,{X^B}_b\}=${\it
combinación de vectores Hamiltonianos en la superficie de
vínculos}'', se garantiza que los vectores ${X^A}_a$ generan una
subvariedad $M$-dimensional, o superficie nula (llamada también
así, pues  éstos son vectores nulos de la 2-forma simpléctica).
Debido a que los vectores ${X^A}_a$ generan las transformaciones
de calibre via (\ref{are21}), éstas superficies también son
llamadas {\it órbitas de calibre}. Entonces, la forma de recuperar
la invertibilidad de $\sigma_{ij}$ en el contexto de una teoría
con vínculos de primera clase, y con vistas a obtener una
estructura de Poisson inducida bien definida, es la de definir un
{\it espacio de fase reducido} como el espacio cociente entre la
superficie de vínculos primarios y las órbitas de calibre (p.ej.:
identificación de los puntos de una órbita de calibre).

En el caso opuesto, es decir cuando el sistema de vínculos es
puramente de segunda clase, $\chi_{\alpha}\approx0$, $\alpha =
1,...,M$, la discusión es más simple ya que no hay órbitas de
calibre. En este tipo de sistemas la invertibilidad de la 2-forma
inducida está garantizada ya que los vectores Hamiltonianos
definidos con las funciones vínculos
\begin{eqnarray}
{X^A}_{\alpha}\equiv \sigma^{AB}\partial_B\chi_{\alpha}
 \, \,\,\, ,
\label{are23}
\end{eqnarray}
no son vectores nulos de la 2-forma simpléctica, pues
\begin{eqnarray}
\sigma_{AB}{X^A}_{\alpha}{X^B}_{\beta}=
\sigma^{AB}\partial_A\chi_{\alpha}\partial_B\chi_{\beta}=
\{\chi_{\alpha},\chi_{\beta}\}\neq 0
 \, \,\,\, ,
\label{are24}
\end{eqnarray}
por lo cual, la 2-forma inducida es no degenerada. Así, la
estructura de \,Poisson inducida está bien definida en la
superficie de vínculos secundarios (pudiese haber un caso mixto,
es decir con vínculos de  primera y segunda clase donde el
\,procedimiento consiste en reducir el espacio de fase con las
órbitas de calibre correspondientes a los vínculos de primera
clase, quedando ahora un problema de vínculos de segunda clase
solamente).

Para finalizar esta muy breve revisión de ideas, podemos resaltar
que existen dos aplicaciones inmediatas de lo anteriormente
discutido, las cuales utilizaremos en buena parte de este trabajo,
enfocando nuestra atención  en sistemas de segunda clase. Por un
lado, se puede probar un teorema \cite{r52},\cite{r53} que
establece:\,''{\it Los corchetes de Dirac asociados a los vínculos
de segunda clase, $\chi_{\alpha}\approx 0 $ son iguales a los
corchetes de Poisson inducidos en la superficie de  estos vínculos
\begin{eqnarray}
\{F,G\}_D\mid_{\chi_{\alpha}\approx 0}=\{f,g\}^*
 \, \,\,\, ,
\label{are25}
\end{eqnarray}
donde $F\mid_{\chi_{\alpha}\approx 0}=f$ y
$G\mid_{\chi_{\alpha}\approx 0}=g$}\,''. Lo cual es un hecho
sumamente poderoso a la hora de calcular (de manera indirecta) los
corchetes de Dirac.

Por otro lado, si la acción de la teoría con vínculos de segunda
clase es escrita como
\begin{eqnarray}
S= \int dt \,(P_n\,\dot{\phi}^n -H^{(c)}-u^{\alpha}\chi_{\alpha})
 \, \,\,\, ,
\label{are26}
\end{eqnarray}
es posible evaluarla sobre la superficie de los vínculos
$\chi_{\alpha}\approx 0$, con las ecuaciones paramétricas
$x^A=x^A(y^i)$ y  $x^A\equiv(\phi^n,P_n)$. Para esto se escoge la
1-forma
\begin{eqnarray}
\rho _A\equiv (P_n,0)  \, \,\,\,, \label{are27}
\end{eqnarray}
con lo cual, la parte cinética de (\ref{are26}) es reescrita como
\begin{eqnarray}
P_n\dot{\phi}^n =\rho _A (x) \dot{x}^A  \, \,\,\,. \label{are27a}
\end{eqnarray}
La 1-forma inducida es entonces
\begin{eqnarray}
\rho _i (y)=\frac{\partial x^A}{\partial y^i}\,\rho
_A(x(y))\equiv\frac{\partial x^n}{\partial y^i}\,P_n(x(y)) \,
\,\,\,. \label{are27b}
\end{eqnarray}
De esto sigue una expresión para la parte cinética en términos de
la 1-forma inducida
\begin{eqnarray}
\rho _i \dot{y}^i =\frac{\partial x^n}{\partial y^i}\,P_n\,
\dot{y}^i =P_n\, \dot{x}^n\equiv P_n\, \dot{\phi}^n\, \,\,\,.
\label{are27c}
\end{eqnarray}
Si además consideramos
\begin{eqnarray}
H^{(c)}\mid_{\chi_{\alpha}= 0} =h(y)\, \,\,\,, \label{are28}
\end{eqnarray}
entonces, esta expresión junto con (\ref{are27c}) nos permiten
decir
\begin{eqnarray}
S[y(t)]\equiv S\mid_{\chi_{\alpha}= 0}= \int dt \,(\rho
_i\dot{y}^i -h(y))
 \, \,\,\, .
\label{are29}
\end{eqnarray}

Seguidamente, examinamos la variación de $S\mid_{\chi_{\alpha}=
0}$ en las coordenadas $y^i$, obteni\'endose
\begin{eqnarray}
\delta_y S[y(t)]= \int dt \,\bigg(\big(\frac{\partial \rho
_j}{\partial y^i}-\frac{\partial \rho_i }{\partial
y^j}\big)\dot{y}^j -\frac{\partial h}{\partial y^i} \bigg)\delta
y^i
 \, \,\,\, ,
\label{are30}
\end{eqnarray}
a menos de un término de borde. Por el lema de Poincaré podemos
pensar  $\frac{\partial \rho _j}{\partial y^i}-\frac{\partial
\rho_i }{\partial y^j}$ como una 2-forma local, con lo cual
decimos que si la 2-forma inducida tiene a $\rho_j$ como la
1-forma de potencial, entonces se propone
\begin{eqnarray}
\sigma_{ij}=\frac{\partial \rho _j}{\partial y^i}-\frac{\partial
\rho_i }{\partial y^j}
 \, \,\,\, .
\label{are31}
\end{eqnarray}
Con esta prescripción local, reescribimos la variación
(\ref{are30}) como
\begin{eqnarray}
\delta_y S[y(t)]= \int dt \,\sigma_{ij}\big( \dot{y}^j -
\{y^j,h\}^* \big)\delta y^i
 \, \,\,\, ,
\label{are32}
\end{eqnarray}
con la ayuda de la definición de los corchetes de Poisson
inducidos (\ref{are18}). Inmedia-tamente podemos ver que la
extremal de $S[y(t)]$ conduce a
\begin{eqnarray}
\dot{y}^j = \{y^j,h\}^*
 \, \,\,\, ,
\label{are33}
\end{eqnarray}
que son las ecuaciones de Hamilton sin vínculos en la superficie
$\chi_{\alpha}\approx 0$. Con esto se dice que hemos resuelto los
vínculos dentro de la acción y la expresión (\ref{are29}) recibirá
el nombre de acción reducida.

\vspace{1cm}

{\bf 2.2.1) El \'algebra de operadores en la teoría autodual} \vskip
.1truein

Con vistas a obtener la acción reducida de la teoría autodual de
espín 2, se puede comenzar  mirando la descomposici\'on $2+1$ de
la acci\'on de la teor\'{\i}a del campo autodual de esp\'{\i}n 2,
ec.(\ref{t16}) en un espacio plano
\begin{eqnarray}
{S_{ad}}^{(2)}=\frac m2 \int d^3 x \{-\epsilon _{ij}\dot{N}_i N_j
+{\dot{h}^{(s)}}\,_{ik}(\epsilon _{ij}{h^{(s)}}_{jk}-\delta
_{ik}V)+\dot{V}{h^{(s)}}_{kk}+\nonumber
\\+\,m({h^{(s)}}^2-{h^{(s)}}_{ij}{h^{(s)}}_{ij}+2V^2)-2n(m{h^{(s)}}_{kk}+\epsilon
_{ij}\partial _i N_j)+\nonumber
\\+\,2M_k(\epsilon _{ij}\partial _i
{h^{(s)}}_{jk}-\partial _k V +mN_k)\}
 \, \, ,
\label{a38}
\end{eqnarray}
donde hemos hecho uso de la notaci\'on \cite{r54}
\begin{eqnarray}
n\equiv h_{00} \, \, , \label{a39}
\end{eqnarray}
\begin{eqnarray}
N_i\equiv h_{i0} \, \, , \label{a40}
\end{eqnarray}
\begin{eqnarray}
M_i\equiv h_{0i} \, \, , \label{a41}
\end{eqnarray}
\begin{eqnarray}
{h^{(s)}}_{ij}\equiv \frac{1}{2}(h_{ij}+h_{ji}) \, \, ,
\label{a42}
\end{eqnarray}
\begin{eqnarray}
V\equiv \frac{1}{2}\,\epsilon _{ij}h_{ij} \, \, . \label{a43}
\end{eqnarray}
Seguidamente, realizamos la descomposici\'on
transverso-longitudinal (TL) me-diante
\begin{equation}
N_i \equiv \epsilon _{ik}\partial _k N^T + \partial _i N^L \, \, ,
\label{a44}
\end{equation}
\begin{equation}
M_i \equiv \epsilon _{ik}\partial _k M^T + \partial _i M^L \, \, ,
\label{a45}
\end{equation}
\begin{equation}
{h^{(s)}}_{ij} \equiv (\delta _{ij}\Delta -\partial _i \partial
_j){h^{(s)TT}} +\partial _i \partial _j {h^{(s)LL}} +( \epsilon
_{ik}\partial _k
\partial _j +\epsilon _{jk}\partial _k \partial _i ){h^{(s)TL}} \, \,,
\label{a46}
\end{equation}
con la cual reescribimos la acci\'on (\ref{a38}) como
\begin{eqnarray}
{S_{ad}}^{(2)}=m \int d^3 x \{\dot{N}^T \Delta {N}^L+  (\Delta
{h^{(s)LL}} -\Delta {h^{(s)TT}} )\Delta \dot{h}^{(s)TL} -V\Delta
\dot{h}^{(s)LL}+\nonumber
\\ -\,V\Delta \dot{h}^{(s)TT} +m[\Delta {h^{(s)TT}}
\Delta {h^{(s)LL}}-(\Delta {h^{(s)TL}})^2 +V^2]+\nonumber
\\+\,
n[\Delta {N}^T-m\Delta {h^{(s)LL}} -m\Delta
{h^{(s)TT}}]+M^T[{\Delta }^2 {h^{(s)TT}}-m\Delta {N}^T ]+\nonumber
\\
+\,M^L[{\Delta }^2 {h^{(s)TL}}+\Delta V -m\Delta {N}^L ]  \}
 \, \, ,
\label{a47}
\end{eqnarray}
donde los campos $n$, $M^T$ y $M^L$ aparecen como multiplicadores
de Lagrange asociados a los vínculos
\begin{eqnarray}
\Delta {N}^T-m\Delta {h^{(s)LL}} -m\Delta {h^{(s)TT}}=0
 \, \, ,
\label{a48}
\end{eqnarray}
\begin{eqnarray}
{\Delta }^2 {h^{(s)TT}}-m\Delta {N}^T =0
 \, \, ,
\label{a49}
\end{eqnarray}
\begin{eqnarray}
{\Delta }^2 {h^{(s)TL}}+\Delta V -m\Delta {N}^L=0
 \, \, .
\label{a50}
\end{eqnarray}

Con la finalidad de obtener la acci\'on reducida, uno puede
comenzar restringi\'endose al espacio f\'{\i}sico de los
v\'{\i}nculos (\ref{a48})-(\ref{a50}), lo cual permite escribir la
acci\'on en t\'erminos de los campos ${h^{(s)TT}}$, ${h^{(s)TL}}$
y $V$, es decir
\begin{eqnarray}
{S_{ad}}^{(2)*}=\int d^3 x \{2m{\Delta }^2
{h^{(s)TL}}\dot{h}^{(s)TT} + \Delta {h^{(s)TT}}(\Delta -m^2
)\Delta {h^{(s)TT}}+\nonumber \\-\,m^2(\Delta {h^{(s)TL}})^2 +mV^2
\}
 \, \, ,
\label{a51}
\end{eqnarray}
de la cual puede observarse que el campo $V$ no propaga en el
espacio f\'{\i}sico (p.ej., su ecuaci\'on es $V=0$). Con esto, e
introduciendo la notaci\'on
\begin{eqnarray}
P\equiv \sqrt 2 \,m \Delta {h^{(s)TL}}
 \, \, ,
\label{a52}
\end{eqnarray}
\begin{eqnarray}
Q\equiv \sqrt 2 \, \Delta {h^{(s)TT}}
 \, \, ,
\label{a53}
\end{eqnarray}
se observa que la acci\'on reducida posee la forma can\'onica
\begin{eqnarray}
{S_{ad}}^{(2)*}=\int d^3 x \{P\dot{Q}-\frac12 P^2 - \frac12
\,Q(-\Delta +m^2)Q\}
 \, \, ,
\label{a54}
\end{eqnarray}
mostrando claramente que la teor\'{\i}a propaga un solo grado de
libertad y la energía es positiva definida.

Esto nos permite considerar, en primer lugar que todos los campos
f\'{\i}sicos pueden ser expresados en t\'erminos de las variables
can\'onicas $Q$ y $P$. Para esto, mostramos las ecuaciones de
movimiento obtenidas a partir de la acci\'on (\ref{a47})
\begin{equation}
\frac{\delta {S_{ad}}^{(2)}}{\delta
{h^{(s)TT}}}=0\,\,\Longrightarrow \,\, \dot{V}=mn +\Delta
(\dot{h}^{(s)TL}-M^T-m{h^{(s)LL}})
 \, \, , \label{a55}
\end{equation}
\begin{equation}
\frac{\delta {S_{ad}}^{(2)}}{\delta
{h^{(s)LL}}}=0\,\,\Longrightarrow \,\, \dot{V}=mn -\Delta
(\dot{h}^{(s)TL}+m{h^{(s)TT}})
 \, \, , \label{a56}
\end{equation}
\begin{equation}
\frac{\delta {S_{ad}}^{(2)}}{\delta
{h^{(s)TL}}}=0\,\,\Longrightarrow \,\,
\dot{h}^{(s)TT}-\dot{h}^{(s)LL}+M^L-2m{h^{(s)TL}}=0
 \, \, , \label{a57}
\end{equation}
\begin{equation}
\frac{\delta {S_{ad}}^{(2)}}{\delta V}=0\,\,\Longrightarrow \,\,
V= \frac1{2m}\Delta (\dot{h}^{(s)TT}+\dot{h}^{(s)LL}-M^L)
 \, \, , \label{a58}
\end{equation}
\begin{equation}
\frac{\delta {S_{ad}}^{(2)}}{\delta n}=0\,\,\Longrightarrow \,\,
N^T=m({h^{(s)TT}}+{h^{(s)LL}})
 \, \, , \label{a59}
\end{equation}
\begin{equation}
\frac{\delta {S_{ad}}^{(2)}}{\delta M^T}=0\,\,\Longrightarrow \,\,
N^T=\frac1m \Delta {h^{(s)TT}}
 \, \, , \label{a60}
\end{equation}
\begin{equation}
\frac{\delta {S_{ad}}^{(2)}}{\delta M^L}=0\,\,\Longrightarrow \,\,
N^L= \frac1m V +\frac1m \Delta {h^{(s)TL}}
 \, \, , \label{a61}
\end{equation}
\begin{equation}
\frac{\delta {S_{ad}}^{(2)}}{\delta N^L}=0\,\,\Longrightarrow \,\,
\dot{N}^T=mM^L
 \, \, , \label{a62}
\end{equation}
\begin{equation}
\frac{\delta {S_{ad}}^{(2)}}{\delta N^T}=0\,\,\Longrightarrow \,\,
\dot{N}^L=n -mM^T
 \, \, , \label{a63}
\end{equation}
y usando las definiciones (\ref{a52}) y (\ref{a53}) obtenemos
\begin{equation}
V=0
 \, \, , \label{a64}
\end{equation}
\begin{eqnarray}
n=\frac{1}{\sqrt 2 \,m^2}\, \Delta Q
 \, \, ,
\label{a65}
\end{eqnarray}
\begin{eqnarray}
N^T=\frac{1}{\sqrt 2 \,m} \,Q
 \, \, ,
\label{a66}
\end{eqnarray}
\begin{eqnarray}
N^L=\frac{1}{\sqrt 2 \,m^2}\, P
 \, \, ,
\label{a67}
\end{eqnarray}
\begin{eqnarray}
M^T=\frac{1}{\sqrt 2 \,m}\, Q
 \, \, ,
\label{a68}
\end{eqnarray}
\begin{eqnarray}
M^L=\frac{1}{\sqrt 2 \,m^2}\, P
 \, \, ,
\label{a69}
\end{eqnarray}
\begin{eqnarray}
{h^{(s)LL}} = \frac1{\sqrt 2 \,m^2}(\Delta -m^2 ){\Delta}^{-1}Q
 \, \, .
\label{a70}
\end{eqnarray}

Con esto y la ayuda de (\ref{a39})-(\ref{a46}), todas las
componentes del campo auto-dual, $h_{\mu \nu}$ pueden ser escritas
en funci\'on de las variables $Q$ y $P$. La forma (\ref{a54}) de
la acci\'on reducida sugiere que las variables $Q$ y $P$ son
can\'onicamente conjugadas, raz\'on por la cual, toda vez que
promovamos la formulaci\'on cu\'antica de la teor\'{\i}a autodual,
postulamos la regla fundamental de conmutaci\'on entre operadores
\begin{eqnarray}
\big[Q(x),P(y) \big]_{t'=t}=i{\delta}^2(\vec{x}-\vec{y})
 \, \, .
\label{a71}
\end{eqnarray}

A partir de esta regla, los conmutadores a tiempos iguales no
nulos son
\begin{eqnarray}
\big[h_{i0}(x),h_{jk}(y) \big]
=\big[h_{0i}(x),h_{jk}(y)\big]=\frac{i}{2m^2}\{
{p^{(m)}}_{ij}\partial _k+{p^{(m)}}_{ik}
\partial _j+\nonumber \\-{p^{(m)}}_{jk}\partial _i \}{\delta}^2(\vec{x}-
\vec{y}) \, \, , \label{a72}
\end{eqnarray}
\begin{eqnarray}
\big[h_{i0}(x),h_{00}(y)
\big]=\big[h_{0i}(x),h_{00}(y)\big]=\frac{i}{2m^4}\,\partial
_i(-\Delta ){\delta}^2(\vec{x}-\vec{y})
 \, \, ,
\label{a73}
\end{eqnarray}
\begin{eqnarray}
\big[h_{i0}(x),h_{0k}(y)\big]=\big[h_{i0}(x),h_{k0}(y)\big]=\big[h_{0i}(x),h_{k0}(y)\big]
=\nonumber \\=\frac{i\,\epsilon _{ik}}{2m^3}\,(-\Delta )
{\delta}^2(\vec{x}-\vec{y})
  \, \, ,
\label{a74}
\end{eqnarray}
\begin{eqnarray}
\big[h_{00}(x),h_{ij}(y) \big]=\frac{i}{2m}\{ \epsilon
_{ki}\,{p^{(m)}}_{kj}+\epsilon _{kj}\,{p^{(m)}}_{ki}\}
{\delta}^2(\vec{x}-\vec{y})
 \, \, ,
\label{a75}
\end{eqnarray}
\begin{eqnarray}
\big[h_{ij}(x),h_{kl}(y)\big] =\frac{i}{4m}\{\epsilon
_{ik}\,{p^{(m)}}_{jl} +  \epsilon _{jk}\,{p^{(m)}}_{il}
 +  \epsilon _{il}\,{p^{(m)}}_{jk}+\nonumber \\+\epsilon
 _{jl}\,{p^{(m)}}_{ik}\}{\delta}^2(\vec{x}-\vec{y}) \, \, ,
\label{a76}
\end{eqnarray}
donde ${p^{(m)}}_{ij}= \delta _{ij} - \frac{\partial _i
\partial _j}{m^2}$ es un proyector transverso en la capa de masas
tiempo constante, $\Delta -m^2=0$. El \'algebra obtenida usando la
acci\'on reducida será la misma que se obtiene si se procede
siguiendo extrictamente el procedimiento de cuantizaci\'on de
Dirac debido a que los corchetes de Dirac en la variedad de los
v\'{\i}nculos coinciden con los de Poisson en el espacio de las
variables reducidas, según lo discutido en la sección 2.2.

\vspace{1cm}
{\bf 2.2.2) Separación de la parte transversal-sin traza de
${h^{(s)}}_{\mu \nu }$}\vskip .1truein

En la sección anterior estudiamos la construcción de la acción
reducida mediante la separación en partes transversales y
longitudinales de los campos, inmediatamente después de realizarse
la descomposición $2+1$ de la acción original. Ahora, queremos
mostrar la equivalencia de la construcción anterior con el
proce-dimiento equivalente, que se fundamenta en la extracción de
la parte tranversal-sin traza (Tt) del campo autodual simétrico,
que más adelante  llamaremos ${h^{(s)Tt}}_{\mu \nu }$.

Para esto, comenzamos por descomponer el campo $h_{\mu \nu }$ en
su parte sim\'etrica (${h^{(s)}}_{\mu \nu }$) y antisim\'etrica
($V^{\lambda}$) como en (\ref{tl25})
\begin{equation}
h_{\mu \nu }\equiv { h^{(s)}}_{\mu \nu }+\epsilon _{\mu \nu
\lambda }V^{\lambda}\, \, , \label{e77}
\end{equation}
y sustituyendo esta definici\'on en (\ref{t16}), se obtiene la ya
conocida (\ref{t17})
\begin{eqnarray}
{S_{ad}}^{(2)}=\frac m2 \int d^3 x \big(\,\epsilon ^{\mu \nu
\lambda }{{h^{(s)}}_\mu }^\alpha
\partial _\nu
{h^{(s)}}_{\lambda \alpha }-m({h^{(s)}}_{\mu \nu }h^{(s){\mu \nu
}}-{h^{(s)}}^2 )+\nonumber \\+\, 2V^\mu (\partial _\mu {h^{(s)}}
-\partial _\nu{{h^{(s)}}_\mu}^\nu)-\epsilon ^{\mu \nu \lambda
}V_\mu
\partial _\nu
V_\lambda -2mV_\mu V^\mu \big) \, \, . \label{e78}
\end{eqnarray}

Seguidamente, introducimos una  descomposici\'on en la que
aislamos la parte Tt de ${h^{(s)}}_{\mu \nu }$, y además
exhibiendo las componentes de espines bajos
\begin{equation}
{h^{(s)}}_{\mu \nu }\equiv {h^{(s)Tt}}_{\mu \nu }+\partial _\mu
{a^T}_\nu +\partial _\nu {a^T}_\mu + \partial _\mu\partial
_\nu\phi +\eta _{\mu \nu} \,\psi \, \, , \label{e79}
\end{equation}
con las condiciones suplementarias
\begin{equation}
{{h^{(s)Tt}}_\mu}^\mu =0 \, \, , \label{e80}
\end{equation}
\begin{equation}
\partial ^\mu{h^{(s)Tt}}_{\mu \nu }=0
\, \, , \label{e81}
\end{equation}
\begin{equation}
\partial _\mu {a^T}^\mu=0
\, \, . \label{e82}
\end{equation}
Adem\'as, tambi\'en descomponemos la parte de antisim\'etrica en
sus componentes de esp\'{\i}n 1 y 0
\begin{equation}
V_\mu\equiv {V^T}_\mu +\partial _\mu V  \, \, , \label{e83}
\end{equation}
con
\begin{equation}
\partial _\mu V^{T\mu}=0\, \, . \label{e84}
\end{equation}

Usando (\ref{e79}) y (\ref{e83}) en (\ref{e78}) se obtiene
\begin{eqnarray}
{S_{ad}}^{(2)}=\frac m2 \int d^3 x (\,\epsilon ^{\mu \nu \lambda
}{{h^{(s)Tt}}_\mu }^\alpha
\partial _\nu
{h^{(s)Tt}}_{\lambda \alpha }-m {h^{(s)Tt}}_{\mu \nu }h^{(s)Tt\mu
\nu }+ \nonumber \\ -\,\epsilon ^{\mu \nu \lambda }
\partial _\nu {a^T}_\lambda \Box{}{a^T}_\mu+2m{a^T}^{\mu}\Box{} {a^T}_{\mu}-2{V^T}^{\nu}
\Box{} {a^T}_{\nu}+ \nonumber \\-\,\epsilon ^{\mu \nu \lambda }
{V^T}_\mu \partial _\nu {V^T}_\lambda  -2m{V^T}^\mu {V^T}_\mu+ 2m
V\Box{}V+ \nonumber \\+\,4m \psi\Box{} \phi +6m \psi ^2
-4\psi\Box{} V) \, \, . \label{e85}
\end{eqnarray}

Ahora examinamos las ecuaciones de movimiento que resultan de la
extremal de la acción $\delta{S_{ad}}^{(2)}$ cuando realizamos
variaciones independientes en los campos ${h^{(s)Tt}}_{\lambda
\alpha }$, ${a^T}_{\nu}$, ${V^T}_{\nu}$, $V$, $\phi$ y $\psi$,
respectivamente
\begin{equation}
\epsilon ^{\mu \nu \lambda }
\partial _\mu{{h^{(s)Tt}}_\nu }^\alpha +\epsilon ^{\mu \nu \alpha
}\partial _\mu{{h^{(s)Tt}}_\nu }^\lambda -2m h^{(s)Tt\lambda
\alpha }=0
 \, \, , \label{e86}
\end{equation}
\begin{equation}
-\epsilon ^{\mu \nu \lambda }\partial _\mu\Box{} {a^T}_\nu+2m
\Box{}a^{T\lambda}- \Box{} V^{T\lambda}=0
 \, \, , \label{e87}
\end{equation}
\begin{equation}
\epsilon ^{\mu \nu \lambda }\partial _\mu {V^T}_\nu +2m
V^{T\lambda}+ \Box{} a^{T\lambda}=0
 \, \, , \label{e88}
\end{equation}
\begin{equation}
m\Box{}V-\Box{}\psi =0
 \, \, , \label{e89}
\end{equation}
\begin{equation}
\Box{}\psi=0
 \, \, , \label{e90}
\end{equation}
\begin{equation}
m\Box{}\phi-\Box{}V+3m\psi=0
 \, \, . \label{e91}
\end{equation}

Es de observarse, que alternativamente se podrían introducir la
función tensorial ${F_1}^\alpha$ y las funciones escalares, $F_2$
y $F_3$,  arbitrarias y derivables, de tal manera que las
ecuaciones (\ref{e86})-(\ref{e88}) fuesen reemplazadas por
\begin{equation}
\epsilon ^{\mu \nu \lambda }
\partial _\mu{{h^{(s)Tt}}_\nu }^\alpha +\epsilon ^{\mu \nu \alpha
}\partial _\mu{{h^{(s)Tt}}_\nu }^\lambda -2m h^{(s)Tt\lambda
\alpha }=\partial^\lambda{F_1}^\alpha+\partial^\alpha{F_1}^\lambda
 \, \, , \label{e86a}
\end{equation}
\begin{equation}
-\epsilon ^{\mu \nu \lambda }\partial _\mu\Box{} {a^T}_\nu+2m
\Box{}a^{T\lambda}- \Box{} V^{T\lambda}=\partial^\lambda F_2
 \, \, , \label{e87a}
\end{equation}
\begin{equation}
\epsilon ^{\mu \nu \lambda }\partial _\mu {V^T}_\nu +2m
V^{T\lambda}+ \Box{} a^{T\lambda}=\partial^\lambda F_3
 \, \, , \label{e88a}
\end{equation}
y además que los objetos introducidos deben satisfacer las
condiciones de consistencia
\begin{eqnarray}
\Box{}{F_1}^\lambda  =0\,\,\,\, ,\,\,\partial_\lambda{F_1}^\lambda
=0\,\,\,\, ,\label{ad1}
\end{eqnarray}
\begin{eqnarray}
\Box{} F_2=0\,\, ,\label{ad2}
\end{eqnarray}
\begin{eqnarray}
\Box{} F_3=0\,\, .\label{ad3}
\end{eqnarray}
Entonces, las relaciones (\ref{ad1})-(\ref{ad3}) nos sugieren que
puede recuperarse el sistema (\ref{e86})-(\ref{e88}) si removemos
las soluciones armónicas de las funciones introducidas. Así,
asumimos el sistema de ecuaciones (\ref{e86})-(\ref{e91}).

las ecuaciones (\ref{e89})-(\ref{e91}) pueden ser reescritas como
\begin{eqnarray}
3\psi+\Box{}\phi=0\,\, ,\label{e92}
\end{eqnarray}
\begin{eqnarray}
\Box{}V=0 \,\, ,\label{e93}
\end{eqnarray}
\begin{eqnarray}
\Box{}\psi =0\,\, .\label{e94}
\end{eqnarray}
La ecuación (\ref{e92}) indica que $h_{\mu \nu }$  no propaga
esp\'{\i}n 0, pues garantiza que ${{h^{(s)}}_\mu}^\mu =0$. Por
otro lado, (\ref{e93}) nos muestra que la parte antisimétrica del
campo auto-dual, $V_\mu$ no propaga esp\'{\i}n 0 masivo, de igual
manera con $\psi$. Con esto, si extraemos las soluciones
armónicas, tendremos
\begin{eqnarray}
\phi=0\,\, ,\label{e92a}
\end{eqnarray}
\begin{eqnarray}
V=0 \,\, ,\label{e93a}
\end{eqnarray}
\begin{eqnarray}
\psi =0\,\, .\label{e94a}
\end{eqnarray}

Las ecuaciones (\ref{e87}) y (\ref{e88}) se pueden desacoplar de
forma que
\begin{eqnarray}
{V^T}_\nu =0\, \, , \label{e95}
\end{eqnarray}
\begin{eqnarray}
\Box{}{a^T}_\nu=0\, \, , \label{e96}
\end{eqnarray}
y a menos de soluciones armónicas, tenemos
\begin{eqnarray}
{a^T}_\nu=0\, \, . \label{e96a}
\end{eqnarray}

Entonces, con la ayuda de  (\ref{e93a}) y (\ref{e95}) vemos que
(\ref{e83}) nos dice que la parte antisimétrica del campo autodual
no propaga, o sea
\begin{eqnarray}
V_\nu =0\, \, , \label{e95a}
\end{eqnarray}
estableciendo que $h_{\lambda \alpha}={h^{(s)}}_{\lambda \alpha
}$. Pero, más aún, las relaciones (\ref{e92a}), (\ref{e94a}) y
(\ref{e96a}) en (\ref{e79}) nos permiten asegurar
\begin{equation}
h_{\mu \nu }= {h^{(s)Tt}}_{\mu \nu } \, \, , \label{e95b}
\end{equation}
obteni\'endose una descripción consistente de una propagación de
espín 2 pura.

La extracción de armónicos ha sido un procedimiento expedito para
establecer la relación (\ref{e95b}), y podría enfocarse de manera
relacionada con el hecho de que para poder definir el operador
$\Box{}^{-1}$ (el cual pudiera haber sido introducido desde el
principio, a nivel de la descomposición (\ref{e79})), no deban
considerarse las soluciones en ondas planas con $p_\mu p^\mu=0$,
donde dicho operador no es regular. En nuestra discusión no hemos
seguido este camino, ya que necesitamos un proce-dimiento que
evite en la medida de lo posible la definición de potencias
distintas de la unidad del D'Alembertiano, teniendo ésto serios
inconvenientes en espacios no-Minkowskianos.

En resumidas cuentas, el procedimiento de separación de la parte
simétrico-tranverso-sin traza, aquí discutido está respaldado por
el hecho de que si hubiésemos partido de la acción decompuesta
según las partes simétrica-antisimétrica, ec.(\ref{e78}), se
hubiera encontrado que el espacio de fase reducido es aquél con
$V_\mu =0$ y el campo simétrico satisfaciendo las condiciones
suplementarias ${{h^{(s)}}_\mu}^\mu =0$ y $\partial
^\mu{h^{(s)}}_{\mu \nu }=0$.

Con todo esto, podemos decir que hemos conseguido una descripción
consistente del campo autodual que propaga espín 2 y que está
descrito por ${h^{(s)Tt}}_{\mu \nu }$. Entoces,  la ecuaci\'on
(\ref{e86}) que describe dinámicamente a este campo, la podemos
 reescribir como
\begin{equation}
\epsilon ^{\mu \nu \lambda }\partial _\mu{{h^{(s)Tt}}_\nu }^\alpha
 -m h^{(s)Tt\lambda \alpha }=0
 \, \, , \label{e97}
\end{equation}
gracias a la propiedad de  transversalidad. Partiendo de
(\ref{e97})  se puede obtener la forma hiperbólica-causal de tipo
''Klein-Gordon'', esto es
\begin{equation}
(\Box{}-m^2){h^{(s)Tt}}_{\mu \nu }=0
 \, \, . \label{ad9}
\end{equation}

De los dos campos libres representados por ${{h^{(s)Tt}}}_{ \mu
\nu}$, se puede mostrar que solo se propaga un grado de libertad.
Para esto, se puede retomar la definici\'on (\ref{tl36}) de las
partes ''+'' y ''$-$'' \cite{r48},\cite{r54}, reescrita como
\begin{equation}
{{{h^{(s)Tt}}}_{ \mu \nu}}^\pm \equiv \frac{1}{2}({\delta ^\alpha
}_\mu{\delta ^\beta}_\nu \pm {\delta ^\alpha }_\mu {\epsilon
_\nu}^{\sigma \beta }\frac{\partial _\sigma
}{{\Box{}}^{\frac{1}{2}}}){{h^{(s)Tt}}}_{ \alpha\beta}
 \, \, , \label{e98}
\end{equation}
y la ecuaci\'on (\ref{ad9}) en la capa de masas equivale a
\begin{equation}
(\Box{} -m^2){{{h^{(s)Tt}}}_{ \mu \nu}}^+=0
 \, \, , \label{e99}
\end{equation}
\begin{equation}
{{{h^{(s)Tt}}}_{ \mu \nu}}^- =0
 \, \, . \label{e100}
\end{equation}

As\'{\i}, con las condiciones ya establecidas,  podemos escribir
la acción reducida en términos exclusivos de la propagación de
espín 2, representada por la parte simétrica-transversa-sin traza
del campo autodual
\begin{eqnarray}
{S_{ad}}^{(2)*}=\frac{m}2\int d^3 x (\,\epsilon ^{\mu \nu \lambda
}{{h^{(s)Tt}}_\mu }^\alpha
\partial _\nu
{h^{(s)Tt}}_{\lambda \alpha }-m {h^{(s)Tt}}_{\mu \nu }h^{(s)Tt\mu
\nu })
 \, \, ,
\label{e101}
\end{eqnarray}
y su descomposici\'on $2+1$ es
\begin{eqnarray}
{S_{ad}}^{(2)*}=\frac m2 \int d^3 x \big(\,-2\epsilon
^{ij}{h^{(s)Tt}}_{00}
\partial _i
{h^{(s)Tt}}_{j0 } +2\epsilon ^{ij}{h^{(s)Tt}}_{k0}
\partial _i
{h^{(s)Tt}}_{jk }+ \nonumber \\+ \,\epsilon ^{ij}{h^{(s)Tt}}_{i0}
{{\dot{h}}^{(s)Tt}}\,_{j0}
 -\epsilon ^{ij}{h^{(s)Tt}}_{ik} {{\dot{h}}^{(s)Tt}}\,_{jk} -m
({h^{(s)Tt}}_{00})^2 + \nonumber \\+\,2m {h^{(s)Tt}}_{0k
}{h^{(s)Tt}}_{0k} -m {h^{(s)Tt}}_{ij }{h^{(s)Tt}}_{ij} \big)
 \, \, .
\label{e102}
\end{eqnarray}

Seguidamente introducimos una forma general TL para las diferentes
componentes de ${{h^{(s)Tt}}}_{ \mu \nu}$
\begin{equation}
{{h^{(s)Tt}}}_{00} \equiv \Phi \, \, , \label{e103}
\end{equation}
\begin{equation}
{{h^{(s)Tt}}}_{ 0i} \equiv \epsilon _{ik}\partial _k u^T +
\partial _i u^L \, \, , \label{e104}
\end{equation}
\begin{equation}
{{h^{(s)Tt}}}_{ ij}\equiv (\delta _{ij}\Delta -\partial _i
\partial _j)\phi ^{TT} +\partial _i \partial _j \phi ^{LL} +( \epsilon
_{ik}\partial _k
\partial _j +\epsilon _{jk}\partial _k \partial _i )\phi ^{TL} \, \,.\label{e105}
\end{equation}
Las propiedades Tt del campo, (\ref{e80}) y (\ref{e81}) establecen
las siguientes condiciones
\begin{equation}
\Phi = \Delta (\phi ^{TT} +\phi ^{LL})\, \, , \label{e106}
\end{equation}
\begin{equation}
{\dot{u}}^L = \Delta \phi ^{LL}\, \, , \label{e107}
\end{equation}
\begin{equation}
{\dot{u}}^T = \Delta \phi ^{TL}\, \, , \label{e108}
\end{equation}
\begin{equation}
\dot{\Phi}= \Delta u^L\, \, , \label{e109}
\end{equation}
con las cuales podemos reescribir las definiciones
(\ref{e103})-(\ref{e105}) como
\begin{equation}
{h^{(s)Tt}}_{00} \equiv \Phi \, \, , \label{e110}
\end{equation}
\begin{equation}
{h^{(s)Tt}}_{0i} \equiv \epsilon _{ik}\partial _k u^T -\partial _i
(-\Delta)^{-1}\dot{\Phi} \, \, , \label{e111}
\end{equation}
\begin{equation}
{h^{(s)Tt}}_{ij} \equiv (\delta _{ij}\Delta -\partial _i \partial
_j)(-\Delta)^{-2}\Box{}\Phi +\partial _i \partial _j
(-\Delta)^{-2}\ddot{\Phi} -( \epsilon _{ik}\partial _k
\partial _j +\epsilon _{jk}\partial _k \partial _i )(-\Delta)^{-1}{\dot{u}}^T
\, \,.\label{e112}
\end{equation}
Ahora la acci\'on reducida $2+1$, ec.(\ref{e102}) es
\begin{eqnarray}
{S_{ad}}^{(2)*}= m \int d^3 x \big(\,2\Phi \Delta u^T -4u^T
\ddot{\Phi}-2(-\Delta)^{-1}{\ddot{u}}^T\ddot{\Phi} -m \Phi ^2+ \nonumber \\
-\,mu^T \Delta u^T+2m \dot{\Phi}(-\Delta)^{-1}\dot{\Phi}
-m\ddot{\Phi}(-\Delta)^{-2}\ddot{\Phi}-m({\dot{u}}^T)^2 \big)
 \, \, .
\label{e113}
\end{eqnarray}

Introduciendo la variable
\begin{equation}
\Psi \equiv \Phi +(-\Delta )^{-1}\ddot{\Phi}\equiv -(-\Delta
)^{-1}\Box{} \Phi\, \, , \label{e114}
\end{equation}
la acci\'on reducida toma la forma compacta
\begin{eqnarray}
{S_{ad}}^{(2)*}= m \int d^3 x \big(\,(2\Psi - mu^T) \Box{}u^T
-m{\Psi }^2 \big)
 \, \, ,
\label{e115}
\end{eqnarray}
con las ecuaciones de movimiento
\begin{eqnarray}
\Box{}u^T-m\Psi  =0
 \, \, ,
\label{e116}
\end{eqnarray}
\begin{eqnarray}
\Box{}(\Psi - mu^T) =0
 \, \, .
\label{e117}
\end{eqnarray}
Usando estas ecuaciones se puede mostrar que la acci\'on reducida
es tambi\'en ${S_{sd}}^{(2)*}= \int d^3 x \, \Psi ( \Box{}
-m^2)\Psi $.

Entonces, uno puede introducir una definici\'on para el grado de
libertad y su momento can\'onico conjugado
\begin{eqnarray}
Q\equiv \sqrt{2}\, \Psi
 \, \, ,
\label{e118}
\end{eqnarray}
\begin{eqnarray}
P\equiv \sqrt{2} \,m \,{\dot{u}}^T
 \, \, ,
\label{e119}
\end{eqnarray}
de manera que la acci\'on (\ref{e115}) toma la forma can\'onica
esperada (p.ej.: $\int d^3 x ( P\dot{Q}-\frac{P^2}{2}-\frac{1}{2}Q
( -\Delta +m^2)Q )$), mostr\'andose que a nivel cl\'asico las
descomposiciones TL y Tt describen el mismo grado de libertad
propagado. M\'as a\'un, siguiendo el programa que parte de
sustituir los campos por sus operadores mec\'anico-cu\'anticos y
la incorporaci\'on de la regla de conmutaci\'on fundamental
(\ref{a71}), es posible promover la mencionada equivalencia a
nivel cu\'antico. El primer paso consiste en reescribir las
definiciones (\ref{e110})-(\ref{e112}) en t\'erminos del grado de
libertad $Q$ y su momento can\'onico conjugado $P$ con la ayuda de
(\ref{e118}) y  (\ref{e119}):
\begin{equation}
{h^{(s)Tt}}_{00} = -\frac{1}{\sqrt{2}}\,\Box{}^{-1}(-\Delta )Q \,
\, , \label{e120}
\end{equation}
\begin{equation}
{h^{(s)Tt}}_{0i} = \frac{1}{\sqrt{2}\,m}\,\epsilon _{ik}\partial
_k Q +\frac{1}{\sqrt{2}}\,\Box{}^{-1}\partial _i P  \, \, ,
\label{e121}
\end{equation}
\begin{eqnarray}
{h^{(s)Tt}}_{ij} = \frac{1}{\sqrt{2}}\,\delta _{ij}\,Q +
\frac{1}{\sqrt{2}}\,\partial _i
\partial _j(-\Delta)^{-1}\,Q +\frac{1}{\sqrt{2}}\,\partial _i
\partial _j\,\Box{}^{-1}\,Q
 +\nonumber \\+\, \frac{m^2}{\sqrt{2}}\,\partial _i
\partial _j(-\Delta)^{-1}\,\Box{}^{-1}\,Q -\frac{1}{\sqrt{2}\,m}( \epsilon _{ik}\partial _k
\partial _j +\epsilon _{jk}\partial _k \partial _i )(-\Delta)^{-1}P\, \,.\label{e122}
\end{eqnarray}

Entonces, seg\'un la regla $\big[Q(x),P(y)
\big]_{t'=t}=i{\delta}^2(\vec{x}-\vec{y})$ se puede mostrar que en
la capa de masas, los conmutadores entre las diferentes
componentes de ${h^{(s)Tt}}_{\mu\nu}$ recuperan el \'algebra
(\ref{a72})-(\ref{a76}). Esto muestra la equivalencia cu\'antica
entre las formulaciones TL y Tt.


\newpage
{\bf 2.3) Generadores del álgebra de Poincaré}\vskip .1truein

La consistencia de una teor\'{\i}a cu\'antica de campo relativista
pasa por la obtenci\'on expl\'{\i}cita de los operadores
mec\'anico-cu\'anticos $\mathcal{P}^{\mu}$ y
$\mathcal{J}^{\mu\nu}$, que satisfacen el \'algebra de Poincar\'e
(\ref{t1})-(\ref{t3}).

Con la finalidad de construir los generadores del \'algebra de
Poincar\'e en t\'erminos de las variables fundamentales ''$Q$'' y
''$P$'', comenzamos determinando el tensor momento-energ\'{\i}a de
Belinfante, $T^{\alpha \beta}$ asociado al campo autodual. Así,
extendemos la definici\'on de la acci\'on autodual al caso en que
el espacio-tiempo est\'a provisto de una m\'etrica general,
$g_{\mu\nu}$ y coordenadas curvil\'ineas
\begin{eqnarray}
{S_{adg}}^{(2)}=\frac m2 \int d^3 x \sqrt {-g}\big( \varepsilon
^{\mu \nu \lambda }{h_\mu }^\alpha \nabla _\nu h_{\lambda \alpha
}-m(h_{\mu \nu }h^{\nu \mu }-h^2 ) \big) \, \,,\label{g123}
\end{eqnarray}
donde $\nabla _\nu $ es la derivada covariante y $ \varepsilon
^{\mu \nu \lambda } \equiv \frac {\epsilon ^{\mu \nu \lambda }}
{\sqrt {-g}}$ es el tensor de Levi-Civita. Esta generalizaci\'on
no incluye t\'erminos de acoplamiento no minimales con la gravedad
debido a que nuestro inter\'es ahora est\'a enfocado en el
l\'{\i}mite plano de $T^{\alpha \beta}$.

Entonces, el tensor momento-energ\'{\i}a sim\'etrico en el
espacio-tiempo plano es
\begin{eqnarray}
T^{\alpha \beta} =\bigg[\frac{2}{\sqrt{-g}}\, \frac{\delta
S}{\delta g_{\alpha \beta}}\bigg]_{g_{\mu \nu}=\eta _{\mu \nu}} =
\frac{m^2}{2}\,\big( h^{\sigma
\alpha}{h^{\beta}}_{\sigma}+h^{\sigma
\beta}{h^{\alpha}}_{\sigma}-h\,h^{\alpha \beta}-h\,h^{\beta
\alpha} +\nonumber \\ -\,\eta ^{\alpha \beta}h_{\mu \nu}h^{\nu
\mu}+ \eta ^{\alpha \beta}h^2 \big) -\frac{1}{2}\,\big(
\partial _{\sigma} t^{\alpha \beta \sigma} +
{h_{\sigma}}^{\alpha}E^{\sigma \beta} +
{h_{\sigma}}^{\beta}E^{\sigma \alpha}\big) \,\,\, \,,\label{g124}
\end{eqnarray}
donde $t^{\alpha \beta \sigma}\equiv  m\epsilon ^{\mu \alpha \nu
}{h_\mu }^\beta {h_\nu }^\sigma  + m\epsilon ^{\mu \beta \nu
}{h_\mu }^\alpha {h_\nu }^\sigma $ y $E^{\mu \rho }\equiv
m\epsilon ^{\mu \nu \lambda }\partial _\nu {h_\lambda}^\rho +
m^2(\eta ^{\mu \rho }h - h^{\rho \mu })$, como en ec.(\ref{tl18}).

Con la ayuda de la descomposici\'on TL discutida en
(\ref{a44})-(\ref{a46}), puede observarse que en el espacio
f\'{\i}sico de los campos (esto es $E^{\mu \rho }=0$ ), los
generadores de translaciones (temporal y espaciales) son
equivalentes al caso del campo escalar. En efecto, el Hamiltoniano
y el momentum vienen dados por
\begin{eqnarray}
\mathcal{H}=\int d^2 x \{ m^2(-\Delta {h^{(s)TL}})^2 - \Delta
{h^{(s)TT}}(-\Delta +m^2 )(-\Delta) {h^{(s)TT}} \}
\nonumber\\=\frac{1}{2}\int d^2 x \{(\dot{Q})^2 +\partial _i Q
\partial _iQ +m^2 Q^2 \}
 \, \,,\label{g125}
\end{eqnarray}
\begin{eqnarray}
\mathcal{P}^i=-2m\int d^2 x \,(-\Delta) {h^{(s)TL}}\partial _i
(-\Delta) {h^{(s)TT}} =-\int d^2 x \,\dot{Q}\partial _i Q
 \, \,.\label{g126}
\end{eqnarray}

De manera id\'entica ocurre con los generadores de rotaciones
$\mathcal{J}^{ij}=\epsilon ^{ij} \mathcal{J}=\int d^2 x \{x^i
T^{0j}-x^j T^{0i}\}$, donde
\begin{eqnarray}
\mathcal{J}\equiv-2m\int d^2 x \,(-\Delta) {h^{(s)TL}}\epsilon
^{ij}x^i\partial _j (-\Delta) {h^{(s)TT}} =-\int d^2 x
\,\dot{Q}\epsilon ^{ij}x^i
\partial _j Q
\, \,.\label{g127}
\end{eqnarray}
Esta coincidencia con el caso de un campo escalar ocurre debido a
que en dos dimensiones espaciales, las rotaciones est\'an
descritas por el grupo $O(2)$, el cual no requiere de un
esp\'{\i}n definido. Sin embargo, la contribuci\'on del esp\'{\i}n
aparece cuando los generadores de {\it boosts} de Lorentz,
$\mathcal{J}^{i0}=\int d^2 x \{x^i T^{00}-x^0 T^{0i}\}$ son
calculados
\begin{eqnarray}
\mathcal{J}^{i0}=\int d^2 x \, x^i\{ m^2(-\Delta {h^{(s)TL}})^2 -
(-\Delta) {h^{(s)TT}}(-\Delta +m^2 )\Delta {h^{(s)TT}} \} -x^0
\mathcal{P}^i +\nonumber \\
 -\,4m^2\int d^2 x \,(-\Delta) {h^{(s)TL}}\epsilon ^{ij}\partial _j
{h^{(s)TT}}\nonumber \\ =\frac{1}{2}\int d^2 x \,x^i \{(\dot{Q})^2
+\partial _i Q
\partial _i Q +m^2 Q^2 \}-x^0 \mathcal{P}^i -2m\epsilon ^{ij}\int
d^2 x \,\dot{Q}\,\frac{\partial _j}{-\Delta}\,Q
\,\,\,\,,\nonumber\\\label{g128}
\end{eqnarray}
donde se puede observar la contribuci\'on de un t\'ermino singular
infrarrojo, el cual dice que el campo  $Q(x)$ no transforma como
un escalar, como era de esperarse. Este t\'ermino singular, como
veremos inmediatamente, representa la contribuci\'on del
esp\'{\i}n de la teor\'{\i}a.

Para remover la singularidad infrarroja recurrimos al bien
conocido proce-dimiento sobre la interpretaci\'on de la
contribuci\'on de esp\'{\i}n \cite{r55},  que parte con la
expansi\'on en ondas planas considerando operadores
creaci\'on-aniquilaci\'on
\begin{eqnarray}
Q(x)=\int \frac{d^2 k}{2\pi \sqrt{2w(\bf{k})}} \, \{
e^{-iKx}a({\bf{k}}) + e^{iKx}a^+(\bf{k}) \} \, \,.\label{g129}
\end{eqnarray}
con $K=(w(\bf{k}),\bf{k})$,\,\,\,
$w({\bf{k}})=\sqrt{{\bf{k}}^2+m^2}$\,\, y
\,\,\,\,$[a({\bf{k}}),a^+({\bf{k'}})]=\delta ^2(\bf{k}-\bf{k'})$.
Con esto, los generadores de translaciones y rotaciones son
\begin{eqnarray}
\mathcal{P}^\mu=\int d^2 k \,k^\mu a^+({\bf{k}})a({\bf{k}}) \,
\,,\label{g130}
\end{eqnarray}
\begin{eqnarray}
\mathcal{J}^{ij}=\int d^2 k \, a^+({\bf{k}})\frac{\epsilon ^{ij}
}{i}\,\frac{\partial}{\partial \theta }\, a({\bf{k}})
 \, \,,\label{g131}
\end{eqnarray}
donde $k_1\,\tan{\theta}=k_2$,  con $\theta \in [0 , 2\pi )$  el
\'angulo polar en el plano 2-dimensional de momentos. En esta
representaci\'on, los generadores de {\it boosts} de Lorentz
tambi\'en exhiben la singularidad infrarroja
\begin{eqnarray}
\mathcal{J}^{i0}=\frac{i}{2}\,\int d^2 k \,w({\bf{k}})\{
a^+({\bf{k}})\overleftrightarrow{\partial _i}\,
a({\bf{k}})\}-2m\int d^2 k \,\frac{\epsilon
^{ij}\,k^j}{{\bf{k}}^2} \, a^+({\bf{k}})a({\bf{k}})
 \, \,,\label{g132}
\end{eqnarray}
donde $a^+({\bf{k}})\overleftrightarrow{\partial _i}\,
a({\bf{k}})\equiv a^+({\bf{k}})\partial _i\, a({\bf{k}})
-a({\bf{k}})\partial _i\, a^+({\bf{k}})$.

Seguidamente, realizamos una transformaci\'on de fase en los
operadores creaci\'on-aniquilaci\'on
\begin{eqnarray}
 a({\bf{k}})\longrightarrow e^{is\,\frac{m}{|m|}\,\theta}a({\bf{k}})
 \, \,,\label{g133}
\end{eqnarray}
siendo $s$ un par\'ametro real desconocido. El mapa (\ref{g133})
está bien definido sebido a que es posible fijar un dominio en el
cual es invertible (p. ej., $\theta \in
[0,\frac{2\pi}{s}\frac{m}{|m|})$).

Los generadores de translaciones y las relaciones de conmutaci\'on
de  $a$ y $a^+$ permanecen invariantes bajo (\ref{g133}), pero los
generadores de {\it boosts} y rotaciones son ahora
\begin{eqnarray}
\mathcal{\overline{J}}\,^{i0}=\frac{i}{2}\,\int d^2 k
\,w({\bf{k}})\{ a^+({\bf{k}})\overleftrightarrow{\partial _i}\,
a({\bf{k}})\}+2\frac{m}{|m|}\,\int d^2 k \,\frac{\epsilon
^{ij}k^j}{w({\bf{k}})+|m|} \,a^+({\bf{k}})a({\bf{k}})
+\nonumber\\+\,(s-2)\,\frac{m}{|m|}\,\int d^2 k
\,w({\bf{k}})\,\frac{\epsilon ^{ij}k^j}{{\bf{k}}^2}
\,a^+({\bf{k}})a({\bf{k}}) \, \,,\nonumber\\\label{g134}
\end{eqnarray}
\begin{eqnarray}
\mathcal{\overline{J}}\,^{ij}=\int d^2 k \,
a^+({\bf{k}})\frac{\epsilon ^{ij}}{i}\,\frac{\partial}{\partial
\theta }\, a({\bf{k}})+s\frac{m}{|m|}\,\int d^2 k \,\epsilon
^{ij}a^+({\bf{k}})a({\bf{k}}) \, \,.\label{g135}
\end{eqnarray}

Inmediatamente podemos ver que la singularidad infrarroja es
removida, si y solo si se fija el valor  $s=2$ para el par\'ametro
libre. M\'as a\'un, el valor del esp\'{\i}n $2\frac{m}{|m|}$ es
recuperado, y su sensibilidad bajo cambios de signo de la masa
reflejan la helicidad de la excitaci\'on propagada. Desde el punto
de vista Lagrangiano, esta expresi\'on de la helicidad proviene
del signo del t\'ermino de masa lineal de la acci\'on (\ref{t16}),
y sea cual sea el caso, tal signo no afecta al Hamiltoniano, como
uno puede deducir directamente de la transformada de Legendre de
la acci\'on reducida (\ref{a54}).

En el estudio de la teor\'ia autodual de esp\'{\i}n 2 hemos
mostrado que la formulaci\'on de la acci\'on reducida constituye
una herramienta poderosa para la construcci\'on de la teor\'{\i}a
cu\'antica correspondiente, evitando el extenuante procedimiento
can\'onico de Dirac. M\'as a\'un, el formalismo reducido que
describe la excitaci\'on masiva propagada, permite pr\'acticamente
de manera directa determinar la contribuci\'on de esp\'{\i}n que
establece el comportamiento no escalar del grado de libertad.
All\'{\i} observamos que es posible evitar la singularidad
infrarroja mediante una transformaci\'on de fase en los operadores
creaci\'on-aniquilaci\'on.

Finalmente, si se desea confirmar el comportamiento relativista
consistente de la teor\'{\i}a, se debe verificar que los
generadores obtenidos satisfagan el \'algebra de Poincar\'e. Si
consideramos los generadores obtenidos antes de realizar la
transformaci\'on de fase (\ref{g133}), es decir los objetos
(\ref{g130}) y (\ref{g131}) puede mostrarse que el \'algebra de
Poincar\'e es satisfecha con excepci\'on del \'algebra de los
generadores de {\it boosts} debido a la singularidad infrarroja, y
la cual exhibe una ''anomal\'{\i}a'', es decir
\begin{eqnarray}
i[\mathcal{J}^{i0}\,,\,\mathcal{J}^{j0}]=\epsilon
^{ij}(\mathcal{J}-\mathcal{A}) \, \,,\label{g136}
\end{eqnarray}
donde $\mathcal{A}\equiv -2m\int d^2 k \,w({\bf{k}})\partial_i
f^i({\overrightarrow{k}})a^+({\bf{k}})a({\bf{k}})$ con
$f^i({\overrightarrow{k}})= \frac{k^i}{{\bf{k}}^2}$ una funci\'on
singular en el origen. En (\ref{g136}), el t\'ermino
''anomal\'{\i}a'' representa el hecho de la asociaci\'on entre la
\'unica excitaci\'on propagada del campo autodual y la
correspondiente al caso de un campo escalar, debido a que \'esta
analog\'{\i}a solo es aparente hasta el momento en que los
generadores de {\it boosts} de lorentz son involucrados. Es obvio
que tal ''anomal\'{\i}a'' deba aparecer pues de lo contrario el
grado de libertad estudiado ser\'{\i}a simplemente de tipo
escalar.

Sin embargo, lo interesante est\'a en que luego de realizar la
transformaci\'on de fase sobre los operadores
creaci\'on-aniquilaci\'on se puede volver a calcular el \'algebra
de los  {\it boosts}, obteni\'endose
\begin{eqnarray}
i[\mathcal{\overline{J}}\,^{i0}\,,\,\mathcal{\overline{J}}\,^{j0}]_{s=2}=\epsilon
^{ij}(\mathcal{J}-2\frac{m}{|m|}\,\int d^2 k
\,a^+({\bf{k}})a({\bf{k}}))= \epsilon
^{ij}\,\mathcal{\overline{J}}_{s=2}\, \,,\label{a137}
\end{eqnarray}
con lo cual uno puede decir que ha removido la ''anomal\'{\i}a'',
satisfaci\'endose el \'algebra de Poincar\'e. Es de subrayarse que
la teor\'{\i}a autodual planteada en (\ref{t16}) es invariante
relativista por construcci\'on, raz\'on por la cual la
anomal\'{\i}a discutida es solo una expresi\'on de la singularidad
infrarroja y no de alguna inconsistencia intr\'{\i}seca en el
car\'acter covariante de la misma.

\newpage
{\large {\bf 3) Teoría autodual de espín 2 en espacios de
curvatura constante}} \vskip .5truein

Como mencionáramos al principio de este trabajo, en el contexo de
la teoría de campos ordinaria, entre otras ha habido un gran
interés sobre el estudio lagrangiano de campos de espín alto con
interacción externa. Tales teorías son solo conocidas bajo cierto
régimen de acoplamiento, ya sean de origen electromagnético,
gravitacional, entre otros.  En el contexto de la interacción
gravitacional, diversos autores han considerado espacios de
curvatura constante \cite{r14}-\cite{r10}, hasta de tipo
no-Einstenianos \cite{r10}.

La razón fundamental de lo anterior, insistimos es que no existe
una teoría de campos general consistente de espines altos con
interacción como consecuencia de la no conservación de los grados
de libertad y la violación de la causalidad. El primer hecho
relacionado con la posibilidad de que los campos auxiliares
propaguen grados de libertad cuando aparece una interacción
arbitraria, y el segundo con que la ecuación de movimiento pudiera
describir propagaciones no causales.

Con respecto a la posible violación de la causalidad, basaremos
nuestra discusión con la siguiente nomenclatura \cite{r10}.
Enfocando nuestro interés en campos con espín entero, $h_{\alpha
_1 \alpha _2...}$, en términos generales es posible obtener
ecuaciones de movimiento  de una formulación lagrangiana dada, las
cuales pueden ser establecidas como $({\mathcal{M}_{\beta _1
...\alpha _1 ...}})^{\mu \nu } {\nabla}_{\mu }{\nabla}_{\nu
}h^{\alpha _1...} +... =0$, con la ayuda de los vínculos
lagrangianos. Seguidamente, sean $n_\mu$ las componentes de
vectores, entonces la matriz característica es definida mediante
$\mathcal{M}_{AB}(n)\equiv {\mathcal{M}_{AB}}^{\mu\nu}n_\mu
n_\mu$, donde $A,B$ son índices compuestos. La ecuación
característica es $\det{\mathcal{M}_{AB}(n)}=0$, cuyas soluciones
definen superficies características que describen los posibles
procesos de propagación. Si la solución de la ecuación
característica proporciona un $n_0$ real, el sistema de ecuaciones
de movimiento se llama hiperbólico. Un sistema hiperbólico se
llama causal si no hay vectores tipo tiempo dentro de las
soluciones de la ecuación característica (de lo contrario, si hay
vectores tipo tiempo, las superficies características
correspondientes son de tipo espacio, con lo cual sus puntos
estarían conectados por procesos superluminales y se violaría la
causalidad).

Entonces, cuando una interacción externa arbitraria es
introducida, la matriz característica, $\mathcal{M}_{AB}(n)$ no
necesariamente define un sistema de ecuaciones hiperbólico-causal.

Aquí, estaremos interesados en estudiar los aspectos cruciales ya
mencionados (conservación de los grados de libertad y causalidad),
haciendo énfasis en la formulación lagrangiana de la teoría del
campo autodual de espín 2 acoplado con la gravitación. En este
sentido, serán discutidas las resticciones físicas que
proporcionan consistencia a la teoría, que no solo se traducen en
condiciones sobre el campo gravitacional (p.ej., espacios de
curvatura constante), si no también en restricciones sobre los
posibles valores permitidos de los parámetros asociados a la masa.

\vspace{1cm}
{\bf 3.1) Vínculos Lagrangianos en un espacio curvo}\vskip
.1truein

Comenzamos por presentar el modelo autodual acoplado no
minimalmente con gravedad. Para esto, introducimos la interacción
gravitacional mediante un conjunto general de términos de
acoplamiento no minimales en la formulación Lagrangiana,
construidos a partir del tensor de Ricci y sus contracciones, ya
que en $2+1$ dimensiones éste describe completamente a la
curvatura de Riemann (Apéndice B). Entonces, en un espacio
Riemanniano nuestro modelo es \cite{r56}
\begin{equation}
{S_{adg}}^{(2)}=\int \frac{d^3 x}{2} \sqrt {-g}( m\,\varepsilon
^{\mu \nu \lambda }{h_\mu }^\alpha \nabla _\nu h_{\lambda \alpha
}+ {\Omega}^{\alpha \beta \sigma \lambda }h_{\alpha \beta
}h_{\sigma \lambda } ) \, \, , \label{c1}
\end{equation}
donde $\nabla _\nu $ es la derivada covariante definida con los
símbolos de Christoffel y $ \varepsilon ^{\mu \nu \lambda } \equiv
\frac {\epsilon ^{\mu \nu \lambda }} {\sqrt {-g}}$. Debido al
hecho de que en un espacio-tiempo $2+1$ dimensional el tensor
conformal de Weyl es idénticamente nulo, el tensor de curvatura de
Riemann puede ser escrito en términos exclusivos de el de Ricci
(p.ej.: $R_{\lambda \mu\nu \rho}=g_{\lambda\nu}R_{\mu
\rho}-g_{\lambda\rho}R_{\mu\nu}-g_{\mu\nu}R_{\lambda\rho}+g_{\mu\rho}R_{\lambda\nu}-\frac
R2 \,(g_{\lambda\nu}g_{\mu \rho}-g_{\lambda\rho}g_{\mu\nu}) $), el
acoplamiento no minimal en la acción (\ref{c1}) está caracterizado
por el tensor ${\Omega}^{\alpha \beta \sigma \lambda }$, cuyo
aspecto más general es
\begin{eqnarray}
{\Omega}^{\alpha \beta \sigma \lambda } \equiv m^2(g^{\sigma
\lambda }g^{\alpha \beta }-
    g^{\sigma \beta }g^{\alpha \lambda})+ a_1 (R^{\sigma \lambda }g^{\alpha \beta
}+
    R^{\alpha \beta }g^{\sigma \lambda})+\nonumber \\+\, a_2 (R^{\sigma \beta }g^{\alpha
\lambda } + R^{\alpha \lambda }g^{\sigma \beta}) + a_3 R^{\alpha
\sigma }g^{\beta \lambda } + a_4 R^{\beta \lambda }g^{\alpha
\sigma } +\nonumber \\+ \, a_5 Rg^{\alpha \beta }g^{\sigma
\lambda} + a_6 Rg^{\sigma \beta }g^{\alpha \lambda } +  a_7
Rg^{\lambda \beta }g^{\sigma \alpha }
 \, \, , \label{c2}
\end{eqnarray}
con la propiedad ${\Omega}^{\alpha \beta \sigma \lambda }
={\Omega}^{\sigma \lambda\alpha \beta  } $ y los parámetros
reales, $a_n$ ($n=1,...,7 $) libres.

Entonces, tomando variaciones arbitrarias del campo autodual, la
extremal de la accion $S_{adg}$ provee las siguientes ecuaciones
de campo (nueve vínculos primarios)
\begin{equation}
\Phi ^{(1) \mu \alpha }\equiv m\,\varepsilon ^{\mu \nu \lambda }
\nabla _\nu {h_\lambda}^\alpha + \Omega ^{\mu  \alpha \sigma
\lambda  } h_{\sigma \lambda } \approx 0 \, \, . \label{c3}
\end{equation}
Tres vínculos más aparecen cuando $ \Phi ^{(1) 0 \rho }\approx 0$
es preservado
\begin{eqnarray}
\Phi ^{(2) \alpha }\approx \nabla _\mu  \Phi ^{(1) \mu \alpha
}\equiv \Omega ^{\mu \alpha \sigma \lambda }\nabla _\mu h_{\sigma
\lambda}+\mathcal{B} ^{\alpha \sigma \lambda }h_{\sigma \lambda}
\approx 0\, \, , \label{c4}
\end{eqnarray}
donde se ha definido el objeto
\begin{eqnarray}
\mathcal{B} ^{\alpha\sigma \lambda }\equiv \frac m2 \,\varepsilon
^{\mu \nu \rho}({R^{\alpha \lambda}}_{\nu \mu}\,{\delta
^{\sigma}}_{\rho} -{R^{\sigma}}_{\rho \nu \mu}\,g^{\alpha
\lambda}) +\nabla _\mu \Omega ^{\mu \alpha \sigma \lambda }\, \, .
\label{c5}
\end{eqnarray}

Por otro lado, la conservación de  $ \Phi ^{(1) k \alpha }\approx
0$ conduce a seis relaciones para las aceleraciones (como en el
caso plano, aquí es demandado   $m\neq 0$)
\begin{eqnarray}
{\nabla _0}^2 {h_j}^\alpha=-\frac{\overline{\varepsilon
}_{kj}}{m}\,\Omega ^{k \alpha \sigma \lambda }\nabla _0 h_{\sigma
\lambda}-\big(\frac{\overline{\varepsilon }_{kj}}{m}\,\nabla
_0\Omega ^{k \alpha \sigma \lambda }+{R^\sigma}_{ 0j0}\,g^{\alpha
\lambda}\big)h_{\sigma \lambda}+\nonumber \\
+ \,\nabla _j\nabla _0{h_0}^\alpha +{R^{\alpha\lambda}}_{
j0}\,h_{0 \lambda}\,\,
 , \label{c6}
\end{eqnarray}
donde la notación significa $\overline{\varepsilon }_{\sigma
\lambda}\equiv \varepsilon _{0 \sigma \lambda}$,
 quedando aún por determinarse las ace-leraciones,  ${\nabla _0}^2
h_{0 \lambda }$.

Hasta este punto, el análisis Lagrangiano con los parámetros de
acoplamiento libres, en un espacio-tiempo arbitrario es
equivalente al caso plano. Sin embargo, siguiendo el siguiente
paso, se puede notar con la ayuda  de (\ref{c6}), que la
preservación de $\Phi ^{(2) \alpha } \approx 0$ toma la forma
\begin{eqnarray}
\Phi ^{(3) \alpha } \equiv \nabla _0 \Phi ^{(2) \alpha }\approx
{\Omega}^{0\alpha 0\lambda}{\nabla _0}^2 h_{0 \lambda}+
\big(\Omega ^{0 \alpha j \lambda }+\Omega ^{j \alpha 0 \lambda }
\big)\nabla _j\nabla _0 h_{0 \lambda}+ \nonumber \\
+\,\big(-\frac{\overline{\varepsilon }_{kl}}{m}\,\Omega ^{0 \alpha
l \rho }{\Omega ^{\mu \lambda k}}_\rho +\nabla_0 \Omega ^{0 \alpha
\mu \lambda}+\mathcal{B} ^{\alpha \mu \lambda } \big)\nabla _0
h_{\mu\lambda}+ \nonumber \\
+\big[ -\frac{\overline{\varepsilon }_{kl}}{m}\,\Omega ^{0 \alpha
l \rho }\nabla_0{\Omega ^{\mu \lambda k}}_\rho - \Omega ^{0 \alpha
l \lambda}{R^\mu}_{0 l0} + \nabla_0\mathcal{B} ^{\alpha \mu
\lambda }+ \nonumber \\-\Omega ^{\nu \alpha \mu \rho
}{R^\lambda}_{\rho \nu 0} -\Omega ^{\sigma \alpha \nu \lambda
}{R^\mu}_{\nu \sigma 0}\big]h_{\mu\lambda}-\Omega ^{0 \alpha l
\rho }{R^\lambda}_{\rho l0}h_{0\lambda}+\nonumber
\\ +\nabla_0 \Omega ^{i
\alpha \mu \lambda}\nabla _i h_{\mu \lambda} + \Omega ^{i \alpha j
\lambda}\nabla _i\nabla_0 h_{j \lambda}\approx 0
  \,
\, , \label{c7}
\end{eqnarray}
con la esperanza de que represente tres nuevos vínculos, de manera
equivalente al caso plano. Esto significa que de la expresión
(\ref{c7}) no deba ser imposible obte-ner ninguna relación para
las aceleraciones ${\nabla _0}^2 h_{0 \lambda }$, aún no
resueltas. Debido a que (\ref{c7}) constituye un sistema completo
para las mencionadas aceleraciones, exigiremos que todas las
matrices $3\times 3$, $2\times 2$ y $1\times
 1$ construidas ${\Omega}^{0\alpha 0\lambda}$, tengan determinante nulo
(p.ej.: ${\Omega}^{0\alpha 0\lambda}$ es totalmente degenerada)
para garantizar la imposibilidad de despejar ${\nabla _0}^2 h_{0
\lambda
 }$. Esta condición significa
\begin{equation}
\Omega ^{0 \alpha 0 \lambda } =0 \, \, , \label{c8}
\end{equation}
la cual al ser usada en (\ref{c2}), proporciona las siguientes
restricciones sobre los parámetros de acoplamiento
\begin{eqnarray}
a_1 = -a_2 \equiv a \,\,, \,\, a_6 = -a_5 \equiv b \,\, , \,\, a_3
= a_4 = a_7 = 0 \,\,, \label{c9}
\end{eqnarray}
quedando solo dos de ellos libres. Entonces, el tensor (\ref{c2})
es ahora
\begin{eqnarray}
{\Omega}^{\alpha \beta \sigma \lambda } = a\,( R^{\alpha \beta
}g^{\sigma \lambda } +R^{\sigma \lambda }g^{\alpha \beta }
-R^{\alpha \lambda }g^{\beta\sigma }- R^{\beta\sigma }g^{\alpha
\lambda})+\nonumber \\+\,
 (m^2 -b R) (g^{\alpha \beta }g^{\sigma \lambda}
- g^{\alpha \lambda }g^{\beta\sigma  } )
 \, \,\,\, , \label{c10}
\end{eqnarray}
con las propiedades de simetría
\begin{eqnarray}
{\Omega}^{\alpha \beta \sigma \lambda } ={\Omega}^{\sigma \lambda
\alpha \beta }=-{\Omega}^{\alpha \lambda\sigma \beta  }
 \, \, \, \,. \label{c11}
\end{eqnarray}
Con esto, el objeto  $\mathcal{B} ^{\alpha\sigma \lambda }$, dado
por (\ref{c5}) puede ser reescrito en términos del tensor de
Einstein, $G_{\mu\nu}=R_{\mu\nu}-\frac{g_{\mu\nu}}{2}\,R$
(apéndice B), exhibiendo la propiedad antisimétrica respecto a los
índices $\alpha \lambda$
\begin{eqnarray}
\mathcal{B} ^{\alpha\sigma \lambda }=  m\,\varepsilon ^{\alpha
\lambda \beta}\,{G^{\sigma}}_{\beta} -(a-2b)(g^{\sigma
\lambda}\nabla ^\alpha G-g^{\sigma \alpha}\nabla ^\lambda G)+
\nonumber
\\+\, a\,{\varepsilon ^{\alpha \lambda}}_\nu \,\varepsilon ^{\mu
\sigma \beta}\nabla_\mu {G^{\nu}}_{\beta}=-\mathcal{B}
^{\lambda\sigma \alpha }\, \,\, \,. \label{c12}
\end{eqnarray}

Entonces, con la ayuda de (\ref{c10}) se pueden escribir los tres
vínculos
 $\Phi ^{(3) \alpha } \equiv \nabla _0 \Phi ^{(2) \alpha }
\approx 0$, de la manera
\begin{eqnarray}
\Phi ^{(3) \alpha } = \mathcal{N}^{\alpha \lambda }\nabla _0h_{0
\lambda} +\nabla _0\Omega ^{ i \alpha 0 \lambda} \nabla _ih_{0
\lambda}+\mathcal{A}^{ \alpha \lambda} h_{0 \lambda}+\Omega ^{ i
\alpha j \lambda} \nabla _i \nabla _0h_{j \lambda}\nonumber
\\+\,\mathcal{C}^{
\alpha j \lambda} \nabla _0h_{j \lambda}+\nabla _0\Omega ^{ i
\alpha j \lambda} \nabla _i h_{j \lambda}+\mathcal{D}^{ \alpha j
\lambda} h_{j \lambda}
 \approx 0 \,
\, , \label{c13}
\end{eqnarray}
donde hemos definido
\begin{eqnarray}
\mathcal{N}^{\alpha \lambda } \equiv \frac 1m
\,\overline{\varepsilon }_{kl}\,\Omega ^{0 \alpha k \rho}{\Omega
^{0 \lambda l}}_\rho + \mathcal{B}^{\alpha 0
\lambda}=-\mathcal{N}^{\lambda \alpha }  \, \, , \label{c14}
\end{eqnarray}
\begin{eqnarray}
\mathcal{A}^{\alpha \lambda } \equiv -\frac 1m \,\Omega ^{0 \alpha
l \rho}\big(\overline{\varepsilon }_{kl}\,\nabla_0{\Omega ^{0
\lambda k}}_\rho +m {R^\lambda}_{ \rho l 0}+m {R^0}_{ 0 l
0}{\delta^\lambda}_\rho \big)\nonumber \\ +\,
\nabla_0\mathcal{B}^{\alpha 0 \lambda}-\Omega ^{\mu \alpha 0
\rho}{R^\lambda}_{ \rho \mu 0} -\Omega ^{\mu \alpha \nu
\lambda}{R^0}_{ \nu \mu 0}\, \, , \label{c14a}
\end{eqnarray}
\begin{eqnarray}
\mathcal{C}^{\alpha j \lambda } \equiv -\frac 1m
\,\overline{\varepsilon }_{kl}\,\Omega ^{0 \alpha l \rho}{\Omega
^{j \lambda k}}_\rho +\mathcal{B}^{\alpha j \lambda}+
\nabla_0\Omega ^{0 \alpha j \lambda}\, \, , \label{c14b}
\end{eqnarray}
\begin{eqnarray}
\mathcal{D}^{\alpha j \lambda } \equiv -\frac 1m
\,\overline{\varepsilon }_{kl}\,\Omega ^{0 \alpha l
\rho}\nabla_0{\Omega ^{j \lambda k}}_\rho
+\nabla_0\mathcal{B}^{\alpha j \lambda}-\Omega ^{\mu \alpha j
\rho}{R^\lambda}_{ \rho \mu 0}\nonumber \\-\Omega ^{\mu \alpha \nu
\lambda}{R^j}_{ \nu \mu 0}-\Omega ^{0 \alpha l \lambda}{R^j}_{ 0l
0}\, \, . \label{c14b}
\end{eqnarray}

Continuando con el procedimiento de análisis Lagrangiano, la
preservación de $\Phi ^{(3) \rho } \approx 0$ debería representar
dos expresiones para las aceleraciones  ${\nabla _0}^2 h_{0 \sigma
}$ y una para el último vínculo (cuya preservación, a su vez
proporcione una relación más para las aceleraciones aún
desconocidas, y así termine el procedimiento). Consideremos las
matrices  $3\times 3$ y $2\times 2$ construidas con el objeto
$\mathcal{N}^{\alpha \lambda }$, entonces el requerimiento
anterior significa que la matriz $\mathcal{N}^{\alpha \lambda }$
tenga rango 2, de la forma
\begin{eqnarray}
\det (\mathcal{N}^{\alpha \lambda } )=0 \,\,, \label{c15}
\end{eqnarray}
 \begin{eqnarray}
\det (\mathcal{N}^{ij } ) \neq 0 \,\,. \label{c16}
\end{eqnarray}
Por un lado, debido a la antisimetría de la matriz impar
$(\mathcal{N}^{\alpha \lambda } )$, la relación (\ref{c15}) es
satisfecha idénticamente. Pero, la relación (\ref{c16}) significa
una restricción sobre el posible campo gravitacional, y conduce a
\begin{eqnarray}
\overline{\varepsilon }_{ij}\,\mathcal{N}^{ij }  \neq 0 \,\,.
\label{c17}
\end{eqnarray}

Puede mostrarse que esta restricción, la cual debe satisfacerse
con la finalidad de mantener la consistencia en el  número
 de grados de libertad podría contener soluciones no Einstenianas.
Por ejemplo, consideremos un espacio vacío no Einsteniano con
curvatura $R_{\lambda \mu\nu \rho}=\frac{f(x)}{6}
\,(g_{\lambda\nu}g_{\mu \rho}-g_{\lambda\rho}g_{\mu\nu})$.
Entonces, la restricción (\ref{c17}) conduce a una relación
diferencial parcial de primer orden para $f(x)$
\begin{eqnarray}
6M^4 - m^2f(x)+m\sigma {\varepsilon ^k}_0
\partial _k f(x)\neq 0 \,\,,
\label{c18}
\end{eqnarray}
con el parámetro $\sigma \equiv \frac{2}{3}\,a - b$.

\vspace{1cm}
{\bf 3.2) Teoría autodual en un espacio de dS/AdS}\vskip .1truein

Aquí, nuestro interés estará enfocado en una solución particular
de la clase $\partial_k f(x)=0$, la cual está relacionada con las
de tipo dS/AdS. Entonces, conside-raremos un espacio-tiempo de
curvatura constante, con constante cosmológica  $\lambda$, que
pudiese caracterizar a un espacio de dS ($\lambda> 0$) o AdS
($\lambda< 0$) via la ecuación de Einstein,
$R_{\mu\nu}-\frac{g_{\mu\nu}}2\,R -\lambda\,g_{\mu\nu}=0 $, donde
los tensores de Riemann y Ricci, y la curvatura escalar en $2+1$
dimensiones estan dados por
\begin{eqnarray}
R_{\lambda \mu\nu \rho}=\frac R6 \,(g_{\lambda\nu}g_{\mu
\rho}-g_{\lambda\rho}g_{\mu\nu})
 \, \, ,\,\,\,\,\,R_{ \mu\nu }=\frac R3 \,g_{\mu\nu}\, \, ,\,\,\,\,\,R=-6\lambda\,\,, \label{d19}
\end{eqnarray}
respectivamente. Con esto, el tensor (\ref{c10}) es
\begin{eqnarray}
\Omega^{\alpha \beta \sigma \lambda } =
 M^2\,(g^{\alpha \beta }g^{\sigma \lambda}
- g^{\sigma \beta }g^{\alpha \lambda } )
 \, \, , \label{d20}
 \end{eqnarray}
donde $M^2 = m^2+\sigma R$.

Como estamos considerando espacios de curvatura constante, y
usando el tensor (\ref{d20}) podemos evaluar los objetos
relevantes de la teoría. Por ejemplo, la acción (\ref{c1}) toma la
forma
\begin{equation}
{S_{ad\lambda}}^{(2)}=\int \frac{d^3 x}{2} \sqrt {-g}(
m\,\varepsilon ^{\mu \nu \lambda }{h_\mu }^\alpha \nabla _\nu
h_{\lambda \alpha }-M^2\,(h_{\mu \nu }h^{\nu \mu }-h^2 ) ) \, \, ,
\label{d21}
\end{equation}
el objeto $\mathcal{B} ^{\alpha 0 \lambda }$ dado por (\ref{c12}),
ahora es
\begin{eqnarray}
\mathcal{B} ^{\alpha 0 \lambda }=  -\frac{mR}6\,\varepsilon
^{0\alpha \lambda }\, \,\, \,, \label{d22}
\end{eqnarray}
y con esto, de (\ref{c14}) escribimos
\begin{equation}
\mathcal{N}^{ij } \equiv \bigg(\frac{6M^4-Rm^2}{6m} \bigg)
\,\varepsilon ^{ij} \, \, , \label{d23}
\end{equation}
con lo cual, la relación de consistencia (\ref{c17}) es ahora
\begin{equation}
6M^4-Rm^2\equiv 6m^4+(12\sigma -1)\,Rm^2+6\sigma^2R^2\neq 0\, \, .
\label{d24}
\end{equation}

Buscando una interpretación, se pudiera pensar esta última
relación como una restricción para $m^2$ en términos del escalar
de curvatura y el parámetro libre $\sigma$, queremos decir
\begin{equation}
m^2\neq {m_{\pm}}^2 \equiv \frac {R}{12}\,\big(1-12\sigma \pm
\sqrt{1-24\sigma }\,\big)
 \, \, . \label{d25}
\end{equation}
Por lo tanto, los valores prohibidos de $m$ son

\begin{center}

\begin{tabular}{|c|c|c|}
 \hline
\textbf { $R$} & \textbf{ $\sigma$} & \textbf{ $m\,\,\,prohibida $} \\
\hline \hline
$>0 \,(AdS)$& $\leq \frac{1}{24}$ & $m_{\pm}$  \\
\hline
$<0 \,(dS)$ & $---$ & $---$  \\
\hline

\end{tabular}
\vspace{0.5cm}

 tabla 1

\end{center}

La existencia de valores prohibidos de masa con la finalidad de
mantener la consistencia del número de grados de libertad es un
hecho bien conocido en el contexto de las  teorías de espines
altos \cite{r2},\cite{r4}. Sin embargo, uno también podría hacer
la interpretación de que (\ref{d24}) representa restricciones
sobre los posibles valores de la curvatura (por tanto de la
constante cosmológica), o sea $R\neq R_{\pm} \equiv \frac
{m^2}{12\sigma^2}\,\big(1-12\sigma \pm \sqrt{1-24\sigma }\,\big)$,
para una masa dada.

Seguidamente, revisamos los vínculos lagrangianos, esta vez en
espacios de dS/AdS. Los nueve vínculos primarios, ec.(\ref{c3})
son
\begin{equation}
\Phi ^{(1) \mu \alpha }\equiv m\,\varepsilon ^{\mu \nu \lambda }
\nabla _\nu {h_\lambda}^\alpha + M^2\,(g^{\mu \alpha }h -
h^{\alpha \mu }) \approx 0 \, \, , \label{d26}
\end{equation}
y los vínculos secundarios (\ref{c4}) los escribimos como
\begin{equation}
\Phi ^{(2) \alpha }\approx M^2\,(\nabla ^{\alpha}h
 -\nabla _{\mu}h^{\alpha \mu })+\frac{mR}{6}\, \varepsilon ^{\alpha
 \sigma
\lambda} h_{\sigma \lambda}
 \approx  0\, \, , \label{d27}
\end{equation}
o tambi\'en como $\Phi ^{(2) \alpha }\approx \nabla _\mu \Phi
^{(1) \mu \alpha }- \frac{M^2}{m}\,\varepsilon ^{\alpha \sigma
\lambda }{\Phi ^{(1)}}_{\mu \alpha
}=\big(\frac{m^2R-6M^4}{6m}\big)\,\varepsilon ^{\alpha
 \sigma
\lambda} h_{\sigma \lambda}\approx 0$, revelando la propiedad de
simetría del campo autodual, en virtud de la restricción
(\ref{d24}).

La preservación de $\Phi ^{(2) \alpha }\approx 0$ provee tres
nuevos vínculos
\begin{equation}
\Phi ^{(3) \alpha }\approx
\bigg(\frac{m^2R-6M^4}{6m}\bigg)\,\varepsilon ^{\alpha
 \sigma
\lambda}\nabla _0 h_{\sigma \lambda}\approx  0\, \, , \label{d28}
\end{equation}
 y particularmente, el vínculo $\Phi ^{(3) 0 }\approx 0$ es
 reescrito como
\begin{equation}
\Phi ^{(3) 0 }\approx \nabla _\mu \Phi ^{(2) \mu
}+\bigg(\frac{6M^4-m^2R}{6m^2}\bigg)\, {\Phi ^{(1) \mu}}_\mu =
\frac{M^2}{3m^2}\,(6M^4-m^2R)h \approx 0\, \, , \label{d29}
\end{equation}
diciendo que el campo autodual no posee traza (obviamente si $M^2
\neq 0$ y $6M^4-m^2R\neq 0$). El último vínculo aparece de la
preservación de $\Phi ^{(3) 0 }\approx 0$, es decir
\begin{equation}
\Phi ^{(4) }\equiv \nabla _0 \Phi ^{(3) 0 }\approx
\frac{M^2}{3m^2}\,(6M^4-m^2R)\nabla _0h \approx 0\, \, ,
\label{d30}
\end{equation}
y expresa el hecho de la conservación de la traza nula.

Queremos notar que las propiedades de transversalidad y traza nula
para una descripción consistente del campo autodual (como
cualquier campo con espín), demandan la condición
\begin{equation}
M^2\neq  0\, \, , \label{d31}
\end{equation}
como consecuencia de (\ref{d27}), (\ref{d29}) y (\ref{d30}). De
otra forma, si tal condición se relaja (es decir, $M^2= 0$), la
transversalidad y la no traza de  $h_{\mu\nu}$ no estarían
aseguradas, y entonces el sistema de vínculos lagrangianos no
proporcionaría el número correcto de grados de libertad, pudiendo
haber propagación de espines bajos.

Entonces, estando garantizadas las propiedades de simetría,
transversalidad y traza nula, podemos escribir la ecuación de
segundo orden ${h^{(s)Tt}}_{\mu \nu}$, partiendo de (\ref{d26})
\begin{equation}
\big( \Box{} -\frac{M^4}{m^2}+\frac{R}{2} \big){h^{(s)Tt}}_{ \mu
\nu}=0 \, \, , \label{d32}
\end{equation}
la cual es claramente hiperbólica-causal, debido a que la podemos
reescribir de la forma
\begin{equation}
({\mathcal{M}^{\beta\sigma}}_{\rho\alpha})^{\mu\nu}\nabla _\mu
\nabla _\nu{h^{(s)Tt}}_{\beta \sigma}+...=0\, \, . \label{d33}
\end{equation}
donde
$({\mathcal{M}^{\beta\sigma}}_{\rho\alpha})^{\mu\nu}=g^{\mu\nu}{\delta^{\beta\sigma}}_
{\rho\alpha}$ y ${\delta ^{\beta\sigma}}_{\rho \alpha}\equiv
\frac{1}{2}({\delta^{\beta}}_{\rho}{\delta^{\sigma}}_{\alpha}+
{\delta^{\sigma}}_{\rho}{\delta^{\beta}}_{\alpha})$. Entonces,
sean $n_\mu$ las componentes de trivectores, definimos la matriz
característica como
\begin{eqnarray}
{\mathcal{M}^{\beta\sigma}}_{\rho\alpha}(n)={\delta
^{\beta\sigma}}_{\rho \alpha}\,n^2\, \, , \label{d34}
\end{eqnarray}
y la ecuación característica es
\begin{eqnarray}
det(\mathcal{M})=(n^2)^6=0 \, \, , \label{d35}
\end{eqnarray}
teniendo un vector nulo como solución.

Los espacios de dS/AdS son conformalmente planos y sus conos de
luz son equivalentes a los de el espacio de Minkowski, pues ellos
estan relacionados vía un mapa de Weyl. Seguido a esto podemos
escribir la ecuación $n^2=0$ en un sistema localmente
''plano-Weyl'' a través de un mapa, usando el hecho de que la
transformación conforme para la métrica es
$g_{\mu\nu}=\Omega^{2}(x)\, \eta_{\mu\nu}$ (Apéndices B y C), como
sigue
\begin{eqnarray}
-(n_0)^2+n_in_i=0 \, \, , \label{d36}
\end{eqnarray}
la cual claramente describe una propagación hiperbólica ($n_0 \in
\mathrm{R}$) y causal ($n^2=0$ implica que no hay trivectores
tipo-tiempo).

Por otro lado, la tabla 1 de valores prohibidos de masa en dS/AdS
debe ser extendida debido a la restricción (\ref{d31})

\begin{center}
\begin{tabular}{|c|c|c|}
 \hline
\textbf { $R$} & \textbf{ $\sigma$} & \textbf{ $m\,\,\,prohibida$} \\
\hline \hline
$>0 \,(AdS)$& $\leq \frac{1}{24}$ & $m_{\pm}$  \\
\hline
$>0 \,(AdS)$& $<0$ & $\sqrt{-\sigma R}$  \\
\hline
$<0 \,(dS)$ & $>0$ & $\sqrt{-\sigma R}$  \\
\hline

\end{tabular}
\vspace{0.5cm}

 tabla 2

\end{center}

Queremos notar que en el estudio de la teoría del campo autodual
acoplado con gravedad, un hecho bien conocido es verificado:
dentro del posible conjunto de soluciones, en las que corresponden
a espacios de curvatura constante es respetado el número de grados
de libertad y la causalidad. Sin embargo, y en contraste con otras
clases de teorías que propagan espín 2 \cite{r10},  la de tipo
autodual no posee límite no masivo (de hecho, $m\neq 0$ es una
condición necesaria), y más aún, la restricción $M^2\neq 0$ es
demandada para poderse garantizar la equivalencia entre el sistema
de vínculos lagrangianos y la existencia de un campo (autodual)
con propiedades de simetría, transversalidad, traza nula, provisto
de una ecuación de movimiento hiperbólica-causal.

Hay otros aspectos del modelo autodual en dS/AdS, relacionados con
la condición $M^2\neq 0$. Por ejemplo, la acción (\ref{d21}) no
posee invariancia conforme (Apéndice C), lo cual es esencialmente
un reflejo de la restricción $M^2\neq 0$, ya que la traza del
tensor momento-energía del campo autodual en la capa de masas es
\begin{eqnarray}
{T^\mu}_\mu=-\frac{M^2}2\,{h^{(s)Tt}}_{\mu \nu}h^{(s)Tt\mu \nu} \,
\, , \label{d37}
\end{eqnarray}
razón por la cual, además no sea posible mediante una
transformación de Weyl, conseguir un marco local en el cual la
propagación sea no masiva. Pero, si se persiste en mantener un
término cuadrático en los campos al estilo Proca en la acción, no
hay una via consistente de ''mejorar'' el término lineal en masa,
de la manera general en la que $m^2$ es reemplazada por $f(N)R$
(Apéndice C), donde $f(N)$ es una función de la dimensión.

Más aún, el valor crítico $M^2=0$ revela la existencia de una
discontinuidad en la teoría autodual. Por un lado, está la
inconsistencia que ocurre en el sistema de vínculos Lagrangianos
cuando es evaluado este límite, siendo esto es una circuns-tancia
no asociada al esquema de acoplamiento con el campo externo, sino
una característica propia de los términos de autointeracción
cuadráticos en los campos. Esto podría ilustrarse desde el punto
de vista de la teoría plana, considerando la acción siguiente con
dos parámetros
\begin{eqnarray}
{S_{m_1,\,m_2}}= \int d^3 x (\frac{m_1}2\,\epsilon ^{\mu \nu
\lambda }{h_\mu }^\alpha
\partial _\nu
h_{\lambda \alpha }-\frac{{m_2}^2}2(h_{\mu \nu }h^{\nu \mu }-h^2
)) \, \, ,\label{d38}
\end{eqnarray}
la cual contiene a la teoría autodual para el caso especial
$m_1=m_2$. En nuestra discusión, es suficiente notar que del
procedimiento de construcción de la acción reducida de (\ref{d38})
(siguiendo el esquema presentado en la sección \S 2.2),  se puede
mostrar que tomando $m_2=0$, se obtiene una teoría que no propaga
grados de libertad locales (de hecho la acción reducida es
idénticamente nula). Pero si en cambio consideramos $m_2\neq 0$
durante todo el procedimiento, se llega a  la expresión esperada:
${S_{m_1,\,m_2}}^{*}=\int d^3 x \{P\dot{Q}-\frac12 P^2 + \frac12
\,Q(\Delta -\mathbb{M}^2)Q\}$, donde $\mathbb{M}\equiv
\frac{{m_2}^2}{m_1}$\,,\,\, $P\equiv \sqrt 2 \,m_2 \Delta
{h^{(s)TL}}$ y $Q\equiv \sqrt 2 \,\frac{m_1}{m_2} \Delta
{h^{(s)TT}}$ es una función singular en $m_2=0$, evi-denciándose
que la teoría no posee un límite bien definido en este punto
crítico. De manera análoga, lo anterior se refleja en la acción
para  espacios de dS/AdS, (\ref{d21}) cuando $M^2 =  0$.

En síntesis, las restricciones (\ref{d24}) y (\ref{d31}),
\begin{equation}
6M^4-Rm^2\neq 0\, \, , \label{d40a}
\end{equation}
\begin{equation}
M^2\neq  0\, \, , \label{d40b}
\end{equation}
con $M^2=m^2+\sigma R$, tienen información sobre el {\it
background} y además coinciden en el límite plano con la condición
de consistencia del modelo autodual: $m\neq 0$. Sin embargo,
insistimos en distinguir las condiciones (\ref{d40a}) y
(\ref{d40b}), pues hemos visto que $M^2$ aparece como una ''masa''
al cuadrado en la acción (\ref{d21}), con lo que esta última
podría pensarse como una versión curvilínea de la acción de dos
parámetros plana, (\ref{d38})(que contiene a la teoría autodual).
La presencia de $M^2$ en la acción (\ref{d21}) garantiza la no
invariancia conforme del modelo autodual y establece el car\'acter
discontínuo o no de la teoría. Este comportamiento imita al
término Lagrangiano en $m^2$ para el caso de la teoría plana.

\vspace{1cm}

{\bf 3.2.1) Acción reducida}\vskip .1truein

Si en el contexto de una teor\'{\i}a en un espacio-tiempo curvo se
desea realizar un procedimiento para obtener la acci\'on reducida
correspondiente, siguiendo un programa similar al desarrolado en
el caso plano, es posible encontrar serios obst\'aculos cuando se
estudian espacios no Minkowskianos, inclu\'{\i}dos los
conformalmente planos.

Teniendo en mente la teor\'{\i}a autodual, podemos decir que,
independientemente de lo extenuante que es el procedimiento de
descomposici\'on $2+1$ en un espacio curvo, lo cual  apunta a
colocar las componentes del campo autodual $h_{\mu\nu}$ en
t\'erminos de una descomposici\'on TL al estilo plano (si uno
ensaya este camino) y en caso de poderse identificar las variables
can\'onicas ''$Q$'' y ''$P$'' involucradas en la acci\'on, nos
encontramos finalmente con la dificultad de definir
consistentemente  potencias arbitrarias del D'Alembertiano en un
espacio no-Minowskiano (hecho relacionado con la imposibilidad de
implementar una descripción de Fourier consistente \cite{r57}),
que permiten expresar las componentes del campo autodual en
funci\'on de las variables can\'onicas. A esta situaci\'on debemos
agregar que, como consecuencia tampoco sea posible desarrollar un
procedimiento con proyectores en las diferentes componentes de
esp\'{\i}n.

Sin embargo, como pasaremos a discutir de inmediato, es posible
abordar cierto an\'alisis que apunta a la descripción del grado de
libertad propagado via la acci\'on reducida, al menos en el
contexto de espacios de curvatura constante.

Adicionalmente, requeriremos de un procedimiento que mantenga la
covariancia expl\'{\i}cita de la teor\'{\i}a, evitando los
inconvenientes adicionales que significa una descomposici\'on
$2+1$ curva. En este sentido, recurrimos a una descomposici\'on
Tt, al estilo plano.

Comenzamos por descomponer al campo autodual en sus partes
sim\'etrica y antisim\'etrica
\begin{equation}
h_{\mu \nu }\equiv { h^{(s)}}_{\mu \nu }+\varepsilon _{\mu \nu
\lambda }V^{\lambda}\, \, , \label{r42}
\end{equation}
que usando en la acci\'on (\ref{d21}), nos proporciona
\begin{eqnarray}
{S_{ad\lambda}}^{(2)}= \int d^3 x \,\sqrt{-g}\big(\,\frac
m2\,\varepsilon ^{\mu \nu \sigma }g^{\beta\alpha}{h^{(s)}}_{\mu
\beta} \nabla _\nu {h^{(s)}}_{\sigma \alpha }-\frac
{M^2}{2}\,({h^{(s)}}_{\mu \nu }h^{(s){\mu \nu }}-{h^{(s)}}^2
)+\nonumber
\\+\, mV^\mu (\nabla _\mu {h^{(s)}} -\nabla
_\nu{{h^{(s)}}_\mu}^\nu)-\frac m2\,\varepsilon ^{\mu \nu \sigma
}V_\mu \nabla _\nu V_\sigma -M^2V_\mu V^\mu \big) \, \, ,\nonumber
\\\label{r43}
\end{eqnarray}
cuyas ecuaciones de movimiento tienen un aspecto similar a las del
caso plano
\begin{eqnarray}
m\varepsilon ^{\mu \nu \lambda } \nabla_\mu{{h^{(s)}}_\nu }^\alpha
+m\varepsilon ^{\mu \nu \alpha }\nabla_\mu{{h^{(s)}}_\nu }^\lambda
-2M^2 h^{(s)\lambda
\alpha }+  2M^2 g^{\lambda \alpha }h^{(s)}+\nonumber \\
-2mg^{\lambda \alpha }\nabla_\mu V^\mu + m(\nabla^\lambda
V^\alpha+ \nabla^\alpha V^\lambda)=0
 \, \, , \label{rb1}
\end{eqnarray}
\begin{equation}
m\varepsilon ^{\mu \nu \lambda }\nabla_\mu V_\nu +2M^2 V^\lambda
-m\nabla^\lambda h^{(s)}+ m\nabla_\mu h^{(s) \mu\lambda }=0
 \, \, . \label{rb2}
\end{equation}
La traza y la divergencia  de (\ref{rb1}) proporcionan
\begin{eqnarray}
 M^2h^{(s)}-m\nabla_\mu V^\mu =0
 \, \, , \label{rb3}
\end{eqnarray}
\begin{eqnarray}
m\varepsilon ^{\mu \nu \lambda } \nabla_\nu
\overline{\mathcal{H}}_\lambda -2M^2\overline{\mathcal{H}}\,^\mu+
 2M^2 \nabla^\mu h^{(s)} +m\Box{}V^\mu +\nonumber \\-\,
m\nabla^\mu \nabla_\alpha V^\alpha-\frac{mR}{3}\,V^\mu=0
 \, \, , \label{rb4}
\end{eqnarray}
donde $\overline{\mathcal{H}}_\lambda \equiv
\nabla_\alpha{{h^{(s)}}_\lambda }^\alpha$. Usando la ecuación
(\ref{rb4}) en el rotacional de (\ref{rb2}), se obtiene
\begin{equation}
(Rm^2-6M^4)V_\sigma =0
 \, \, , \label{rb5}
\end{equation}
que gracias a la restricción (\ref{d24}) significa $V_\sigma =0$.
Ésto,  junto con (\ref{rb3}) conduce a la relación suplementaria
$h^{(s)}=0$, y  la ecuación de movimiento (\ref{rb2}) asegura que
$\overline{\mathcal{H}}_\lambda
\equiv\nabla_\alpha{{h^{(s)}}_\lambda }^\alpha =0$. Así, como en
el caso plano, de manera idéntica en los espacios de curvatura
constante se describe un campo autodual simétrico-transverso-sin
traza que propaga espín 2, y el espacio de fase reducido está
descrito únicamente por ${h^{(s)Tt}}_{\mu \nu }$, obteniéndose la
acción reducida de la forma
\begin{eqnarray}
{S_{sd\lambda}}^{(2)*}= \int d^3 x\,\sqrt{-g}\, \big(\,\frac
{m}2\,\varepsilon ^{\mu \nu \sigma }{{h^{(s)Tt}}_\mu }^\alpha
\nabla _\nu {h^{(s)Tt}}_{\sigma \alpha }-\frac {M^2}2\,
{h^{(s)Tt}}_{\mu \nu }h^{(s)Tt\mu \nu } \big) \, \, , \label{r65}
\end{eqnarray}
que tiene un aspecto similar a la del caso plano. Gracias a la
propiedad Tt, la ecuaci\'on de movimiento que se deriva de esta
acción  la podemos escribir como
\begin{eqnarray}
m\,\varepsilon ^{\sigma\mu \nu } \nabla _\mu{{h^{(s)Tt}}_\nu
}^\beta -M^2 h^{(s)Tt\sigma \beta }=0
 \, \, . \label{r66}
\end{eqnarray}
De esta ecuación  concluimos que se trata de una propagaci\'on
masiva, pues si actuamos sobre ella con $\varepsilon
_{\sigma\alpha \rho }\nabla^\rho$ obtenemos la ecuaci\'on
(\ref{d32}), es decir $\big( \Box{} -\frac{M^4}{m^2}+\frac{R}{2}
\big){h^{(s)Tt}}_{ \mu \nu}=0 $. En el espacio tangente, $T_p(M)$
con coordenadas locales planas $\xi^a$, esta última es
\begin{equation}
\big( \Box{}_{(\xi)} -\frac{M^4}{m^2}+\frac{R}{2}
\big){h^{(s)Tt}}_{ ab}(\xi)=0 \, \, . \label{r67}
\end{equation}
Seguidamente, definimos localmente las partes ''+'' y ''$-$'' de
${h^{(s)Tt}}_{ab}(\xi)$ de la forma
\begin{equation}
{{{h^{(s)Tt}}}_{ab}}^\pm \equiv \frac{1}{2}\bigg(\frac{{\delta
^d}_a{\delta ^c}_b}{\mathcal{A}}  \pm {\delta ^d }_a {\epsilon
_b}^{rc }\frac{\partial _r
}{{\Box{}_{(\xi)}}^{\frac{1}{2}}}\bigg){{h^{(s)Tt}}}_{dc}
 \, \, , \label{r68}
\end{equation}
donde $\mathcal{A}\equiv \sqrt{ 1- \frac{Rm^2}{2M^4}}$. Con esto,
en la capa de masas local (\ref{r67}) se obtiene
\begin{equation}
\big( \Box{}_{(\xi)} -\frac{M^4}{m^2}+\frac{R}{2}
\big){{{h^{(s)Tt}}}_{ab}}^+(\xi)=0 \, \, , \label{r69}
\end{equation}
\begin{equation}
{{{h^{(s)Tt}}}_{ab}}^-(\xi)=0 \, \, , \label{r70}
\end{equation}
indicando que solo se propaga localmente ${{{h^{(s)Tt}}}_{ab}}^+$
(${{{h^{(s)Tt}}}_{ab}}^-$ ), según el signo de la helicidad.

Es de observarse que la expresión (\ref{r68}) puede reescribirse
como ${{{h^{(s)Tt}}}_{ab}}^\pm \equiv {P^{\pm dc
}}_{ab}{{h^{(s)Tt}}}_{dc}$, donde
\begin{equation}
{P^{\pm dc }}_{ab}\equiv
\frac{1}{4}\bigg(\frac{1}{\mathcal{A}}\big({\delta ^d}_a{\delta
^c}_b+{\delta ^c}_a{\delta ^d}_b\big) \pm \big({\delta ^d }_a
{\epsilon _b}^{rc }+{\delta ^c }_a {\epsilon _b}^{rd
}\big)\,\frac{\partial _r }{{\Box{}_{(\xi)}}^{\frac{1}{2}}}\bigg)
 \, \, , \label{r71}
\end{equation}
no es un proyector, ya que $\mathcal{A}\neq 1$.

\newpage
{\large {\bf 4) Formulación $GL(N,R)$ de calibre de la gravedad}}
\vskip .5truein

El problema relacionado con la construcción de una teoría de
calibre para la gravitación abarca una considerable cantidad de
aproximaciones. Partiendo con Utiyama \cite{r21}, quien fuese uno
de los primeros en reconocer el carácter de ''calibre'' del campo
gravitacional al presentar una formulación para la gravedad basada
en la visión del grupo homogéneo de Lorentz como grupo de calibre,
pasando por construcciones en las que se relaja la propiedad de
simetría de la conexión (p.ej.: espacio-tiempo de Riemann-Cartan)
\cite{r30},  incluso donde se remueve la condición
métrico-compatible o metricidad ($\nabla_\alpha g_{\mu\nu}
\neq0$), arribando a teorías basadas en geometrías no-Riemannianas
\cite{r59}.

Con la finalidad de comparar y explorar la consistencia entre las
soluciones  de la teoría de Einstein con las provenientes de una
formulación de calibre de la gravedad,  enfocamos nuestra atención
en la construcción basada en un subgrupo afín, $GL(N,R)$
\cite{r25},\cite{r38},  la cual considera como campo de calibre a
la conexión afín. Obviamente, escoger este subgrupo produce
limitaciones relacionadas con el contexto de teorías
supersimétricas que demandan las simetrías de translación. Pero
teniendo en mente la discusión de la consistencia bajo un dado
esquema  de acoplamiento con materia, es suficiente abordar la
construcción de calibre $GL(N,R)$ en un espacio Riemanniano.

Abordaremos una densidad Lagrangiana de tipo Yang-Mills con el
grupo de calibre $GL(N,R)$, que estará relacionada con una de tipo
cuadrático en la curvatura de Riemann-Christoffel. Este tipo de
lagrangianos poseen gran interés, debido a que, desde el punto de
vista de la teoría de campos estándar conducen a teorías en las
que los problemas de renormalización son menos severos \cite{r60};
desde el punto de vista de la teoría de cuerdas, este tipo de
términos aparecen en el límite de bajas energías del Lagrangiano
efectivo de gravedad (ver \cite{r61}, por ejemplo).

El propósito principal es el de explorar un esquema covariante
general (o invariante de calibre) para el acoplamiento no minimal
con campos materiales en $N$ dimensiones, via la conexión de
calibre $GL(N,R)$. Así, podremos estudiar la consistencia entre
este tipo de teorías y la de Einstein. Como mostraremos, en el
caso del vacío, el último requerimiento demanda que en el
Lagrangiano se introduzca un término proporcional al cuadrado de
la constante cosmológica.

Por otra parte, cuando los campos materiales son introducidos, se
observa que la consistencia exigida produce restricciones sobre
estos campos y los posibles va-lores de la constante cosmológica.
Seguidamente, se encuentra que la introducción de términos
Lagrangianos de acoplamiento asociados a campos auxiliares,
permite remover las restricciones sobre los campos materiales
\cite{r38}. Allí, y como escenario útil debido a sus propiedades
geométricas particulares, abordaremos el caso $2+1$ dimensional.

Finalmente, en dimensión $2+1$ introducimos la formulación de
calibre $GL(3,R)$ de la gravedad topológicamente masiva con
constante cosmológica \cite{r62}, observando su límite de torsión
nula, el cual es consistente con la gravedad topológica masiva con
constante cosmológica \cite{r67}.

\vspace{1cm}

{\bf 4.1) Teoría libre}\vskip .1truein

Con la finalidad de motivar de una manera heurística la
introducción de una formulación de calibre basada en el grupo de
transformaciones generales de coordenadas de la Relatividad
General, comenzamos por establecer algunos de los elementos
formales desde el punto de vista geométrico-diferencial \cite{r63}
de tal cons-trucción.

Consideremos que nuestro espacio-tiempo $N$-dimensional es una
variedad dife-rencial, $M$  provista con métrica $g_{\mu \nu }$ de
signatura Lorentziana, coordenadas curvas $x^\mu $, y localmente
planas $\xi ^a $. En cada punto $p \in M$ se define un espacio
tangente, $T_p(M)$ como aproximación lineal de $M$ en una vecindad
de $p$. Esto nos permite representar objetos tensoriales en $M$
mediante tensores del espacio tangente.

Sean  ${V_{\mu}}^a$, las componentes de un objeto tensorial mixto
con índices curvo y plano local, que toman valores en la variedad,
entonces introducimos la conexión de espín, ${\omega _{\mu b}}^a $
y la afín, ${\Gamma ^\lambda }_{\mu \nu }$ con las cuales se
define la derivada covariante $ D_\mu $ que actúa sobre objetos
mixtos (apéndice B). De aquí, la condición de metricidad o
condición métrico-compatible de la conexión afín ($\nabla_\mu
g_{\alpha\beta }=0$), la garantizaremos demandando la propiedad
$D_\mu {e_\nu }^a =0$, donde ${e_\nu }^a$ son los vielbeins que
satisfacen $g_{\mu \nu } = {e_\mu }^a {e_\nu }^b \eta _{ab}$. De
esto sigue que el tensor de torsión puede ser escrito como
\begin{equation}
{T^\lambda }_{\mu \nu } \equiv {\Gamma ^\lambda }_{\mu \nu } -
{\Gamma ^\lambda }_{\nu \mu } = {e^\lambda }_a (\partial_\mu
{e_\nu }^a - \partial_\nu {e_\mu }^a + {\omega _{\mu \nu }}^a -
{\omega _{\nu \mu }}^a)
 \, \, , \label{eq1}
\end{equation}
donde la notación es ${\omega _{\mu \nu }}^a \equiv {e_\nu }^b
\,{\omega _{\mu b }}^a$, etc.

Si los elementos de matriz de $GL(N,R)$  y los de las
transformaciones de Lorentz locales son denotados como
 ${(U)^\alpha }_\mu \equiv \frac{\partial {x^\prime
}^\alpha }{\partial x^\mu}$ y ${(L)^a}_b \equiv \frac{\partial
{\xi^\prime }^a }{\partial \xi^b}$, respectivamente, el
comportamiento covariante de la derivada $D_{\mu}$ demanda las
siguientes reglas de transformación para las conexiones
\begin{equation}
{\omega ^\prime }_\mu  =L \omega _\mu L^{-1} + L \partial _\mu
L^{-1} \, \, , \label{eq2}
\end{equation}
\begin{equation}
{\mathbb{A}_a}^\prime =U\mathbb{A}_a U^{-1} + U \partial _a U^{-1}
\, \, , \label{eq3}
\end{equation}
con la notación ${(\mathbb{A}_a)^\mu }_\nu \equiv {e^\alpha }_a
{(\mathbb{A}_{\alpha })^\mu }_\nu$, siendo
\begin{equation}
{(\mathbb{A}_{\alpha })^\mu }_\nu \equiv  {\Gamma ^\mu }_{\alpha
\nu }
 \, \, , \label{eq4}
\end{equation}
observándose que ${(\mathbb{A}_a)^\mu }_\nu $ se comporta como una
conexión $GL(N,R)$ que transforma como un vector lorentziano
(local) en el índice plano. Esto sugiere la idea de que subyace
alguna estructura de fibrado construido a partir del espacio base
$M$.

Para dilucidar esto, comencemos recordando que toda variedad
$N$-dimensional diferenciable $M$ (que también llamaremos variedad
base), por definición está provista de una familia de cartas,
$\{(U_n\,,\,\varphi_n)\}$ que la cubren. Aquí, $U_n$ es una
vecindad del punto $p_n \in M$ (tal que $\bigcup_{todos\, los
\,n}U_n=M$) y $\varphi_n$ es la función de coordenadas
representadas por $N$ funciones reales
$\{x^0(p),\,x^1(p),...,x^{N-1}(p)\}$. El cubrimiento de $M$
también puede ser enfocado desde el punto de vista de una
colección de espacios tangentes definiendo la variedad
\begin{equation}
T(M)\equiv \bigcup_{p\in M}T_p(M)
 \, \, , \label{f1}
\end{equation}
y cuya definición podemos especializar para una vecindad
cualquiera de $M$
\begin{equation}
T(U_n)\equiv \bigcup_{p\in U_n}T_p(M)
 \, \, . \label{f2}
\end{equation}

Los elementos de $T(U_n)$ estan caracterizados por un arreglo del
tipo $(p\,,V(p))$ donde $V(p)=V^\mu (p)\,\frac{\partial}{\partial
x^\mu}\,\in\, T_p(M)$ es un vector representado en la base del
espacio tangente. Entonces, si observamos que la vecindad  $U_n$
es homeomórfica a un subconjunto de $R^N$ y que cada $T_p(M)$ a
$R^N$, se concluye que $T(U_n)$ puede ser identificado con
$R^N\times R^N$.

Seguidamente, se introduce un mapa sobreyectivo llamado {\it
proyección}, que manda puntos  de $T(U_n)$ a $U_n$, $\pi :
T(U_n)\longrightarrow U_n$. Esto indica que para cualquier punto
$u \in T(U_n)$, $\pi (u)$ es un punto $p \in U_n$ en donde está
definido un vector $V(p) \in T_p(M)$. Entonces, la {\it fibra} en
el punto $p$ se define como $F_p\equiv \pi^{-1}(p)=T_p(M)$, razón
por la cual esta construcción  recibe el nombre de {\it fibrado
tangente}.

Es fundamental notar que la identificación de $T(U_n)$ con
$R^N\times R^N$ nos dice que $T(M)$ posee esta estructura
localmente. En el caso del fibrado trivial $T(M)=R^N\times R^N$ (o
localmente en un caso no trivial), se puede ver que un elemento
$V(p)$ del espacio tangente a un punto $p \in U_{mn}\equiv
U_{m}\bigcup U_{n}$ posee simultáneamente dos representaciones
relacionadas por
\begin{equation}
{V'^\mu}_{(x'_{(p)})}={{(U)^\mu
}_\nu}_{(x_{(p)})}\,{V^\nu}_{(x_{(p)})}
 \, \, , \label{f3}
\end{equation}
donde ${(U)^\mu }_\nu$ es un elemento no singular de $GL(N,R)$.
Esto indica que los elementos de la fibra son transformados
mediante $\mathcal{G}\equiv GL(N,R)$, el cual recibe el nombre de
{\it grupo de estructura} del fibrado tangente, $T(M)$.

La construcción restante, es decir, el mapa de {\it trivialización
local} que manda a $\pi^{-1}(U_n)\longrightarrow U_n \times F$ y
las {\it funciones de transición}, $t_{mn}\in \mathcal{G}$, que
relacionan dos elementos de la fibra en $p \in U_{mn}$ (esto es
$f_m=t_{mn}(p)f_n$ con $f_m$ y $f_n$ $\in F_p$), seran asumidos
como ya establecidos.

Dada la formulación del fibrado tangente, si mantenemos la
variedad base, el grupo de estructura, $\mathcal{G}$ y las
funciones de transición, podemos reemplazar la fibra, $T_p(M)$ por
$\mathcal{G}$ y obtener, así el {\it fibrado principal}, $P(E)$.
Este tipo de fibrado, asociado a cualquier fibrado $E$ juega un
rol fundamenal en el criterio que se usa para discernir la
trivialidad o no de éste último. En nuestro caso, es bien sabido
que el fibrado principal asociado a $E\equiv T(M)$ es el {\it
fibrado de referenciales}  ({\it frame bundle}), $P(E)=F(M)$, cuya
fibra consiste en la colección ordenada de todas las bases del
espacio tangente. Como un par de bases cualesquiera estan
conectadas por la acción del grupo $\mathcal{G}=GL(N,R)$, la fibra
en cuestión es identificada con la variedad asociada a $GL(N,R)$.

Los términos {\it conexión} y {\it curvatura} estan estrechamente
vinculados. Cuando decimos que nuestra variedad $M$ es curva nos
referimos a que el espacio tangente cambia en la medida en que nos
movemos de un punto a otro en la variedad. A la par de esto, la
conexión se encarga de establecer un ''puente'' entre distintos
espacios tangentes, y ésta es introducida formalmente cuando se
 realiza el transporte paralelo.

Desde el punto de vista de la estructura geométrica discutida, el
transporte paralelo requiere  que el espacio tangente $T_u(P)$ al
fibrado principal $P(E)\equiv F(M)$, sea separado (suma directa)
en un sub-espacio vertical y uno horizontal
\begin{equation}
T_u(P)=V_u(P)\oplus H_u(P)\,\,\,\,\,\,,\,\,\,u\in P(E)
 \, \, , \label{f4}
\end{equation}
con lo cual un campo vectorial bien definido en $P(E)$ puede ser
descompuesto en sus partes pertenecientes a $V_u(P)$ y $H_u(P)$.
Además es demandado que los subespacios $H_u(P)$ y $H_{u'}(P)$,
con $u'=ug\,,\,\,g\in \mathcal{G}$, esten relacionados por un mapa
lineal, asegurándose el transporte paralelo a lo largo de la fibra
(todos los $H_{u'}(P)$ de una fibra quedan determinados por
$H_u(P)$). Entonces, con todo esto y si la separación
$T_u(P)=V_u(P)\oplus H_u(P)$ es única, y se dice que se ha
definido una conexión.

Esta construcción geométrica es realizada mediante la introducción
de una uno-forma de conexión, $\mathcal{A}_n$ perteneciente al
espacio cotangente (${T_p}^*(P)$) en un punto $p$ de $U_n$, y que
toma valores en el álgebra del grupo $\mathcal{G}$ (p.ej.:
$\mathcal{A}=\mathcal{A}_{\mu}dx^{\mu}={\mathcal{A}^a}_{\mu}\,t^a
dx^{\mu}$ donde $t^a$ son los generadores del grupo
$\mathcal{G}$). Esta conexión local debe satisfacer la condición
de compatibilidad $\mathcal{A}_n ={t_{mn}}^{-1}
\mathcal{A}_mt_{mn}+{t_{mn}}^{-1} dt_{mn}$, en los puntos
pertenecientes al solapamiento de las cartas $U_{mn}\equiv
U_{m}\cup U_{n}$. Tal condición esencialmente encierra el concepto
de transformación de calibre, ya que escogiendo dos secciones
locales, $\sigma_1(p)$ y $\sigma_2(p)$ (definidas por mapas de $p
\in M$ en la fibra de $P(E)$) sobre una carta de $M$, y sabiendo
que éstas estan relacionadas mediante la acción del grupo de
estructura ($\sigma_2(p)=\sigma_1(p)g(p)\,,\,g\in \mathcal{G}$),
entonces las uno-forma de conexión locales, $\mathcal{A}_1$ y
$\mathcal{A}_2$ se relacionan mediante (en componentes)
\begin{equation}
{\mathcal{A}_{2\mu}}_{(p)} ={g^{-1}}_{(p)}
{\mathcal{A}_{1\mu}}_{(p)}g_{(p)}+{g^{-1}}_{(p)}
\partial_\mu g_{(p)}\,\,\,\,\,\,,\,\,\,p\in U_1\cup U_2
 \, \, , \label{f5}
\end{equation}
que salvo un cambio de nomenclatura es la misma relación
(\ref{eq3}).

Queremos insistir en notar que, siguiendo la filosofía de Cartan y
Palatini uno puede asumir una descripción en la que tanto la
metricidad (estructura métrica asociada a la variedad), como el
paralelismo (estructura afín) sean conceptos independientes, sin
establecer a priori ninguna relación funcional entre la conexión y
la métrica.

Siguiendo con la construcción de los objetos que necesitaremos
para la formulación Lagrangiana, introducimos el tensor de
curvatura de Riemann, ${R^\alpha}_{\sigma \mu \nu }$ a través de
la aplicación del conmutador $[\nabla_\mu ,\nabla_\nu]\equiv
\nabla_\mu \nabla_\nu - \nabla_\nu \nabla_\mu $ sobre algún tensor
de rango 1 (Apéndice B), en donde pueden identificarse las formas
del tensor de curvatura y torsión. Usando esta definición, las
componentes del tensor de curvatura son reescritas como
\begin{equation}
{R^\sigma}_{\alpha \mu \nu }\equiv {(\mathbb{F}_{\nu \mu})^\sigma
}_\alpha
 \, \,, \label{eq5}
\end{equation}
donde
\begin{eqnarray}
{(\mathbb{F}_{\mu \nu})^\sigma}_\alpha \equiv {\big (\partial _\mu
\mathbb{A}_\nu -
\partial _\nu \mathbb{A}_\mu+
[\mathbb{A}_\mu\,,\,\mathbb{A}_\nu]\big)^\sigma}_\alpha \qquad
,\label{eq5a}
\end{eqnarray}
son las componentes de la curvatura de tipo Yang-Mills.

\vspace{1cm}
{\bf 4.1.1) Ecuaciones de campo}\vskip .1truein

Con la finalidad de estudiar la relación entre las soluciones
obtenidas de una formulación lagrangiana invariante de calibre
construida con $\mathbb{F}_{\alpha \beta}$, y las correspondientes
a la gravedad de Einstein con constante cosmológica en un espacio
sin torsión ($EG\lambda$), consideraremos que las soluciones
físicas del tensor de Ricci en el contexto de la  $EG\lambda$, son
las que satisfacen la ecuación de campo
\begin{equation}
R^{\alpha \beta }-\frac{g^{\alpha \beta }}2R-\lambda g^{\alpha
\beta}=-8\pi G T^{\alpha \beta } \,\, , \label{eq6a}
\end{equation}
donde $T^{\alpha \beta }$ es el tensor momento-energía asociado a
los campos materiales y $\lambda$ la constante cosmológica.

Un primer paso será el de presentar el modelo libre (sin materia)
con torsión no necesariamente nula. Así, sea la acción invariante
de calibre para $T^{\alpha \beta } =0$
\begin{equation}
S_o =\kappa^{4-N} \int d^N x \sqrt{-g} \,\, (-\frac 14 \, tr\,
\mathbb{F}^{\alpha \beta}\mathbb{F}_{\alpha \beta}+ \Lambda
(\lambda) ) \,\, , \label{eq7}
\end{equation}
donde  $\Lambda(\lambda) $ está relaciona con la constante
cosmológica y $\kappa$ tiene unidades de longitud (o simplemente
decimos, $[\kappa] \sim l$). Obsérvese que la acción (\ref{eq7})
coincide con una teoría Lagrangiana cuadrática del tensor de
Riemann de la forma $S_o =\kappa^{4-N} \int d^N x \sqrt{-g} \,\,
(-\frac 14 \,  R^{\alpha \beta \sigma \rho}R_{\alpha \beta
\rho\sigma}+ \Lambda (\lambda) )$.

La primer variación de $S_o$ en la conexión, a menos de un término
de borde es $\delta _\mathbb{A} S_o=\kappa^{4-N}\int d^N x
\sqrt{-g}\,\,tr\, \mathbb{E}^\lambda \delta \mathbb{A}_\lambda$,
con lo que la ecuación de movimiento es
\begin{equation}
\mathbb{E}^\lambda \equiv \frac 1{\sqrt{-g}}\,\,\partial _\mu
(\sqrt{-g}\,\,\mathbb{F}^{ \mu \lambda}) +
[\mathbb{F}^{\lambda\mu} ,\mathbb{A}_\mu ]=0\,\, , \label{eqmm6}
\end{equation}
y si variamos (\ref{eq7}) respecto a los vielbeins (o la métrica)
obtenemos
\begin{equation}
T_{\mu\nu}(\mathbb{F})+ \kappa^{4-N}\Lambda(\lambda) \,g
_{\mu\nu}=0\,\, , \label{eqmm1}
\end{equation}
donde
$T_{\mu\nu}(\mathbb{F})=\kappa^{4-N}\,tr\,(\mathbb{F}_{\mu\beta}{\mathbb{F}_\nu}^\beta
- \frac{g_{\mu\nu}}4 \,\mathbb{F}^{\alpha \beta}\mathbb{F}_{\alpha
\beta})$, es el tensor momento-energía de Belinfante asociado a la
curvatura $GL(N,R)$. La ecuación (\ref{eqmm1}) nos dice, en
términos generales que en esta formulación sin materia, la fuente
de energía y momento gravitacional proviene de la constante
cosmológica (via $\Lambda (\lambda)$), lo cual es una suerte de
visión compartida con la formulación original de Hilbert-Einstein
en el caso del vacío con constante cosmológica, pues en este caso,
esta última podría ser vista como fuente de curvatura del
espacio-tiempo.

En un espacio sin torsión, el miembro izquierdo de (\ref{eqmm6})
se reduce a $\mathbb{E}^\lambda = \nabla_\mu \mathbb{F}^{\lambda
\mu}$, el cual puede ser reescrito en términos del tensor de Ricci
con la ayuda de las identidades de Bianchi,  de la forma
\begin{equation}
(\mathbb{E}_\lambda )_{\sigma \alpha } = \nabla_\alpha R_{\lambda
\sigma } - \nabla_\sigma R_{\lambda \alpha } \,\,. \label{eqmm7}
\end{equation}
Las soluciones triviales sin torsión para $\delta _A S_o=0$ son
las de tipo dS/AdS. Tomando la solución
 $R_{\alpha \beta }=-(2\lambda /(N-2))
g_{\alpha \beta}$, la curvatura $\mathbb{F}_{\alpha\beta}$ puede
ser evaluada usando (\ref{eq5}), o sea
\begin{equation}
{(\mathbb{F}_{\mu\nu})^\alpha }_\beta =
\frac{2\lambda}{(N-2)(N-1)} ({\delta^\alpha }_\nu g_{\beta
\mu}-{\delta^\alpha }_\mu g_{\beta \nu})
 \, \,, \label{eq10}
\end{equation}
y puede mostrarse por sustitución, que esta última satisface
idénticamente a la ecuación (\ref{eqmm1}) si
\begin{eqnarray}
\Lambda(\lambda) =-\frac{2(N-4)}{(N-2)^2(N-1)}\,\,\lambda ^2 \,\,
. \label{eq12}
\end{eqnarray}
En otras palabras, esta condición garantiza que los espacios de
dS/AdS son soluciones triviales de la extremal de la acción $S_o$.
Puede notarse que la constante $\Lambda$ no distingue entre dS o
AdS en este tipo de formulación Lagrangiana, en contraste con lo
ocurrido en la de Hilbert-Einstein. Sin embargo, $\Lambda$
establece otro tipo de clasificación. Cuando $N=3$, $\Lambda $
toma los valores $\lambda ^2 \geq 0$, mientras que la constante
cosmológica no aparece explícitamente en la acción para  $N=4$.
Por otro lado, si $N
> 4$ se tiene $\Lambda \leq 0$. Inmediatamente uno puede decir
que el contenido físico de las clases de $\Lambda$ está
relacionado no solo con la curvatura del espacio-tiempo sino con
una corrección del Hamiltoniano asociado.

En el análisis variacional que hemos asumido, consecuentemente
hemos pensado en un principio de tipo Palatini (p.ej.: variaciones
independientes de la conexión $GL(N,R)$ y los vielbeins o la
métrica), pensando en la configuración general ${T^\lambda }_{\mu
\nu }\neq 0$, para las variables definidas.

Ahora bien, debemos notar que el resultado obtenido  puede
recuperarse a partir de un procedimiento más general. Un punto de
partida alternativo podría ser el de redefinir una nueva
lagrangiana en la que se incluyan $\frac{N^2(N-1)}{2}$ vínculos
sobre la conexión, los cuales tienen la forma
\begin{equation}
\frac{1}{2}\,{T^\lambda }_{\mu \nu }= \frac{1}{2}\,[
{(\mathbb{A}_\mu)^\lambda}_\nu -{(\mathbb{A}_\nu)^\lambda}_\mu] =
0
 \, \,, \label{e1}
\end{equation}
y  los  llamaremos ''vínculos de torsión''. Para esto,
introducimos $\frac{N^2(N-1)}{2}$ multiplicadores de Lagrange,
$\mathcal{C}_{\alpha\mu\nu}$ que satisfacen la condición de
antisimetría
\begin{equation}
\mathcal{C}_{\alpha\mu\nu} = -\mathcal{C}_{\alpha\nu\mu}
 \, \,. \label{e2}
\end{equation}
Con esto, la nueva acción con vínculos es
\begin{equation}
S'_o = S_o+\kappa^{1-N}\int d^N x \sqrt{-g}
\,{\mathcal{C}_\alpha}^{\lambda\sigma}
{(\mathbb{A}_\lambda)^\alpha}_\sigma \,\, . \label{eqam2}
\end{equation}

Las ecuaciones de movimiento asociadas a la conexión, evaluadas
sobre los vínculos de torsión ahora tienen la forma, $\nabla_\mu
{(\mathbb{F}^{\lambda \mu })^\sigma}_\alpha +
\kappa^{-3}{\mathcal{C}_\alpha}^{\lambda\sigma}=0$, que en
términos del tensor de Ricci son reescritas como
\begin{equation}
\nabla_\alpha R_{\lambda \sigma } - \nabla_\sigma R_{\lambda
\alpha }+ \kappa^{-3}\mathcal{C}_{\alpha\lambda\sigma}=0
 \, \,. \label{eqamm3}
\end{equation}

Las variaciones en los vielbeins, evaluadas sobre los vínculos de
torsión, siguen proporcionando (\ref{eqmm1}). Los vínculos
(\ref{e1}), junto con la condición de metricidad significan
$(\mathbb{A}_\mu)_{\lambda\nu}\equiv{\Gamma_\lambda}_{\mu\nu}=\frac{1}{2}\,(\partial_\mu
g_{\lambda\nu} +\partial_\nu g_{\lambda\mu}- \partial_\lambda
g_{\mu\nu})$.

Así, manipulando a (\ref{eqamm3}), se obtienen los muliplicadores
y una condición para la curvatura escalar, es decir
\begin{eqnarray}
\mathcal{C}_{\alpha\lambda\sigma}=0 \,\, , \label{eqamm5}
\end{eqnarray}
\begin{eqnarray}
R=constante \,\, . \label{eqamm6}
\end{eqnarray}
Obsérvese que (\ref{eqamm5}) indica que, al menos en el caso
libre, el límite de torsión nula se puede obtener mediante la
evaluación directa de las ecuaciones de campo con torsión,
(\ref{eqmm6}) y (\ref{eqmm1}), tomando la conexión como los
símbolos de Christoffel. En la sección $\S$ 4.3.3, mostraremos que
éste no es el caso de la formulación de calibre de la gravedad
topológicamente masiva con constante cosmológica ya que, para
obtener un límite de torsión nula consistente deben incluirse
necesariamente los vínculos de torsión. La restricción sobre la
curvatura, (\ref{eqamm6}) significa una relación de consistencia
con la solución de tipo dS/AdS de la gravedad de Einstein.

Más aún, aquí es importante notar que ésta construcción es
consistente con la formulación original de Einstein, incluso
concluyendo que en una variedad contractible sin constante
cosmológica, la ecuación $\mathbb{F}_{\alpha\beta }=0$ posee
solución nula, en concordancia con el hecho bien conocido de que
en tal caso la gravitación libre no propaga grados locales de
libertad.

\vspace{1cm}
{\bf 4.2) Acoplamiento con materia}\vskip .1truein

Inspirados en un modelo de acoplamiento minimal, más un término de
auto-interacción, como el caso del campo de Proca acoplado con una
corriente: $S=\int d^N\xi (-\frac 14\,f^{\mu\nu}f_{\mu\nu}+j_\mu
A^\mu+\frac{m^2}{2}\,A_\mu A^\mu)$, aquí realizamos un primer
ensayo de un esquema de acoplamiento con materia a través de la
conexión $\mathbb{A}_{\mu}$ \cite{r38}, como una extensión no
abeliana de la idea mencionada. Así, exploraremos un posible
esquema general y covariante bajo $GL(N,R)$, para el acoplamiento
no minimal con campos materiales, de la forma
\begin{equation}
S=\kappa^{4-N}\int d^N x \sqrt{-g}\,\big( -\frac 14 \, tr\,
\mathbb{F}^{\alpha \beta}\mathbb{F}_{\alpha \beta}+ \Lambda
(\lambda)+\ell(g,\psi) + 4\pi
G\,\,tr\,\mathbb{M}^{\alpha\beta}(g,\psi)\mathbb{F}_{\alpha\beta}
\big)
 \, \,, \label{eq13}
\end{equation}
donde el término invariante de calibre  $\ell(g,\psi)$ y el tensor
antisimétrico $\mathbb{M}^{\alpha\beta}(g,\psi) = -
\mathbb{M}^{\beta\alpha}(g,\psi)$ dependen de la métrica y los
campos materiales. De estas definiciones sigue que, después de una
integración por partes de la acción $S$, pueden obte-nerse un
término tipo ''corriente'' (p.ej.: acoplamiento minimal) y un
término de ''masa'' (p.ej.: acoplamiento de tipo Proca).
Obviamente, nuestro problema es el de explorar la forma de los
objetos $\ell(g,\psi)$ y $\mathbb{M}^{\alpha\beta}(g,\psi)$,
requeriendo la consistencia con las soluciones de la gravedad de
Einstein.

Aquí, siguiendo la idea de obtener un término de ''masa''
proponemos un acoplamiento via la fuente de momento y energía de
los campos materiales (que la tomaremos como el tensor
momento-energía de Belinfante de éstos, $T_{\sigma \rho}$), es
decir
\begin{equation}
{(\mathbb{M}^{\alpha \beta})^\mu }_\nu = {(N^{\alpha \beta \sigma
\rho})^\mu}_\nu \,\,T_{\sigma \rho} + {(n^{\alpha \beta})^\mu}_\nu
\, \,, \label{eq14}
\end{equation}
donde los objetos $N^{\alpha \beta \sigma \rho}$ y $n^{\alpha
\beta}$ dependen solo de la métrica y pediremos que tengan las
propiedades de simetría siguientes
\begin{equation}
{(N^{\alpha \beta \sigma \rho})^\mu}_\nu =-{(N^{\beta\alpha \sigma
\rho})^\mu}_\nu ={(N^{\alpha \beta \rho\sigma })^\mu}_\nu \, \,,
\label{eq14a}
\end{equation}
\begin{equation}
{(n^{\alpha \beta})^\mu}_\nu =-{(n^{\beta\alpha })^\mu}_\nu\, \,,
\label{eq14b}
\end{equation}
Las formas generales de éstos, construidas con la mética, y en
consistencia con las propiedades de simetría esperadas son
\begin{eqnarray}
{(N^{\alpha \beta \sigma \rho})^\mu}_\nu \equiv c_1 \big(g^{\mu
\alpha } {\delta ^\beta }_\nu  -g^{\mu \beta}{\delta ^\alpha }_\nu
\big)g^{\sigma \rho}
 + c_2\big(g^{\rho \alpha } {\delta ^\beta }_\nu -g^{\rho \beta}{\delta
^\alpha }_\nu \big)g^{\sigma \mu} \nonumber \\+ \,c_3\big(g^{ \mu
\alpha } g^{\rho \beta } -g^{\mu \beta}g^{\rho \alpha }
\big){\delta ^\sigma }_\nu
 \, \,, \label{eq15}
\end{eqnarray}
\begin{equation}
{(n^{\alpha \beta})^\mu}_\nu \equiv a \big(g^{\mu \alpha } {\delta
^\beta }_\nu - g^{\mu \beta}{\delta ^\alpha }_\nu \big)\, \,,
\label{eq16}
\end{equation}
siendo $c_1$, $c_2$, $c_3$ y $a$  parámetros reales.

Como ilustración de esta primer aproximación del esquema de
acoplamiento, consideraremos un sistema de campos materiales cuyo
tensor momento-energía no depende explícitamente en  la conexión,
solo por simplicidad. Este tipo de sistemas se puede insertar
dentro de una clase caracterizada por un tensor momento-energía,
cuyo aspecto es de la forma
\begin{equation}
T_{\mu \nu } = (\alpha \,{\delta ^\lambda }_\mu {\delta ^\rho
}_\nu + \beta \,g^{\lambda \rho}g_{\mu \nu})\psi _{\lambda \rho}
 \, \,, \label{eq17}
\end{equation}
con $\alpha $ y $\beta$ reales, y $\psi _{\lambda \rho}$ un tensor
simétrico que contiene la información acerca de los campos
materiales (tabla 3). El tensor (\ref{eq17}) puede describir
algunos sistemas interesantes, como los que mostramos a
continuación
\begin{center}
\begin{tabular}{|c|c|c|c|c|}
 \hline
\textbf { $Fuente$}  & \textbf{ $\psi_{\mu\nu}$}& \textbf{ $\alpha$}& \textbf{ $\beta$} \\
\hline \hline
$campo\,\,escalar\,\,no\,\,masivo$& $\partial_\mu\phi\partial_\nu\phi$ & $1$& $-\frac{1}{2}$\\
\hline
$fluido\,\,homog$\'e$neo$& $g_{\alpha\mu}g_{\beta\nu}U^\alpha U^\beta$ & $p+\rho$ & $p$ \\
\hline
$campo\,\,electromagn$\'e$tico$& $g^{\lambda\rho}f_{\mu\lambda}f_{\nu\rho}$ & $1$ & $\frac{1}{4}$ \\
\hline
\end{tabular}
\vspace{0.5cm}

 tabla 3

\end{center}

En la tabla anterior, además del campo escalar real no masivo,
hemos considerado un fluido homog\'eneo perfecto con densidad
$\rho$, presión $p$ y velocidad $U^{\lambda}$, y también un campo
electromagn\'etico con campo de Maxwell, $f_{\mu\nu}\equiv
\nabla_\nu A_\mu -\nabla_\mu A_\nu$. Como es de observarse, en
estos dos ultimos casos, los campos $\psi_{\mu\nu}$ tienen
dependencia en la métrica, y en el de Maxwell, incluso en la
conexión afín (si la torsión no es nula). De ahora en adelante
estudiaremos el caso particular en que $\psi_{\mu\nu}$ no depende
de la métrica (p. ej.: campo escalar real no masivo), sin
modificar notablemente las generalidades de los resultados que
discutiremos, al menos en relación con el caso de dependencia
exclusiva en la métrica y campos materiales \cite{r71}.

Así, realizando variaciones en la conexión de la acción $ S$, se
obtiene
\begin{eqnarray}
\delta _\mathbb{A} S =\kappa^{4-N} \int d^N x
\sqrt{-g}\,\,tr\bigg[ \mathbb{E}^\lambda   - 8\pi G\,\big( \frac
1{\sqrt{-g}}\,\,\partial _\mu (\sqrt{-g}\,\mathbb{M}^{\lambda \mu
})+ [\mathbb{M}^{\lambda \mu} ,\mathbb{A}_\mu ] \big) \bigg]
\delta
\mathbb{A}_\lambda \, \,,\nonumber \\
  \label{eq18}
\end{eqnarray}
a menos de un término de borde. Cuando las ecuaciones de
movimiento provenientes de $\delta _A S =0$ son escritas en un
espacio-tiempo sin torsión, con la ayuda de (\ref{eqmm7}),
(\ref{eq14}) y la condición de metricidad, se tiene
\begin{eqnarray}
\nabla_{\nu}\big(R^{\alpha \mu}-8\pi G c_1 g^{\alpha \mu} T -8\pi
G c_2T^{\alpha \mu}+c_4 \lambda g^{\alpha \mu} \big)+\nonumber \\-
\nabla^{\mu}\big({R^{\alpha}}_{\nu}-8\pi G c_1
{{\delta}^{\alpha}}_{\nu}T -8\pi G c_3{T^{\alpha}}_{\nu}+c_4
\lambda {{\delta}^{\alpha}}_{\nu} \big)+\nonumber \\+\,8\pi G
\big(c_2 {{\delta}^{\alpha}}_{\nu} \nabla_\beta T^{\mu \beta}
-c_3g^{\alpha \mu} \nabla_\beta {T^{\beta}}_{\nu}\big)=0
 \, \,, \label{eq19}
\end{eqnarray}
donde hemos usado la libertad de introducir un término cosmológico
con parámetro  $c_4$. La ecuación (\ref{eq19}) dice que las
soluciones de la gravedad de Einstein con cons-tante cosmológica
($EG \lambda$) siguen siendo triviales (como ocurría con las de
dS/AdS para el vacío, como nuestra condición de consistencia), si
los parámetros toman los valores
\begin{equation}
c_4=2c_1=\frac2{N-2}\,\,\,,\,\,\,c_2=c_3=-1
 \, \,, \label{eq20}
\end{equation}
pues con esto, los argumentos en las dos primeras derivadas
covariantes en (\ref{eq19}), son simplemente $R^{\alpha \beta
}-\frac{g^{\alpha \beta }}2R-\lambda g^{\alpha \beta}+8\pi G
T^{\alpha \beta }$, los cuales se anulan sobre las soluciones de
la $EG \lambda$.

Así, sobre las soluciones de la  $EG \lambda$, la ecuación
(\ref{eq19}) se reduce a
\begin{eqnarray}
-{{\delta}^{\alpha}}_{\nu} \nabla_\beta T^{\mu \beta} +g^{\alpha
\mu} \nabla_\beta {T^{\beta}}_{\nu} =0
 \, \,, \label{eq19a}
\end{eqnarray}
siendo ésta equivalente a demandar que $T^{\mu \nu}$ sea
conservado
\begin{eqnarray}
\nabla_\nu T^{\mu \nu}=0 \, \,. \label{eq19aa}
\end{eqnarray}

La elección de los parámetros de acoplamiento, (\ref{eq20}) (hecha
en la referencia \cite{r38}) en busca de la consistencia con las
soluciones de la $EG\lambda$, no es única. Esto puede mostrarse a
partir de la inclusión de los vínculos de torsión, (\ref{e1}) en
la acción (\ref{eq13}), de manera que ahora se tiene
\begin{eqnarray}
S'=\kappa^{4-N}\int d^N x \sqrt{-g}\,\big( -\frac 14 \, tr\,
\mathbb{F}^{\alpha \beta}\mathbb{F}_{\alpha \beta}+ \Lambda
(\lambda)+\ell(g,\psi) \nonumber \\+\, 4\pi
G\,\,tr\,\mathbb{M}^{\alpha\beta}(g,\psi)\mathbb{F}_{\alpha\beta}
+\kappa^{-3}{\mathcal{C}_\alpha}^{\lambda\sigma}
{(\mathbb{A}_\lambda)^\alpha}_\sigma\big) \, \,.
  \label{eq19b}
\end{eqnarray}

Esta nueva acción proporciona las ecuaciones de movimiento de la
conexión, que pueden ser escritas en analogía al caso libre de
materia, como
\begin{equation}
\nabla_\alpha{\Re^{(2)}}_{\lambda \sigma } -
\nabla_\sigma{\Re^{(3)}}_{\lambda \alpha }+ \nabla_\beta
{\Theta^\beta}_{\lambda\alpha\sigma}+\kappa^{-3}\mathcal{C}_{\alpha\lambda\sigma}=0
 \, \,, \label{eq19c}
\end{equation}
donde hemos definido los objetos
\begin{equation}
{\Re^{(n)}}_{\lambda \sigma }\equiv R_{\lambda \sigma }-8\pi G c_1
T g_{\lambda \sigma }-8\pi G c_n T_{\lambda \sigma }+c_4 \lambda
g_{\lambda \sigma }
 \, \,,\,\,\,\,n=2,3\,\,\,, \label{eq19d}
\end{equation}
con $c_4$ un parámetro libre y
\begin{equation}
\Theta_{\beta\alpha\nu\mu}\equiv 8\pi G
(c_2g_{\alpha\nu}T_{\mu\beta}-c_3g_{\alpha\mu}T_{\nu\beta})
 \, \,. \label{eq19e}
\end{equation}

Manipulando la ecuación (\ref{eq19c}), tomando trazas y
contracciones con el tensor de Levi-Civita, pueden obtenerse los
multiplicadores, más ciertas restricciones sobre los parámetros de
acoplamiento y la curvatura, como se muestra a conti-nuación
\begin{equation}
\mathcal{C}_{\alpha\lambda\sigma}=0 \,\, , \label{eq20a}
\end{equation}
\begin{equation}
c_2=c_3 \,\, , \label{eq20b}
\end{equation}
\begin{equation}
\partial_\mu\big(-R+16\pi G[(N-1)c_1+c_2]T \big)+16\pi G(N-2)c_2\nabla_\alpha
 {T^\alpha}_\mu=0\,\, . \label{eq20c}
\end{equation}
Si imponemos la condición de que el tensor momento-energía de los
campos materiales sea conservado, ec. (\ref{eq19aa}), la relación
(\ref{eq20c}) se reduce a
\begin{equation}
R-16\pi G[(N-1)c_1+c_2]T =c\,\, , \label{eq20d}
\end{equation}
con $c=constante$. Tomando la elección particular
\begin{equation}
(N-1)c_1+c_2=\frac{1}{N-2}\,\, , \label{eq20e}
\end{equation}
\begin{equation}
c=-\frac{2N\lambda}{N-2}\,\, , \label{eq20f}
\end{equation}
puede mostrarse que  (\ref{eq20d}) coincide con la traza de las
ecuaciones de Einstein, lo cual para nuestro interés constituye
una relación de consistencia de la formulación de calibre. La
fijación (\ref{eq20}) es un caso particular de (\ref{eq20e}).

Existen más restricciones sobre los campos materiales, además de
las que surgen al pedirse que éstos posean un tensor
momento-energía conservado. Esta vez cuando las variaciones en los
vielbeins o la métrica, son realizadas. Obviamente necesitamos
decir algo sobre la forma de $\ell(g,\psi)$. Para esto
demandaremos algunas propiedades. Por un lado, esta densidad
Lagrangiana debe ser consistente con el límite de vacío material
de la teoría, es decir $\ell(g,\psi) \rightarrow 0$ si $\alpha $ y
$ \beta $ van a cero. Además,  si la gravitación fuese
''apagada'', habremos de esperar que la acción de los campos
materiales, $\int d^Nx\,\sqrt{-g}\,\ell(g,\psi)$ se redujese a la
acción estándar de éstos en un espacio-tiempo plano, que
llamaremos $\int d^N\xi\,L(\psi)|_{\eta_{\mu \nu}}$,  y cuya
densidad Lagrangiana en un  espacio curvo  ya mostraremos. Sólo
por cuestiones de nomenclatura esta situación la denominaremos
como el {\it límite de no acoplamiento gravitacional}, y lo
realizaremos tomando el límite $G \rightarrow 0$ y $g_{\mu
\nu}\rightarrow \eta_{\mu \nu}$ con $\lambda = 0$.

Con esto en mente, y considerando que nuestra formulación es de
tipo cuadrática en el tensor de curvatura de Riemann-Christoffel,
proponemos una forma general, también cuadrática  para la
contribución de estos campos
\begin{equation}
\ell(g,\psi)\equiv  L(\psi) +b_1 T^2 + b_2 T_{\mu \nu}T^{\mu
\nu}=q+k\,\psi +b\,{\psi}^2
 \, \,, \label{eq21}
\end{equation}
donde hemos usado (\ref{eq17}) y que la densidad Lagrangiana de
los campos materiales en un espacio-tiempo curvo es
$L(\psi)=k_{(\alpha , \beta)}\psi +q_{(\alpha , \beta)}$, donde
los parámetros $k$ y $q$  dependen de cuales sean los campos que
estemos considerando. En el caso del campo escalar real no masivo,
mostrado en la tabla 3, éstos parámetros son $k=-\frac{1}{2}$ y
$q=0$. Además, $b=b_1 (\alpha + N\beta )^2 + b_2 (\alpha
^2+2\alpha \beta +N\beta ^2)$ es otro parámetro real. Más adelante
examinaremos el límite de no acoplamiento gravitacional para el
parámetro  $b$, el cual demanda el valor $b\rightarrow 0$ para que
$\ell(g,\psi)\rightarrow L(\psi)|_{g_{\mu\nu}=\eta_{\mu\nu}}$.

La variación de los vielbeins en la acción $S$ puede escribirse en
términos del tensor de Ricci, Weyl y los campos materiales. Con
esto, las ecuaciones de campo son
\begin{equation}
{P^\sigma }_d
\big[\psi_{\alpha\beta},{e^\mu}_a,R_{\alpha\beta}\big]+{Q^\sigma
}_d \big[{e^\mu}_a,R_{\alpha\beta}\big]+{S^\sigma }_d
\big[\psi_{\alpha\beta},{e^\mu}_a,R_{\alpha\beta},
C_{\alpha\beta\mu\nu}\big]=0
 \, \,, \label{eq22}
\end{equation}
donde ${P^\sigma }_d$ y ${Q^\sigma }_d $ son polinomios
cuadráticos en $\psi _{\alpha \beta}$ y el tensor de Ricci,
definidos por
\begin{eqnarray}
{P^\sigma }_d \equiv (b\psi^2+ k\psi + q){e^{\sigma}}_d-2k
{{\psi}^{\sigma}}_d -4b\psi {{\psi}^{\sigma}}_d+ 16\pi G\alpha\,
\frac{N-3}{N-2}\,R^{\mu\nu}\psi_{\mu\nu}{e^{\sigma}}_d \nonumber
\\+\,8\pi G\alpha\,
\frac{8-3N}{N-2}\,(R_{\mu d }\psi^{\mu\sigma}+
R^{\mu\sigma}\psi_{\mu d }) +16\pi G\beta\,
\frac{4-N}{N-2}\,{R^\sigma}_d \psi \nonumber
\\+\,8\pi G\,
\frac{(3-N)\alpha-(N-1)(4-N)\beta}{(N-2)(N-1)}\,R\psi{e^{\sigma}}_d
\nonumber
\\+\,16\pi G\,
\frac{(N-2)\alpha+(N-1)(4-N)\beta}{(N-2)(N-1)}\,R{\psi^{\sigma}}_d\,
\,, \nonumber
\\ \label{eq22a}
\end{eqnarray}
\begin{eqnarray}
{Q^\sigma }_d \equiv \frac{2(4-N)}{(N-2)^2}\,R^{\mu\sigma}R_{\mu d
}-\frac{4-N}{(N-2)^2}\,R^{\mu\nu}R_{\mu \nu } {e^{\sigma}}_d
-\frac{4}{(N-2)^2(N-1)}\,R {R^{\sigma}}_d \nonumber
\\-\frac{N-6}{2(N-2)^2(N-1)}\,R^2 {e^{\sigma}}_d -8\pi G a R
{e^{\sigma}}_d+16\pi G a {R^{\sigma}}_d +\Lambda
(\lambda){e^{\sigma}}_d \, \,, \nonumber
\\ \label{eq22b}
\end{eqnarray}
y
\begin{eqnarray}
{S^\sigma }_d\equiv  C^{\mu \nu\lambda\sigma}C_{\mu \nu \lambda d
}-\frac{{e^\sigma}_d}{4} \,C^{\mu \nu\lambda\rho}C_{\mu \nu
\lambda \rho }-C^{\mu \nu\lambda\sigma}R_{\mu \nu \lambda d
}-R^{\mu \nu\lambda\sigma}C_{\mu \nu \lambda d }\nonumber \\+
\frac{{e^\sigma }_d}{2} \,C^{\mu \nu\lambda\rho}R_{\mu \nu \lambda
\rho }+16\pi G\alpha \, {C^{\sigma\mu \nu}}_d \psi_{\mu \nu}\,
\,,\nonumber
\\  \label{eq23}
\end{eqnarray}
con $C_{\mu \nu  \lambda \alpha}$ el tensor de Weyl.

En este punto, uno puede explorar el caso particular en el cual
$N=3$, debido a que allí el tensor de Weyl es idénticamente nulo y
el tratamiento se aligera notablemente. Con esto, la ecuación de
movimiento de los dreibeins es
\begin{equation}
\bigg({P^\sigma }_d
\big[\psi_{\alpha\beta},{e^\mu}_a,R_{\alpha\beta}\big]+{Q^\sigma
}_d \big[{e^\mu}_a,R_{\alpha\beta}\big]\bigg)_{N=3}=0
 \, \,. \label{eq22a}
\end{equation}

Entonces, si se espera que la  extremal de la acción $S$  sea
consistente con las ecuaciones de Einstein en un espacio sin
torsión, necesitaremos evaluar (\ref{eq22a}) sobre éstas. Así,
usando (\ref{eq17}) en (\ref{eq6a}), escribimos
\begin{equation}
R_{\alpha \beta }(\psi)=-2\lambda g_{\alpha \beta}+8\pi G(\alpha
+2\beta)\psi g_{\alpha \beta }-8\pi G\alpha \psi_{\alpha \beta }
 \, \,, \label{eq22b}
\end{equation}
con la cual evaluamos (\ref{eq22a}), obteniendo
\begin{eqnarray}
\big(-4b+4(8\pi G)^2({\alpha }^2 +2\alpha \beta
+3{\beta}^2)\big)\psi {\psi ^\sigma}_d \nonumber \\
+\big(b-(8\pi G)^2({\alpha }^2 +2\alpha \beta
+3{\beta}^2)\big){\psi}^2{e^\sigma}_d    \nonumber \\
-\big(2k+16\pi G[3\lambda(\alpha +2\beta) +8\pi
G \alpha a] \big){{\psi}^\sigma}_d \nonumber \\
+\big(k +16\pi G[\lambda(\alpha +2\beta) -8\pi G \beta a]\big)\psi
{e^\sigma}_d \nonumber \\
+(q+16\pi G a\lambda ){e^\sigma}_d =0 \, \,. \label{eq24}
\end{eqnarray}

Si no se considera ningún tipo de restricción sobre los campos
$\psi_{\alpha \beta }$, el término independiente de campos
materiales en (\ref{eq24}), por ejemplo significa una restricción
sobre la constante cosmológica, esto es $\lambda= -\frac{q}{16\pi
G a}$. Pero, desde otro punto de vista, de no esperarse ninguna
obstrucción en los posibles valores de $\lambda$ (a diferencia,
por supuesto del caso del estudio dinámico de los campos de
espines altos en interacción con un campo gravitacional, visto
como un objeto externo), la ecuación (\ref{eq24}) necesariamente
representa una restricción para los campos materiales. Aquí
imponemos una primer condición sobre las posibles configuraciones
de los campos, es decir
\begin{equation}
\psi =constante  \, \,, \label{eq25}
\end{equation}
pues obsérvese que la traza de  (\ref{eq24}) es un polinomio
cuadrático en $\psi$, con  coeficientes constantes. Eventualmente,
el tipo de restricciones como la (\ref{eq25}) no sería severa en
el caso de un fluido perfecto (p.ej.: $\psi = U^\mu U_\mu =-1$).
De todas formas, para todo ${\psi ^\sigma}_d$ con $\psi
=constante$, la ecuación (\ref{eq24}) proporciona un sistema de
dos relaciones para $a$ y $b$
\begin{equation}
2 \psi b + (8\pi G)^2 \alpha a =2(8\pi G)^2({\alpha }^2 +2\alpha
\beta +3{\beta}^2)\psi -24 \pi G\lambda(\alpha +2\beta ) -k \, \,,
\label{eq26}
\end{equation}
\begin{equation}
\psi ^2 b +  16\pi G(\lambda-8\pi G \beta \psi )a =(8\pi
G)^2(\alpha ^2 +2\alpha \beta +3\beta ^2)\psi ^2 -(k+16\pi
G\lambda(\alpha +2\beta ) ) \psi  -q\, \,. \label{eq27}
\end{equation}
el cual posee soluciones regulares si se  demanda que el
determinante sea no singular, dando a lugar una nueva restricción
\begin{equation}
\psi (4\lambda - 8\pi G(\alpha +4\beta )\psi ) \neq 0 \, \,.
\label{eq28}
\end{equation}
Ahora podemos estudiar el límite de no acoplamiento gravitacional
para el parámetro  $b$
\begin{eqnarray}
b\mid _{\lambda =0}=(8\pi G)^2\bigg(\frac{\alpha ^3 +6{\alpha }^2
\beta+11\alpha \beta ^2 +12\beta ^3}{\alpha +4\beta}
\bigg)-\frac{(\alpha
 +2 \beta )k\psi +q\alpha}{(\alpha
+4\beta)\psi ^2}\, \, ,\label{eq29}
\end{eqnarray}
siendo consistente, por ejemplo en el caso de un campo escalar no
masivo, donde $\alpha =1$, $\beta =-1/2$ y $q=0$, pues  $b\mid
_{\lambda =0} =\frac{3}{4}(8\pi G)^2$.

Queremos subrayar que la ecuaciones (\ref{eq25}) y (\ref{eq28})
significan que el esquema de acoplamiento no minimal, presentado
en (\ref{eq13}) es consistente con la gravedad de Einstein, solo
bajo ciertas condiciones relacionadas con la clase de distribución
de los campos materiales.

\vspace{1cm}
{\bf 4.2.1) Inclusión de campos auxiliares}\vskip .1truein

Las restricciones sobre los campos materiales, mencionadas en la
sección anterior no son un aspecto sorprendente. De hecho, desde
el punto de vista de campos de espines altos dinámicos, acoplados
con gravedad como interacción externa, puede encontrarse que tal
teoría solo es consistente en ciertos espacio-tiempos
\cite{r14}-\cite{r10}. Así, ensayaremos la introducción de nuevos
términos de interacción \cite{r38}, hasta el orden cuadrático y
que involucren campos auxiliares en la acción $S$, con la
esperanza de reducir las restricciones sobre los campos
materiales.

Cuando en el contexto de la electrodinámica masiva se estudia la
acción de Proca, cuya densidad Lagrangiana es de la forma
$\mathcal{L}_P=-\frac{1}{4}\,F^{\mu\nu}F_{\mu\nu}-\frac{m^2}{2}\,A^{\mu}A_{\mu}$,
existe una forma de recuperar la invariancia de calibre mediante
la incorporación de un campo escalar auxiliar de
St$\ddot{u}$ckelberg, $\omega(x)$ de manera que la versión
invariante de calibre de la acción de Proca  se consigue ahora con
la nueva densidad definida como
\begin{equation}
\mathcal{L}_{P\omega}=-\frac{1}{4}\,F^{\mu\nu}F_{\mu\nu}
-\frac{m^2}{2}\,(A^{\mu}+\partial^\mu \omega)(A_{\mu}+\partial_\mu
\omega)
 \, \, , \label{sm3}
\end{equation}
donde se demandanda que el campo vectorial transforme como
$A_{\mu}\longrightarrow A'_{\mu}=A_{\mu}+\partial_{\mu}\theta$, y
el escalar como $\omega \longrightarrow\omega'=\omega- \theta$,
para garantizar que la acción $S_{P\omega}=\int
d^4\xi\,\mathcal{L}_{P\omega}$ sea invariante de calibre.

Siguiendo esta idea, introduciremos una nueva acción, $S'$  con
las definiciones de $\mathbb{J}^\alpha$ y $\mathbb{H}^{\alpha
\beta}$ como funcionales de la métrica y los campos materiales, y
${(\mathbb{W}_\alpha)^{\mu}}_{\nu}$ las componentes de un campo
auxiliar que transforma como la conexión $GL(3,R)$,  es decir
\begin{eqnarray}
S'=\kappa \int d^3 x \sqrt{-g}\big(-\frac 14 \, tr\,
\mathbb{F}^{\alpha \beta}\mathbb{F}_{\alpha \beta}+ \Lambda
(\lambda)+\ell(g,\psi) + 4\pi
G\,tr\,\mathbb{M}^{\alpha \beta}(g,\psi)\mathbb{F}_{\alpha \beta}  \nonumber \\
+\,tr\,\mathbb{J}^\alpha(\mathbb{A}_\alpha-\mathbb{W}_\alpha)+tr\,\mathbb{H}^{\alpha
\beta}(\mathbb{A}_\alpha-\mathbb{W}_\alpha
)(\mathbb{A}_\beta-\mathbb{W}_\beta)\big)
 \,\, \,. \nonumber \\\label{eq30}
\end{eqnarray}
La forma en como se acopla el campo auxiliar, garantiza por
construcción la simetría de calibre, ya que la diferencia de dos
conexiones sobre la fibra $GL(3,R)$,
''$\mathbb{A}_\alpha-\mathbb{W}_\alpha$'' transforma según la
representación adjunta del grupo. Esta idea ha sido manejada en
otros contextos, como por ejemplo el estudio de la geometría
singular del espacio de configuración en teorías de Yang-Mills, en
donde es introducido el {\it calibre de fondo} \cite{r64}.

Seguidamente ensayamos una propuesta simple para las componentes
de $\mathbb{J}^\alpha$ y $\mathbb{H}^{\alpha\beta}$ en términos
exclusivos de la métrica y los campos materiales
\begin{equation}
(\mathbb{J}_{\beta})_{\mu \nu }\equiv (d_1 + d_2 \psi){\varepsilon
}_{\beta \mu \nu} \, \,, \label{eq31}
\end{equation}
\begin{equation}
(\mathbb{H}^{\alpha \beta})^{\mu \nu} \equiv a_1
\,g^{\alpha\beta}g^{\mu\nu}+a_2 \,g^{\alpha\mu}g^{\beta\nu}+ a_3
\,g^{\alpha\nu}g^{\beta\mu}
 \, \,, \label{eq32}
\end{equation}
con los parámetros reales $d_1$, $d_2$, $a_1$, $a_2$ y $a_3$.

De la acción (\ref{eq30}), se obtienen las ecuaciones de
movimiento para $\mathbb{W}_\alpha$
\begin{equation}
\mathbb{J}^\beta + \mathbb{H}^{\alpha
\beta}(\mathbb{A}_\alpha-\mathbb{W}_\alpha)+(\mathbb{A}_\alpha-\mathbb{W}_\alpha)
\mathbb{H}^{\beta\alpha}=0
 \, \, ,\label{eq33}
\end{equation}
estableciendo que las ecuaciones de movimiento de la conexión
$\mathbb{A}_\alpha$ se mantienen iguales a las obtenidas de
(\ref{eq13}) con $N=3$, o sea sin los términos de campos
auxiliares. La ecuación (\ref{eq33}) sugiere además un ansatz para
los campos auxiliares:
\begin{equation}
(\mathbb{A}_{\alpha}-\mathbb{W}_{\alpha})_{\mu \nu}= (\theta _1 +
\theta _2 \psi ) \,{\varepsilon }_{\alpha\mu\nu}\, \,,
\label{eq34}
\end{equation}
con $\theta _1$ y $\theta _2$ parámetros reales. Este ansatz, aún
cuando no posea la dependencia más general posible en el campo
$\psi _{\sigma \beta}$,  es suficiente y consistente con la
ecuación (\ref{eq33}). Sustituyendo el ansatz (\ref{eq34}) en
(\ref{eq33}), se obtienen las siguientes relaciones entre
parámetros
\begin{equation}
d_1 = 2(a_2-a_1)\theta _1\, \,, \label{eq34a}
\end{equation}
\begin{equation}
d_2 = 2(a_2-a_1)\theta _2\, \,. \label{eq34b}
\end{equation}

Si se quiere explorar el límite de torsión nula, se deben
introducir los vínculos de torsión (\ref{e1}) en la acción
(\ref{eq30}), de manera que ahora escribimos
\begin{eqnarray}
S''=\kappa \int d^3 x \sqrt{-g}\big(-\frac 14 \, tr\,
\mathbb{F}^{\alpha \beta}\mathbb{F}_{\alpha \beta}+ \Lambda
(\lambda)+\ell(g,\psi) + 4\pi
G\,tr\,\mathbb{M}^{\alpha \beta}(g,\psi)\mathbb{F}_{\alpha \beta}  \nonumber \\
+\,tr\,\mathbb{J}^\alpha(\mathbb{A}_\alpha-\mathbb{W}_\alpha)+tr\,\mathbb{H}^{\alpha
\beta}(\mathbb{A}_\alpha-\mathbb{W}_\alpha
)(\mathbb{A}_\beta-\mathbb{W}_\beta)+\kappa^{-3}{\mathcal{C}_\alpha}^{\mu\nu}
{(\mathbb{A}_\mu)^\alpha}_\nu\big)
 \,\, \,. \nonumber \\\label{eq34c}
\end{eqnarray}

Las ecuaciones de movimiento de la conexión y los campos
auxiliares, conducen nuevamente a las condiciones
(\ref{eq20a})-(\ref{eq20c}). Así, tomando la fijación particular
(\ref{eq20}) y que los campos materiales poseen un tensor
momento-energía conservado, ec. (\ref{eq19aa}), la ecuación de
movimiento para los dreibeins provenientes de (\ref{eq34c}), es
\begin{eqnarray}
{P^\sigma }_d
\big[\psi_{\alpha\beta},{e^\mu}_a,R_{\alpha\beta}\big]+{Q^\sigma
}_d \big[{e^\mu}_a,R_{\alpha\beta}\big]+tr \mathbb{H}^{\alpha
\beta}(\mathbb{A}_{\alpha}-\mathbb{W}_{\alpha})(\mathbb{A}_{\beta}-\mathbb{W}_{\beta})
\,\,{e^\sigma}_d +\nonumber
\\+ \,\frac{1}{\sqrt{-g}}
\,tr\frac{\delta \sqrt{-g}\mathbb{J}^{\beta}}{\delta
{e_\sigma}^d}(\mathbb{A}_{\beta}-\mathbb{W}_{\beta})+tr
\frac{\delta \mathbb{H}^{\alpha \beta}}{\delta
{e_\sigma}^d}(\mathbb{A}_{\alpha}-\mathbb{W}_{\alpha})
(\mathbb{A}_{\beta}-\mathbb{W}_{\beta})=0\, \,, \nonumber
\\\label{eq35}
\end{eqnarray}
que, evaluando sobre las ecuaciones de Einstein (con la ayuda de
(\ref{eq22b})) y usando (\ref{eq31}), (\ref{eq32}) y (\ref{eq34}),
puede encontrarse un sistema de ecuaciones para los parámetros
libres, para todo $\psi_{\alpha\beta}$, es decir
\begin{equation}
16\pi G a \lambda +6(a_2 -  a_1){\theta _1}^2=-q \,, \label{eq36}
\end{equation}
\begin{equation}
 2(8\pi G)^2\beta a -16\pi G(\alpha +2\beta)\lambda =k\,, \label{eq37}
\end{equation}
\begin{equation}
(8\pi G)^2\alpha a +24\pi G(\alpha +2\beta)\lambda =-k \,,
\label{eq38}
\end{equation}
\begin{equation}
b=(8\pi G)^2(\alpha ^2 +2\alpha \beta + 3\beta ^2) \,,
\label{eq39}
\end{equation}
\begin{equation}
\theta _2=0 \,. \label{eq40}
\end{equation}

La consistencia con el límite de no acoplamiento gravitacional
para el parámetro  $b$ es verificada, ya que es proporcional a
$G^2$. En el caso de un campo escalar real no masivo, el sistema
(\ref{eq36})-(\ref{eq40}) es resoluble para los parámetros libres
$a$, $b$, $(a_1 - a_2){\theta _1}^2$ y ${\theta _2}$, y quedando
el parámetro de acoplamiento $a_3$ libre. En cualquier otro caso
(p. ej., $\alpha +2\beta\neq 0$) ocurrirían restricciones sobre la
constante cosmológica.

Debe notarse que, aún cuando los términos de acoplamiento
presentados en (\ref{eq30}), con (\ref{eq31}) y (\ref{eq32}), no
poseen la forma más general posible, la idea de que las
restricciones fuertes sobre los campos materiales ${\psi }_{\alpha
\beta}$ puedan ser evitadas introduciendo campos auxiliares, ha
sido dilucidada.

Si reunimos todas las relaciones de los parámetros involucrados en
la introducción de los campos auxiliares en (\ref{eq30}), para el
caso particular del campo escalar real no masivo ($\alpha=1$,
$\beta=k=-\frac{1}{2}$ y $q=0$), reescribimos las relaciones
(\ref{eq34a}), (\ref{eq34b}), (\ref{eq36})-(\ref{eq40}), como
\begin{equation}
d_1 = 2(a_2-a_1)\theta _1\, \,, \label{eq34aa}
\end{equation}
\begin{equation}
d_2 = 0\, \,, \label{eq34ba}
\end{equation}
\begin{equation}
8\pi G a \lambda +9(a_2 -  a_1){\theta _1}^2=0 \,, \label{eq36a}
\end{equation}
\begin{equation}
 2(8\pi G)^2 a  =1\,, \label{eq37a}
\end{equation}
\begin{equation}
b=\frac{3}{4}\,(8\pi G)^2 \,, \label{eq39a}
\end{equation}
\begin{equation}
\theta _2=0 \,, \label{eq40a}
\end{equation}
el cual es un sistema de seis ecuaciones para nueve parámetros, de
los cuales siete ($a$, $b$, $a_1$, $a_2$, $a_3$, $d_1$ y $d_2$)
son de acoplamiento, siendo $a_3$ libre. Si fijamos por comodidad,
$a_2=a_3=0$, los términos de acoplamiento con campos auxiliares
evaluados sobre estos parámetros adquieren cierta forma familiar
\begin{eqnarray}
tr\,\mathbb{J}^\alpha(\mathbb{A}_\alpha-\mathbb{W}_\alpha)=d_1\,
{\varepsilon^{\alpha\mu}}_\nu\,{(\mathbb{A}_\alpha-\mathbb{W}_\alpha)^\nu}_\mu
 \,\, \,,\label{eq41a}
\end{eqnarray}
\begin{eqnarray}
tr\,\mathbb{H}^{\alpha \beta}(\mathbb{A}_\alpha-\mathbb{W}_\alpha
)(\mathbb{A}_\beta-\mathbb{W}_\beta)=a_1\,tr\,(\mathbb{A}_\mu-\mathbb{W}_\mu
)(\mathbb{A}^\mu-\mathbb{W}^\mu)
 \,\, \,,\label{eq41b}
\end{eqnarray}
con una relación de restricción para los parámetros $d_1$ y $a_1$,
dada por
\begin{eqnarray}
36\pi G {d_1}^2=-\lambda a_1
 \,\, \,.\label{eq41c}
\end{eqnarray}

Las relaciones (\ref{eq41a}) y (\ref{eq41b}), sugieren que en
lugar de la acción (\ref{eq30}), pudiésemos haber comenzado con
otra, de una forma similar a la del modelo de Proca
\begin{eqnarray}
S'=\kappa^{4-N} \int d^3 x \sqrt{-g}\big(-\frac 14 \, tr\,
\mathbb{F}^{\alpha \beta}\mathbb{F}_{\alpha \beta}+ \Lambda
(\lambda)+\ell(g,\psi) + 4\pi
G\,tr\,\mathbb{M}^{\alpha \beta}(g,\psi)\mathbb{F}_{\alpha \beta}  \nonumber \\
+\,tr\,\mathbb{J}^{\alpha}\,(\mathbb{A}_\alpha-\mathbb{W}_\alpha)
+\frac{\mu^2}{2}\,tr\,(\mathbb{A}_\mu-\mathbb{W}_\mu
)(\mathbb{A}^\mu-\mathbb{W}^\mu)\big)
 \,\, \,, \nonumber \\\label{eq42}
\end{eqnarray}
con
\begin{eqnarray}
{(\mathbb{J}^\alpha)^\mu}_\nu=J_0\, {\varepsilon^{\alpha\mu}}_\nu
 \,\, \,,\,\,\,\, \,\,\,\, \,J_0=constante\,\,\, \,,\label{eq42a}
\end{eqnarray}
exhibiendo un término de acoplamiento minimal con una corriente
idénticamente conservada, $\mathbb{J}^{\alpha}$ y otro cuadrático,
al estilo  de  el de Proca con ''masa'' $\mu$ (obsérvese que, en
virtud de (\ref{eq41b}) y (\ref{eq41c}), en espacios de dS se
podría tener $\mu^2<0$).

\newpage
{\bf 4.3) La gravedad topológicamente masiva}\vskip .1truein

La incorporación de términos masivos en teorías de gravedad ha
sido implementada desde distintos puntos de vista. Una
construcción posible es la de agregar a la acción de
Hilbert-Einstein, $S_{HE}$ un término de Fierz-Pauli \cite{r66},
el cual es cuadrático en la fluctuación de la métrica
$h_{\mu\nu}=g_{\mu\nu}-\eta_{\mu\nu}$, es decir $S=S_{HE}+m^2\int
d^4x\,\sqrt{-g}\,(h_{\mu\nu}h^{\mu\nu}-h^2)$. Otra via posible es,
la llamada gravedad topológica masiva \cite{r55} (GTM), en la que
el término de Fierz-Pauli es reemplazado por uno de tipo
Chern-Simons (CS) no abeliano, construido con los símbolos de
Christoffel. Es bien sabido que en 2+1 dimensiones, tanto la
acción de Hilbert-Einstein como la acción de CS no propagan grados
de libertad por separado, pero al combinarlas en la GTM se obtiene
una teoría que describe exitaciones masivas en 2+1.

Si se agrega un término de tipo cosmológico a la GTM, se tiene la
gravedad topológica masiva con constante cosmológica \cite{r67}
(GTM$\lambda$), cuya acción es de la forma
\begin{eqnarray}
S=\frac{1}{\kappa^2} \int d^3 x \sqrt{-g}(R+\lambda)+
\frac{1}{\kappa^2\mu}\,S_{CS}\,\, \,, \label{eqt1}
\end{eqnarray}
donde $S_{CS}$ es la acción de CS. En un espacio-tiempo
Riemanniano, la extremal de la acción (\ref{eqt1}) respecto a
variaciones en la métrica, conduce a la ecuación de movimiento
\begin{equation}
R^{\mu \nu }-\frac{g^{\mu \nu}}2R-\lambda g^{\mu
\nu}+\frac{1}{\mu}\,C^{\mu \nu }=0 \,\, , \label{eqt2}
\end{equation}
donde
\begin{equation}
C^{\mu \nu}=\varepsilon^{\mu\alpha\beta}\,\nabla_\alpha ({R^\nu}_
\beta -\frac{1}{4}\,{\delta^\nu}_ \beta R) \,\, , \label{eqt3}
\end{equation}
es el tensor de Cotton. La traza de la ecuación (\ref{eqt2})
proporciona una condición de consistencia para el tensor de Ricci
\begin{equation}
R=-6\lambda  \,\, . \label{eqt2b}
\end{equation}

Si se manipula (\ref{eqt2}) mediante contracciones con el tensor
de levi-Civita y tomando la divergencia, es posible escribir una
ecuación de propagación masiva para el tensor de Ricci, o sea
\begin{eqnarray}
(\Box{} -
\mu^2)\,R_{\mu\nu}-R^{\alpha\beta}R_{\alpha\beta}g_{\mu\nu}
+3{R^\alpha }_\mu R_{\alpha\nu}
+\frac{\mu^2}{3}\,Rg_{\mu\nu}-\frac{3}{2}\,RR_{\mu\nu}
+\frac{1}{2}\, R^2 g_{\mu\nu} =0
 \, \, ,\nonumber \\ \label{eqt2bb}
\end{eqnarray}
donde $\Box{}=\nabla_\mu \nabla^\mu$.

A la expresión (\ref{eqt2}) la llamaremos la ecuación de la
GTM$\lambda$, y será nuestra condición de consistencia con vistas
a explorar la versión topológicamente masiva de la formulación
$GL(3,R)$ de gravedad, lo cual será seguidamente abordado.

\vspace{1cm}


{\bf 4.3.1) Formulación de calibre $GL(3,R)$ topológica masiva
}\vskip .1truein

El modelo libre consistente con la gravedad de Einstein en el
límite de torsión nula, está representado por la acción invariante
de calibre $GL(3,R)$
\begin{equation}
S^{(2+1)}_o = \kappa\int d^3 x \sqrt{-g} \,\, (-\frac 14 \, tr\,
\mathbb{F}^{\alpha\beta}\mathbb{F}_{\alpha\beta} + \lambda^2 )
\,\, , \label{eqm3}
\end{equation}
donde  $\lambda $ es la constante cosmológica.

Existen otras formas, además de agregar una constante cosmológica,
para obtener soluciones no planas de la gravitación en dimensión
2+1. En el contexto de la formulación de calibre discutida en este
trabajo, se implementó un procedimiento ($\S$4.2) que copia la
idea del presentado en el estudio de la acción del modelo de Proca
invariante de calibre, el cual es clásica y cuánticamente
equivalente al modelo original (\ref{sm3}) (por ejemplo ver la
referencia \cite{r66a}), y se extiende la idea de los campos de
St$\ddot{u}$ckelberg al caso no abeliano con la introducción de la
conexión de fondo $GL(3,R)$ o campo auxiliar, $\mathbb{W}_\alpha$.

No obstante, en esta sección estamos interesados en ensayar la
introducción  de un término de origen topológico. Entonces, el
modelo de la formulación de cali-bre $GL(3,R)$ de la gravedad
topológicamente masiva con constante cosmológica (CGTM$\lambda$),
es introducido de la forma
\begin{equation}
S_{CGTM\lambda} = S^{(2+1)}_o+\frac{m\kappa}{2}\int
d^3x\,\epsilon^{\mu\nu\lambda}\,tr\big(
\mathbb{A}_{\mu}\partial_\nu \mathbb{A}_\lambda +\frac{2}{3}\,
\mathbb{A}_{\mu}\mathbb{A}_\nu \mathbb{A}_\lambda\big) \,\, ,
\label{eqm10}
\end{equation}
la cual, por construcción es invariante de calibre a menos de un
término de borde proporcional al {\it winding number}, $W(U)\equiv
-\frac{1}{3}\int d^3x\,\epsilon^{\mu\nu\lambda}\,tr
U^{-1}\partial_{\mu} UU^{-1}\partial_{\nu}
UU^{-1}\partial_{\lambda} U$, con $U\in GL(3,R)$.

Siguiendo la idea de Palatini, consideramos variaciones
independientes en la conexión $\mathbb{A}_\mu$ y los dreibeins
${e^\mu}_a$ (o la métrica) sobre la acción (\ref{eqm10}). La
extremal de ésta conduce a las siguientes ecuaciones de movimiento
\begin{equation}
\frac 1{\sqrt{-g}}\,\,\partial _\mu
(\sqrt{-g}\,\,\mathbb{F}^{\lambda \mu }) + [\mathbb{F}^{\mu
\lambda  } ,\mathbb{A}_\mu
]-\frac{m}{2}\,\varepsilon^{\mu\nu\lambda}\mathbb{F}_{\mu\nu}=0\,\,
, \label{eqm11}
\end{equation}
\begin{eqnarray}
T_{\mu\nu}(\mathbb{F})+\kappa \lambda^2 \,g_{\mu\nu}=0 \,\, .
\label{eqm12}
\end{eqnarray}

Exploremos el límite de torsión nula. Para esto, recurrimos a los
vínculos  de torsión (\ref{e1}) y consideraremos que en dimensión
$N=3$, los multiplicadores de Lagrange
$\mathcal{C}_{\alpha\mu\nu}$ con la propiedad de antisimetría
(\ref{e2}), pueden ser reemplazados por otros de dos índices,
$\mathcal{C}_{\alpha\mu}$. Entonces, la acción CGTM$\lambda$ con
vínculos es
\begin{equation}
S'_{CGTM\lambda} = S_{CGTM\lambda}+\kappa^{-2}\int d^3 x \sqrt{-g}
\,\mathcal{C}_{\alpha\beta}\,\varepsilon^
{\beta\lambda\sigma}{(\mathbb{A}_\lambda)^\alpha}_\sigma \,\, .
\label{eqam7}
\end{equation}

Con esta acción, la  ecuación de movimiento correspondiente a la
conexión evaluada sobre los vínculos de torsión, es
\begin{equation}
\nabla_\mu {(\mathbb{F}^{\lambda \mu })^\sigma}_\alpha
-\frac{m}{2}\,\varepsilon^{\mu\nu\lambda}{(\mathbb{F}_{\mu\nu})^\sigma}_\alpha+
\kappa^{-3}\mathcal{C}_{\alpha\beta}\,\varepsilon^{\beta\lambda\sigma}=0\,\,
, \label{eqam8}
\end{equation}
y escrita en términos del tensor de Ricci, es
\begin{equation}
\nabla_\mu R_{\sigma\lambda}-\nabla_\lambda R_{\sigma\mu}
-m\,{\varepsilon^{\nu\rho}}_\sigma(g_{\lambda\nu}R_{\mu\rho}-g_{\mu\nu}R_{\lambda\rho}
-\frac{R}{2}\,g_{\lambda\nu}g_{\mu\rho})+
\kappa^{-3}\mathcal{C}_{\mu\beta}\,{\varepsilon^{\beta}}_{\sigma\lambda}=0\,\,
. \label{eqam8a}
\end{equation}
La ecuación de los dreibeins evaluada sobre los vínculos de
torsión sigue siendo, obviamente (\ref{eqm12}). Así, las
ecuaciones (\ref{eqam8a}) y (\ref{eqm12}) pueden ser vistas como
un sistema de 9+6=15 ecuaciones para las  6 componentes del tensor
de Ricci y los 9 multiplicadores de Lagrange.

Procediendo de forma similar al caso libre, usamos la ecuación de
movimiento en términos del tensor de Ricci, (\ref{eqam8a}) la
manipulamos (tomando trazas y contrayendo con el tensor de
Levi-Civita) y obtenemos los multiplicadores y una condición para
la curvatura escalar,  o sea
\begin{eqnarray}
\kappa^{-3}\mathcal{C}_{\mu\nu}=\frac{mR}{6}\,g_{\mu\nu} \,\, ,
\label{eqam9}
\end{eqnarray}
\begin{eqnarray}
R=constante \,\, , \label{eqam10}
\end{eqnarray}
siendo esta ultima relación consistente con la GTM$\lambda$.
Sustituyendo (\ref{eqam9}) en (\ref{eqam8}), obtenemos
\begin{equation}
\nabla_\mu \mathbb{F}^{\lambda \mu }
-\frac{m}{2}\,\varepsilon^{\mu\nu\lambda}\mathbb{F}_{\mu\nu}+
\frac{mR}{6}\,\mathcal{E}^\lambda=0
 \, \, , \label{eqm13}
\end{equation}
donde hemos definido el objeto matricial antisimétrico,
$\mathcal{E}^\lambda$ con componentes
${(\mathcal{E}^\lambda)^\sigma}_\alpha \\ \equiv
{\varepsilon_\alpha}^{\lambda\sigma}$. Equivalentemente, también
podemos evaluar (\ref{eqam8a}) en (\ref{eqam9}), de manera que
\begin{equation}
\nabla_\mu R_{\sigma\lambda}-\nabla_\lambda R_{\sigma\mu}
-m\,{\varepsilon^{\nu\rho}}_\sigma(g_{\lambda\nu}R_{\mu\rho}-g_{\mu\nu}R_{\lambda\rho}
-\frac{2}{3}\,Rg_{\lambda\nu}g_{\mu\rho})=0\,\, . \label{eqam8ab}
\end{equation}

En (\ref{eqm13}) observamos  un término dependiente en la
curvatura escalar, que como veremos,  existe una solución para
(\ref{eqm12}) que expresa que tal término puede ser de origen
cosmológico. Para confirmar esto, escribimos el tensor de
Belinfante en función del tensor de Riemann, considerando la
relación (\ref{eq5}), es decir
\begin{equation}
T_{\mu\nu}(\mathbb{F})=\kappa({R^\sigma}_{\rho\mu\beta}{{R^\rho}_{\sigma\nu}}^\beta
-\frac{g_{\mu\nu}}{4}{R^\sigma}_{\rho\alpha\beta}{{R^\rho}_{\sigma}}^{\alpha\beta})
 \, \, , \label{eqam11}
\end{equation}
y como el tensor de Riemann en 2+1 dimensiones está expresado en
términos exclusivos del tensor de Ricci, mediante
$R_{\lambda\mu\nu \rho}=g_{\lambda\nu}R_{\mu
\rho}-g_{\lambda\rho}R_{\mu\nu}-g_{\mu\nu}R_{\lambda\rho}+g_{\mu\rho}R_{\lambda\nu}-\frac
R2 \,(g_{\lambda\nu}g_{\mu \rho}-g_{\lambda\rho}g_{\mu\nu}) $, se
puede hacer lo propio con $T_{\mu\nu}(\mathbb{F})$. Entonces, la
ecuación de los dreibeins, (\ref{eqm12}) se reduce a
\begin{eqnarray}
2R_{\sigma\mu}{R^\sigma}_\nu
-2RR_{\mu\nu}-g_{\mu\nu}R_{\sigma\rho}R^{\sigma\rho}+\frac{3}{4}\,
g_{\mu\nu}R^2+g_{\mu\nu}\lambda^2=0\,\, . \label{eqam12}
\end{eqnarray}
Podemos explorar soluciones de tipo curvatura escalar constante
(condición de consistencia), y particularmente de la forma
$R_{\mu\nu}=\frac{R}{3}\,g_{\mu\nu}$. Sustituyendo ésta en
(\ref{eqam12}) se obtiene
\begin{eqnarray}
R=\pm 6|\lambda |\,\, , \label{eqam13}
\end{eqnarray}
es decir se tienen soluciones de tipo dS/AdS, las cuales
satisfacen idénticamente (como es de esperarse) a la ecuación de
movimiento (\ref{eqam8ab}). Obsérvese que el caso particular
$\lambda=0$ conduce necesariamente a $R=0$, lo cual es la
condición de consistencia del tensor de Riemann en el contexto de
la GTM \cite{r55}.

Finalmente, comentaremos sobre dos aspectos interesantes del
modelo de la formulación de calibre $GL(3,R)$ de la gravedad
topológicamente masiva, los cuales estan relacionados con la
consistencia de la ecuación de movimiento (\ref{eqam8ab}) con la
de la GTM$\lambda$, ec. (\ref{eqt2}), y por otro lado, la
obtención de la ecuación de propagación masiva para el tensor de
Ricci. Inmediatamente podemos ver que, contrayendo la ecuación
(\ref{eqam8ab}) con el tensor de Levi-Civita, se obtiene
\begin{equation}
R_{\mu \nu }-\frac{g_{\mu
\nu}}3R+\frac{1}{m}\,{\varepsilon_\mu}^{\alpha\beta}\,\nabla_\alpha
R_{\nu\beta}=0 \,\, , \label{eqam13a}
\end{equation}
la cual es consistente con (\ref{eqt2}), si fijamos
\begin{equation}
m=\mu\,\, , \label{eqam13aa}
\end{equation}
y a (\ref{eqam10}) como
\begin{equation}
R=-6\lambda \,\, . \label{eqam13ab}
\end{equation}

Esta consistencia  sugiere que, de haber soluciones no triviales,
éstas deben  propagar de una manera causal y masiva, como en
GTM$\lambda$. En efecto, tomando la divergencia de
(\ref{eqam8ab}),  escribimos
\begin{eqnarray}
(\Box{} -
m^2)\,R_{\mu\nu}-R^{\alpha\beta}R_{\alpha\beta}g_{\mu\nu}
+3{R^\alpha }_\mu R_{\alpha\nu}
+\,\frac{m^2}{3}\,Rg_{\mu\nu}-\frac{3}{2}\,RR_{\mu\nu}
+\frac{1}{2}\, R^2 g_{\mu\nu} =0
 \, \, , \nonumber \\ \label{eqam13b}
\end{eqnarray}
la cual es equivalente a la ecuación de propagación del tensor de
Ricci en la GTM$\lambda$, (\ref{eqt2bb}), bajo (\ref{eqam13aa}) y
(\ref{eqam13ab}).

\newpage
{\large {\bf 5)Conclusiones}} \vskip .5truein

El procedimiento que conduce a  la acción reducida de una teoría
dada, ha resultado ser una herramienta muy útil con miras a no
solo examinar los grados físicos de libertad que se propagan, sino
hacia la obtención de los corchetes de Dirac via los de Poisson
inducidos en la superficie de los vínculos, sin pasar por el
extenuante procedimiento estándar del formalismo de Dirac.

En el  contexto anterior, el proceso de reducción de la teoría
autodual de espín 2 en dimensión 2+1 (Capítulo 2), se realizó
mediante las descomposiciones transverso-longitudinal o
tranversal-sin traza, que en cualquier caso, se recupera la forma
canónica esperada, correspondiente a una excitación masiva:
${S_{ad}}^{(2)*}=\int d^3 x \{P\dot{Q}-\frac12 P^2 + \frac12
\,Q(\Delta -m^2)Q\}$. Es importante notar que aún cuando existe el
procedimiento de descomposición covariante de los proyectores,
éste no se ha implementado en toda su amplitud, pues teniendo en
mente el problema de un espacio-tiempo curvo, es conocido el
obstáculo que representa la imposibilidad de definir las potencias
arbitrarias del D'Alembertiano.

La prueba directa de la consistencia de una teor\'{\i}a cu\'antica
de campo relativista, pasa por la obtenci\'on expl\'{\i}cita de
los generadores $\mathcal{P}^{\mu}$ y $\mathcal{J}^{\mu\nu}$ que
deben satisfacer el \'algebra de Poincar\'e. Estos generadores son
calculados en t\'erminos de las variables canónicas fundamentales,
$Q$ y $P$,  asociadas al \'unico grado de libertad propagado. El
esp\'{\i}n aparece a través de una contribución singular
infrarroja, cuando los generadores de {\it boosts} de Lorentz
($\mathcal{J}^{i0}$) son calculados. Sin embargo, esta
singularidad es evitable mediante el procedimiento bien conocido
de realizar un cambio de fase en los operadores
creación-aniquilación y la fijación del parámetro de espín, $s=2$.

A nivel del álgebra de Poincaré, la singularidad infrarroja se
manifiesta a través de una ''anomalía'' ($\mathcal{A}$) en la
misma, cuando se calcula el corchete de los {\it boosts} de
Lorentz, $i[\mathcal{J}^{i0}\,,\,\mathcal{J}^{j0}]=\epsilon
^{ij}(\mathcal{J}-\mathcal{A})$. Esto debe ser así, pues en algún
momento debe reflejarse el hecho de que no se trata de un campo de
espín 0. No obstante, como ocurre con los generadores, la
transformación de fase que permite remover la singularidad
infrarroja de los {\it boosts}, aquí se encarga de eliminar la
''anomalía'', teniéndose como es de esperarse que
$i[\mathcal{\overline{J}}\,^{i0}\,,\,\mathcal{\overline{J}}\,^{j0}]=
\epsilon ^{ij}\,\mathcal{\overline{J}}$, con la fijación $s=2$. Es
de insistirse en que la teor\'{\i}a autodual de espín 2, descrita
por la acción  ${S_{ad}}^{(2)}=\frac m2 \int d^3 x (\epsilon ^{\mu
\nu \lambda }{h_\mu }^\alpha
\partial _\nu
h_{\lambda \alpha }-m(h_{\mu \nu }h^{\nu \mu }-h^2 ))$ es
invariante relativista por construcci\'on, raz\'on por la cual la
''anomal\'{\i}a'' discutida es solo una expresi\'on de la
singularidad infrarroja y no de alguna inconsistencia
intr\'{\i}seca en el car\'acter covariante de la misma. En este
sentido, queda por evaluarse el álgebra de Schwinger \cite{r66b},
lo cual para el caso de espín 2 resulta ser un proceso laborioso.

Seguidamente (Capítulo 3), extendimos la teoría autodual de espín
2 al caso de un espacio-tiempo curvo, introduciendo términos de
acoplamiento no minimales con la gravedad en la acción:
$$
{S_{adg}}^{(2)}=\int \frac{d^3 x}{2} \sqrt {-g}\big[
m\,\varepsilon ^{\mu \nu \lambda }{h_\mu }^\alpha \nabla _\nu
h_{\lambda \alpha }+ {\Omega}^{\alpha \beta \sigma \lambda
}(R_{\mu\nu},g_{\rho\omega})\,h_{\alpha \beta }h_{\sigma \lambda }
\big]\,\,.
$$
Allí, mostramos que si se demanda la consistencia en el conteo de
grados de libertad, en estrecha analogía con el caso plano, no
solo aparecen restricciones sobres los parámetros libres
($a_1,...,a_7$), sino sobre el mismo espacio-tiempo. Como ejemplo
de esto último, los espacios de dS/AdS son un caso particular en
los que tanto el número de grados de libertad es consistente, como
la propagación es causal.

No obstante, ocurre un efecto (ya reportado en otros contextos
\cite{r2},\cite{r4}) como lo es el de la existencia de valores
prohibidos de la masa, con la finalidad de mantener una
descripción consistente de un campo simétrico, transverso y sin
traza. Ocurren dos tipos de valores prohibidos de masa. Por un
lado, en el contexto de la teoría plana se tiene
$$
m\neq 0\,\,,
$$
el cual  proviene del hecho de que  el límite de masa cero del
modelo autodual conduce a una teoría sin grados de libertad
locales. Adicionalmente, la condición $m\neq 0$ hace que no haya
una versión invariante conforme de la teoría autodual.

Por otro lado, en el contexto de los espacios de dS/AdS aparecen
las restricciones
$$
6M^4-Rm^2\neq 0\,\,,
$$
$$
M^2\equiv m^2+\sigma R\neq 0\,\,,
$$
que pueden ser interpretadas como que el parámetros $m$ del campo
autodual en dS/AdS posee valores prohibidos. Estas restricciones
deben ocurrir para que los vínculos Lagrangianos en estos tipos de
espacio-tiempo, describan consistentemente un campo simétrico,
transverso y sin traza, que propaga causalmente una sola
excitación. Además, ambas restricciones tienen información del
{\it background} y  consistentemente equivalen a $m\neq 0$ en el
límite plano. No obstante, la restricción $M^2\neq 0$ en el modelo
autodual acoplado con dS/AdS juega un rol más parecido a $m^2\neq
0$ en la versión plana, pues además de mantener la no invariancia
conforme, el valor crítico $M^2=0$ revela el carácter discontínuo
de la teoría.

Es posible examinar el único grado de libertad descrito por la
teoría,  mediante una aproximación local del procedimiento de la
acción reducida, en el cual se realiza una descomposición
transverso-sin traza de la parte simétrica del campo autodual, y
posteriormente se ''proyecta'' este campo en el espacio plano
tangente en sus partes ${{{h^{(s)Tt}}}_{ab}}^\pm (\xi)$,
correspondientes a la propagaciones de espín $\pm 2$.

En el Capítulo 4, revisamos una posible formulación de tipo
Yang-Mills para la gravitación, basada  en el fibrado de
referenciales con espacio base un espacio-tiempo $N$-dimensional
métrico compatible con torsión. Allí, mostramos que esta
formulación libre es consistente con las soluciones de tipo
dS/AdS, si la constante cosmológica contribuye de manera
cuadrática en la acción. Esto significa, al menos a nivel clásico
la equivalencia entre la formulación original de Einstein y una de
tipo Yang-Mills.

Seguidamente, ensayamos un esquema covariante de acoplamiento no
minimal con campos materiales, observándose que en general (esto
es, sin imponer más condiciones sobre los campos materiales que la
que significa  pedir que el tensor momento-energía asociado a
éstos, sea conservado) las ecuaciones de campo obtenidas son
consistentes con las de Einstein si se imponen ciertas
restricciones sobre la materia. Este hecho pudiese sugerir la idea
de que estamos viendo ''la otra cara de la moneda'', en el sentido
de que en el Capítulo 3 se discutió cómo hay que restringir al
campo gravitacional para que una teoría de espín alto sea
consistente. Pero aquí, siendo la gravitación el objeto dinámico,
deben restringirse la forma de los campos materiales vistos como
fuentes.

No obstante, como discutimos en el escenario 2+1 dimensional,  la
introducción de campos auxiliares remueve las restricciones  sobre
los campos materiales. Quedarían por explorarse otros esquemas de
acoplamiento donde se incluyan otras clases de campos (fluidos,
campos electromagnéticos, etc.) que involucren dependencia en la
métrica y la conexión, así como la incorporación de objetos
extendidos como cuerdas. De igual manera, en el contexto de la
teoría libre de materia podría emprenderse la búsqueda de
soluciones de tipo monopolo, donde posiblemente las soluciones de
tipo Schwarszchild fuesen modeladas por ansatz de tipo Wu-Yang
como ocurre en el caso de las teorías  de Yang-Mills con el grupo
$SU(2)$ \cite{r69}.

Por último, presentamos brevemente el modelo correspondiente a la
formulación de calibre $GL(3,R)$ topológicamente masiva con
constante cosmológica (CGTM$\lambda$), cuya acción es
$$
S_{CGTM\lambda} =\kappa\int d^3 x \sqrt{-g} \,\, \big(-\frac 14 \,
tr\, \mathbb{F}^{\alpha\beta}\mathbb{F}_{\alpha\beta} + \lambda^2
+\frac{m}{2}\,\epsilon^{\mu\nu\lambda}\,tr\big(
\mathbb{A}_{\mu}\partial_\nu \mathbb{A}_\lambda +\frac{2}{3}\,
\mathbb{A}_{\mu}\mathbb{A}_\nu \mathbb{A}_\lambda\big) \big)\,\, ,
$$
con ${(\mathbb{A}_{\mu})^\alpha}_\beta =
{\Gamma^\alpha}_{\mu\beta}$ y
${(\mathbb{F}_{\alpha\beta})^\mu}_\nu = {R^\mu}_{\nu\alpha\beta}
$. Allí observamos que el límite de torsión nula posee no solo
soluciones (triviales) de tipo dS/AdS, sino que las soluciones no
triviales son consistentes con las de la gravedad topológica
masiva  cosmológicamente extendida (GTM$\lambda$), dada por
$$
S_{GTM\lambda}=\frac{1}{\kappa^2} \int d^3 x \sqrt{-g}(R+\lambda)+
\frac{1}{\kappa^2\mu}\,S_{CS}\,\,\,,
$$
y que, por tanto la teoría posee propagaciones masivas y causales.
En pocas pala-bras, esta consistencia dinámica entre una teoría de
tipo lineal en la curvatura escalar con otra de tipo cuadrático en
el tensor de Riemann-Christoffel, radica en que la forma en como
las  primeras derivadas del tensor de Ricci contribuyen en la
ecuación de la GTM$\lambda$ (via el tensor de Cotton), es
reproducida con el límite de torsión nula por la ecuación de campo
de la CGTM$\lambda$, aquí expuesta. El estudio de la linealización
de la CGTM$\lambda$, la obtención de las cargas conservadas a
ésta,  así como el de la construcción de un esquema consistente de
acoplamiento con campos materiales, son tareas pendientes.

Finalizamos diciendo que, a diferencia de simplemente tratarse de
una teoría de Yang-Mills con grupo de calibre $GL(3,R)$ más un
término de tipo Chern-Simons y otro cosmológico, aquí es posible
recuperar el límite de torsión nula, lo cual proporciona cierto
grado de extensión conceptual con respecto a lo que sería,
simplemente un caso no abeliano de la electrodinámica topológica
masiva.

\newpage
{\large {\bf 6) Apéndices}} \vskip .5truein

\begin{center}
\large{{\bf Apéndice A}: Proyectores.}
\end{center}
\vskip .1truein

Los operadores de proyección \cite{r54} proyectan al campo
$h_{\mu\nu}$ en sus diferentes partes irreducibles, y ellos
constituyen una partición de la ''unidad'', como veremos.

Comenzamos con el caso de tensores de rango 1, introduciendo los
operadores
$$
\widehat{\partial }^\mu\equiv \frac{\partial ^\mu}{\Box{}
^{\frac{1}{2}} } \, \,, \eqno{(A.1)}
$$
que satisfacen
$$
\widehat{\partial }^\mu\widehat{\partial }_\mu=1\, \,.
\eqno{(A.2)}
$$
Puede entenderse como actúa $\widehat{\partial }^\mu$ en el
espacio de Fourier, si tomamos
$$
\phi (x)=\frac1{(2\pi)^3}\int d^3k\,e^{-ik^\nu
x_\nu}\,\bar{\phi}(k)\, \,, \eqno{(A.3)}
$$
entonces
$$
\widehat{\partial }_\mu\phi (x)=-\frac{i}{(2\pi)^3}\int
d^3k\,e^{-ik^\nu x_\nu}\,\frac{k_\mu}{\sqrt{-k^\alpha
k_\alpha}}\,\bar{\phi}(k) \, \,. \eqno{(A.4)}
$$
Seguidamente, introducimos el proyector transverso
$$
{P_\mu}^\nu \equiv {\delta_\mu}^\nu - {w_\mu}^\nu \, \,.
\eqno{(A.5)}
$$
donde ${w_\mu}^\nu \equiv \widehat{\partial }_\mu\widehat{\partial
}^\nu$. El proyector ${P_\mu}^\nu$ mapea cualquier campo vectorial
en su parte transversa
$$
{P_\mu}^\nu V_\nu \equiv {V^T}_\mu \,\Longrightarrow
\,\widehat{\partial }^\nu {V^T}_\nu=0 \, \,. \eqno{(A.6)}
$$
Es bien conocido que la parte de espín 1 de un campo vectorial
está contenida en su parte transversa  ${V^T}_\nu $, la cual tiene
dos posibles helicidades. Podemos proyectar ${V^T}_\nu $ sobre
éstas con los proyectores ${P_{\pm\mu}}^\nu$, definidos como sigue
$$
{P_{\pm\mu}}^\nu\equiv \frac12({P_\mu}^\nu  \pm {\xi_\mu}^\nu )\,
\,, \eqno{(A.7)}
$$
donde hemos introducido el operador
$$
 {\xi_\mu}^\nu \equiv {\epsilon _\mu}^{\lambda \nu}\widehat{\partial}_\lambda\,
\,, \eqno{(A.8)}
$$
siendo la ''raíz cuadrada'' de ${P_\mu}^\nu $ (ver (A.9b)) y es
sensible bajo cambios de paridad. Así, los proyectores
${P_{\pm\mu}}^\nu$ también son paridad dependientes.

Los operadores ${P_\mu}^\nu$, ${P_{\pm\mu}}^\nu$ y ${\xi_\mu}^\nu
$, verifican el  álgebra
$$
{P_\mu}^\nu  {P_\nu}^\sigma = {P_\mu}^\sigma\, \,, \eqno{(A.9a)}
$$
$$
{\xi_\mu}^\nu  {\xi_\nu}^\sigma = {P_\mu}^\sigma\, \,,
\eqno{(A.9b)}
$$
$$
{P_\mu}^\nu  {\xi_\nu}^\sigma = {\xi_\mu}^\nu  {P_\nu}^\sigma
={\xi_\mu}^\sigma\, \,, \eqno{(A.9c)}
$$
$$
{P_{\pm\mu}}^\nu  {P_\nu}^\sigma = {P_\mu}^\nu {P_{\pm\nu}}^\sigma
= {P_{\pm\mu}}^\sigma\, \,, \eqno{(A.9d)}
$$
$$
{P_{\pm\mu}}^\nu  {\xi_\nu}^\sigma = {\xi_\mu}^\nu
{P_{\pm\nu}}^\sigma =\pm{P_{\pm\mu}}^\sigma\, \,. \eqno{(A.9e)}
$$

Entonces, la descomposición de la unidad para los campos
vectoriales es
$$
{1_\mu}^\nu ={P_{+\mu}}^\nu +{P_{-\mu}}^\nu + {w_\mu}^\nu \, \,,
\eqno{(A.10)}
$$
donde ${w_\mu}^\nu $ representa el proyector de la parte de  espín
0.

Ahora veamos la prescripción para los proyectores de tensores de
rango 2.  Una descomposición  simétrico-antisimétrico de la
unidad, ${1_{\alpha\beta}}^{\mu\nu}\equiv
{\delta_\alpha}^{\mu}{\delta_\beta}^{\nu}$ es realizada
$$
1=S+A  \, \,, \eqno{(A.11)}
$$
donde
$$
{S_{\alpha\beta}}^{\mu\nu}=\frac12({\delta_\alpha}^{\mu}
{\delta_\beta}^{\nu}+{\delta_\alpha}^{\nu}{\delta_\beta}^{\mu}) \,
\,, \eqno{(A.12a)}
$$
$$
{A_{\alpha\beta}}^{\mu\nu}=\frac12({\delta_\alpha}^{\mu}
{\delta_\beta}^{\nu}-{\delta_\alpha}^{\nu}{\delta_\beta}^{\mu}) \,
\,. \eqno{(A.12b)}
$$

La parte de espín 2 de  $h_{\mu\nu}$ está en la componente
simétrica, transverso y sin traza,  ${h^{Tt}}_{\mu\nu}$
(${h^{Tt}}_{\mu\nu}={h^{Tt}}_{\nu\mu}$, ${h^{Tt\mu}}_{\mu}=0$ y
$\partial ^\mu {h^{Tt}}_{\mu\nu}=0$). La parte de espín 1 se
encuentra tomando la divergencia sobre cualquier índice, y la de
espín 0 a través de la traza o una doble divergencia. Así,
$h_{\mu\nu}$ tiene nueve componentes, de las cuales extraemos seis
con  $S$ y tres con $A$, quedando repartidas como
\begin{center}
\begin{tabular}{|c|}
 \hline
\textbf { $S$\,:\,  parte\,simétrica\, (6)}  \\
\hline \hline
2 con espín 2 \\
\hline
2 con espín 1 \\
\hline
2 con espín 0  \\
\hline

\end{tabular}
\vspace{0.5cm}
\end{center}
\begin{center}
\begin{tabular}{|c|}
 \hline
\textbf { $A$\,:\, parte\,antisimétrica\, (3)}  \\
\hline \hline
2 con espín 1 \\
\hline
1 con espín 0 \\
\hline
\end{tabular}
\vspace{0.5cm}
\end{center}

Entonces, para la parte simétrica, los proyectores que extraen las
partes de espín 2, espín 1 y espín 0 de
${h^{(s)}}_{\alpha\beta}\equiv
{S_{\alpha\beta}}^{\mu\nu}h_{\mu\nu}$, son
$$
{{{P_S}^2}_{\alpha\beta}}^{\mu\nu}=\frac12({P_\alpha}^{\mu}
{P_\beta}^{\nu}+{P_\alpha}^{\nu}{P_\beta}^{\mu}-P_{\alpha\beta}P^{\mu\nu})
\, \,, \eqno{(A.13a)}
$$
$$
{{{P_S}^1}_{\alpha\beta}}^{\mu\nu}=\frac12({P_\alpha}^{\mu}
{w_\beta}^{\nu}+{P_\alpha}^{\nu}{w_\beta}^{\mu}+{P_\beta}^{\mu}
{w_\alpha}^{\nu}+{P_\beta}^{\nu}{w_\alpha}^{\mu}) \, \,,
\eqno{(A.13b)}
$$
$$
{{{P_S}^0}_{\alpha\beta}}^{\mu\nu}=\frac12\,P_{\alpha\beta}P^{\mu\nu}
\, \,, \eqno{(A.13c)}
$$
$$
{{{P_W}^0}_{\alpha\beta}}^{\mu\nu}=w_{\alpha\beta}w^{\mu\nu} \,
\,, \eqno{(A.13d)}
$$
respectivamente. Puede observarse que ${{P_S}^2}$ es transverso y
sin traza, como es requerido para una parte de espín 2.
${{P_S}^1}$ es una construcción simétrica donde una divergencia es
tomada y luego es proyectad en la parte transversa. Para espín 0,
el objeto ${{P_W}^0}$ toma una doble divergencia y  ${{P_S}^0}$
está relacionado con la traza. Finalmente, (A.13) es la
descomposición de $S$
$$
S={P_S}^2+{P_S}^1+{P_S}^0+{P_W}^0 \, \,. \eqno{(A.14)}
$$

De una manera idéntica, se puede obtener la descomposición de $A$.
Los proyectores de espín 1 y 0 son
$$
{{{P_E}^1}_{\alpha\beta}}^{\mu\nu}=\frac12({P_\alpha}^{\mu}
{w_\beta}^{\nu}-{P_\alpha}^{\nu}{w_\beta}^{\mu}-{P_\beta}^{\mu}
{w_\alpha}^{\nu}+{P_\beta}^{\nu}{w_\alpha}^{\mu}) \, \,,
\eqno{(A.15a)}
$$
$$
{{{P_B}^0}_{\alpha\beta}}^{\mu\nu}=\frac12({P_\alpha}^{\mu}
{P_\beta}^{\nu}-{P_\alpha}^{\nu}{P_\beta}^{\mu}) \, \,,
\eqno{(A.15b)}
$$
y la unidad, $1=S+A$ es escrita así
$$
1={P_S}^2+{P_S}^1+{P_S}^0+{P_W}^0 +{P_E}^1+{P_B}^0\, \,.
\eqno{(A.16)}
$$

La proyección en cada parte de $h_{\mu\nu}$ puede particularizarse
aún más si consideramos la helicidad debido a que cada componente
de espín no nulo tiene dos helicidades posibles. De hecho, la
traza de cualquier proyector de espín no nulo da 2, lo que
corresponde a la dimensión del subespacio en que se proyecta. En
particular, para la parte de espín 2 se reconocen sus dos partes

$$
{{h^{Tt}}_{\mu \nu }}^\pm \equiv \frac{1}{2}\big({h^{Tt}}_{\mu \nu
} \pm \frac{1}{2}({\xi _\mu }^\alpha {\delta _\nu }^\beta+{\xi
_\nu }^\beta {\delta _\mu }^\alpha){h^{Tt}}_{\alpha \beta}\big)
 \, \, . \eqno{(A.17)}
$$

Por lo tanto, se tiene
$$
{{{P_{\pm S}}^2}_{\alpha\beta}}^{\mu\nu}
=\frac14({P_{\pm\alpha}}^{\mu}
{P_\beta}^{\nu}+{P_{\pm\alpha}}^{\nu}{P_\beta}^{\mu}+{P_{\pm\beta}}^{\mu}
{P_\alpha}^{\nu}+{P_{\pm\beta}}^{\nu}{P_\alpha}^{\mu}-P_{\alpha\beta}P^{\mu\nu})
\, \,, \eqno{(A.18)}
$$
y ya que ${P_\mu}^\nu ={P_{+\mu}}^\nu +{P_{-\mu}}^\nu$, se muestra
$$
{{{P_{\pm
S}}^1}_{\alpha\beta}}^{\mu\nu}=\frac12({P_{\pm\alpha}}^{\mu}
{w_\beta}^{\nu}+{P_{\pm\alpha}}^{\nu}{w_\beta}^{\mu}+{P_{\pm\beta}}^{\mu}
{w_\alpha}^{\nu}+{P_{\pm\beta}}^{\nu}{w_\alpha}^{\mu}) \, \,,
\eqno{(A.19a)}
$$
$$
{{{P_{\pm
E}}^1}_{\alpha\beta}}^{\mu\nu}=\frac12({P_{\pm\alpha}}^{\mu}
{w_\beta}^{\nu}-{P_{\pm\alpha}}^{\nu}{w_\beta}^{\mu}-{P_{\pm\beta}}^{\mu}
{w_\alpha}^{\nu}+{P_{\pm\beta}}^{\nu}{w_\alpha}^{\mu}) \, \,,
\eqno{(A.19b)}
$$
y
$$
{P_S}^2={P_{+ S}}^2+{P_{- S}}^2\, \,, \eqno{(A.20a)}
$$
$$
{P_S}^1={P_{+S}}^1+{P_{-S}}^1\, \,, \eqno{(A.20b)}
$$
$$
{P_E}^1={P_{+E}}^1+{P_{-E}}^1\, \,. \eqno{(A.20c)}
$$
Así, en dimensión $2+1$ se tiene un proyector para cada parte
irreducible de $h_{\mu\nu}$.

\newpage
\begin{center}
\large{{\bf Apéndice B}: Gravedad en 3D y espacios de dS/AdS.}
\end{center}

Consideremos una variedad $N$-dimensional, $M$ provista con
métrica $g_{\mu\nu}$ y signatura Lorentziana $(-1,+1,...,+1)$.
Además, a los puntos de la variedad le podremos asignar cartas con
coordenadas curvilíneas $x^\mu$ y localmente planas $\xi^a$.

Mediante la introducción de la conexión afín (${\Gamma ^\lambda
}_{\mu \nu }$) y la de espín (${\omega _{\mu b}}^a$), se define la
derivada covariante de objetos mixtos, como sigue
$$
D_\mu {V_\nu }^a =
\partial _\mu {V_\nu }^a  +{\omega _{\mu b}}^a {V_\nu }^b -
{\Gamma ^\lambda }_{\mu \nu } {V_\lambda }^a\, \,, \eqno{(B.1a)}
$$
$$
 D_\mu V^{\nu a} =
\partial _\mu V^{\nu a}  +{\omega _{\mu b}}^a V^{\nu b} +
{\Gamma ^\nu }_{\mu \lambda } V^{\lambda a}\, \,, \eqno{(B.1b)}
$$
$$
 D_\mu V_{\nu a} =
\partial _\mu V_{\nu a} - {\omega _{\mu a}}^b V_{\nu b} -
{\Gamma ^\lambda }_{\mu \nu } V_{\lambda a}\, \,, \eqno{(B.1c)}
$$
$$
 D_\mu {V^\nu }_a =
\partial _\mu {V^\nu}_a  -{\omega _{\mu a}}^b {V^\nu}_b +
{\Gamma ^\nu }_{\mu \lambda } {V^\lambda}_a\, \,. \eqno{(B.1d)}
$$

En el caso de la derivación covariante sobre objetos con índices
puramente curvilíneos usaremos el símbolo $\nabla _\mu$, indicando
la no inclusión de la corrección de conexión de espín.

Seguidamente, el tensor de curvatura de Riemann
(${R^{\sigma}}_{\alpha \nu \mu } $) y el de torsión (${T^\lambda
}_{\mu \nu }$) pueden ser introducidos via
$$
[\nabla _\mu ,\nabla _\nu]V_\alpha \equiv-{R^{\sigma}}_{\alpha \nu
\mu } V_\sigma - {T^\lambda }_{\mu \nu } \nabla _\lambda V_\alpha
\, \,, \eqno{(B.2)}
$$
con el tensor de torsión dado por ${T^\lambda }_{\mu \nu } \equiv
{\Gamma ^\lambda }_{\mu \nu } - {\Gamma ^\lambda }_{\nu \mu } =
{e^\lambda }_a (\partial_\mu {e_\nu }^a - \partial_\nu {e_\mu }^a
+ {\omega _{\mu \nu }}^a - {\omega _{\nu \mu }}^a)$. Para todo
$V^\alpha $, la relación (B,2) implica que el tensor de curvatura
tiene la forma
$$
{R^\sigma}_{\alpha \mu \nu } =
\partial _\nu {\Gamma ^\sigma }_{ \mu \alpha } - \partial _\mu
{\Gamma ^\sigma }_{ \nu \alpha} + {\Gamma ^\lambda }_{ \mu\alpha}
{\Gamma ^\sigma }_{\nu \lambda }
 -  {\Gamma ^\lambda }_{ \nu\alpha} {\Gamma ^\sigma }_{ \mu \lambda }\, \,, \eqno{(B.3)}
$$
y el  de Ricci viene dado por la contracción ${R^\alpha}_{ \mu
\alpha\nu }=R_{ \mu\nu }$.

Es bien conocido que el tensor de Riemann-Christoffel puede ser
descompuesto en términos del tensor de Ricci, su contracción y el
tensor conformal de Weyl ($C_{\lambda\mu\nu\sigma}$), como sigue
$$
R_{\lambda\mu\nu\sigma}=\frac{1}{N-2}\,(g_{\lambda\nu}R_{\mu
\sigma}-g_{\lambda\sigma}R_{\mu\nu}-g_{\mu
\nu}R_{\lambda\sigma}+g_{\mu\sigma}R_{\lambda\nu})+\,\,\,\,\,\,\,\,\,\,\,\,\,\,
$$
$$
-\frac{R}{(N-1)(N-2)}\, (g_{\lambda\nu}g_{\mu
\sigma}-g_{\lambda\sigma}g_{\mu\nu}) +C_{\lambda\mu\nu\sigma}\,
\,. \eqno{(B.4)}
$$
Si enfocamos nuestro interés en 3D, donde el tensor de Weyl es
idénticamente nulo, vemos que el tensor de curvatura queda
expresado en términos exclusivos de el de Ricci y su contracción.
Particularmente miraremos los espacios de dS ($\lambda>0$) y AdS
($\lambda<0$), los cuales están gobernados por la ecuación de
campo de Einstein
$$
G_{ \mu\nu }-\lambda\,g_{\mu\nu}=0\, \,, \eqno{(B.5)}
$$
donde
$$
G_{ \mu\nu }\equiv R_{ \mu\nu }-\frac{g_{\mu\nu}}{2}\,R\, \,,
\eqno{(B.6)}
$$
es el tensor de Einstein. Con esto, escribimos el tensor de
curvatura en dS/AdS en 3D
$$
R_{\lambda\mu\nu\sigma}= \frac{R}{6}\, (g_{\lambda\nu}g_{\mu
\sigma}-g_{\lambda\sigma}g_{\mu\nu}) \, \,, \eqno{(B.7)}
$$
con
$$
R= -6\lambda\, \,. \eqno{(B.8)}
$$

Si nos restringimos a espacios sin torsión recuperamos la forma
simétrica de la conexión afín (símbolos de Christoffel), es
posible hallar una solución para la ecuación (B,5), pensando en
una métrica estático-estacionaria en coordenadas ''polares'' de la
forma $dS^2=-e^{A(r)}dt^2+e^{B(r)}dr^2+r^2d\theta^2$, que al ser
utilizada en la definición del tensor de Ricci, la ecuación (B,5)
se traduce en un sistema de ecuaciones diferenciales parciales
para la funciones $A(r)$ y $B(r)$. Imponiendo las condiciones de
contorno adecuadas (como la del límite Minkowskiano cuando
$\lambda \rightarrow 0$), se obtiene la forma
$$
dS^2=-(1-\lambda r^2)dt^2+\frac{1}{1-\lambda
r^2}\,dr^2+r^2d\theta^2\, \,. \eqno{(B.9)}
$$

Es bien conocido que esta métrica puede inducirse a partir de un
hiperboloide embebido en un espacio-tiempo 3+1 dimensional.
Particularmente, el hiperboloide de de Sitter es
$$
-(X^o)^2+(X^1)^2+(X^2)^2+(X^3)^2=\frac{1}{\lambda}\, \,,
\eqno{(B.10)}
$$
donde $X^A$ con $A=0,...,3$ son las coordenadas del espacio-tiempo
3+1 al cual se le provee una métrica de tipo
$$
dS^2=\eta_{AB}dX^AdX^B=-(dX^o)^2+(dX^1)^2+(dX^2)^2+(dX^3)^2\, \,,
\eqno{(B.11)}
$$
cuya signatura refleja el hecho de que el grupo isometrías en dS
es isomorfo a $SO(3,1)$.

Entonces, si se ensaya la transformación (consistente con (B,10))
$$
X^3\pm X^o=f(r)\,e^{\pm \sqrt{\lambda}\,t}\, \,, \eqno{(B.12a)}
$$
$$
X^1=r\,cos\theta\, \,, \eqno{(B.12b)}
$$
$$
X^2=r\,sen\theta\, \,, \eqno{(B.12c)}
$$
donde $f(r)= \sqrt{\frac{1}{\lambda}-r^2}$, se puede mostrar
fácilmente que partiendo de (B,11) se recupera la métrica (B,9).
(En el caso de Anti de Sitter 2+1, la discusión parte considerando
una métrica 3+1 de tipo $diag(-1,+1,+1,-1)$, ya que el grupo de
simetrías en AdS es isomorfo a $SO(2,2)$ en 3D).

Finalmente, es posible realizar una proyección estereográfica de
las coordenadas $X^A$ del hiperboloide con la cual es posible
escribir la métrica en su forma conforme (p.ej.:
$g_{\mu\nu}=\Omega ^2(x)\,\eta_{\mu\nu} $). Tal proyección puede
establecerse como
$$
X^\mu=\frac{x^\mu}{1+\lambda\,x^2}\, \,, \eqno{(B.13a)}
$$
$$
X^3=\pm \frac{1}{\sqrt{\lambda}}\,\frac{1}{1+\lambda\,x^2}\, \,,
\eqno{(B.13b)}
$$
donde $x^\mu$ son las coordenadas estereográficas y
$x^2=\eta_{\mu\nu}x^\mu x^\nu$. Usando (B,13) reescribimos la
métrica (B,11) como
$$
dS^2=\Omega ^2(x)\,\eta_{\mu\nu}dx^\mu dx^\nu\, \,, \eqno{(B.14)}
$$
con $\Omega ^2(x)\equiv (1+\lambda\,x^2)^{-1}$.

\newpage
\begin{center}
\large{{\bf Apéndice C}: Elementos de la geometría conformal.}
\end{center}

{\bf C.1) Mapa de Weyl}

Establecemos las transformaciones conformes de la métrica
$g_{\mu\nu}$ y  los campos $\phi_A$, con $A$ un índice múltiple
como los mapas reales ($\Omega (x)\in R$) definidos por
$$
g'_{\mu\nu}=\Omega^2(x)g_{\mu\nu}\, \,\, \,, \eqno{(C.1a)}
$$
$$
g'^{\mu\nu}=\Omega^{-2}(x)g^{\mu\nu}\, \,\, \,, \eqno{(C.1b)}
$$
$$
\sqrt{-g'}=\Omega^N(x)\sqrt{-g}\, \,\, \,, \eqno{(C.1c)}
$$
$$
\phi'_A=\Omega^W(x)\phi_A\, \,\, \,, \eqno{(C.1d)}
$$
donde $N$ es la dimensión del espacio y $W$ es el ''peso'' del
campo $\phi_A$.

A partir de (C,1ab) se obtienen las reglas de transformación
conforme de los símbolos de Christoffel y las contracciones del
tensor de curvatura
$$
{\Gamma'^\rho}_{\mu\nu}={\Gamma^\rho}_{\mu\nu} + {\delta^\rho}_\nu
\partial_\mu \ln \Omega + {\delta^\rho}_\mu
\partial_\nu \ln \Omega -
g_{\mu\nu}g^{\rho\lambda}\partial_\lambda \ln \Omega\, \,\, \,,
\eqno{(C.2)}
$$
$$
R'_{\mu\nu}=R_{\mu\nu}+2(2-N)\Omega^{-2}
\partial_\mu  \Omega \partial_\nu  \Omega+ (N-2)\Omega^{-1}
\nabla_\mu  \partial_\nu \Omega\,\,+
$$
$$
+(N-3)g_{\mu\nu}\Omega^{-2}
\partial_\lambda  \Omega \partial^\lambda  \Omega+g_{\mu\nu}\Omega^{-1}
\Box{} \Omega \, \,\, \,, \eqno{(C.3)}
$$
$$
R'=\Omega^{-2}R+2(N-1)\Omega^{-3} \Box{} \Omega+
(N-1)(N-4)\Omega^{-4} \partial_\lambda  \Omega \partial^\lambda
\Omega \, \,\, \,. \eqno{(C.4)}
$$

Un método que permite formular la versión invariante conforme de
una acción dada, es ilustrado por el procedimiento que
generalmente es realizado en el caso del campos escalar real
masivo, cuya acción original es
$$
S(\phi , m)=-\frac{1}2\,\int
d^Nx\,\sqrt{-g}\,(g^{\mu\nu}\partial_\mu \phi \,\partial_\nu \phi
+m^2\phi^2)\, \,\, \,. \eqno{(C.5)}
$$
Aquí, se realiza el cambio
$$
m^2\longrightarrow \zeta (N)R \, \,\, \,, \eqno{(C.6)}
$$
donde es bien conocido que, para este caso, $\zeta (N)$ es una
función de la dimensión
$$
\zeta (N)=-\frac{1}4\,\bigg(\frac{N-2}{N-1}\bigg) \, \,\, \,,
\eqno{(C.7)}
$$
garantizándose la invariancia conforme de $S(\phi, \zeta (N)R)$
bajo (C,1) y (C,4), con $W=\frac{2-N}{2}$ en (C,1d).

\vspace{1cm} {\bf C.2) El criterio de Polchinski} \cite{r70}

Consideremos una acción tal que la densidad Lagrangiana depende de
los campos, sus derivadas y la métrica del espaco-tiempo, de la
forma
$$
S=\int d^Nx\,\sqrt{-g}\,\mathcal{L}(\phi_A,\partial \phi_A,
g_{\mu\nu})\, \,\, \,. \eqno{(C.8)}
$$
Seguidamente realizamos una transformación conforme infinitesimal
$\Omega (x)=1+\frac{\omega (x)}{2}$, con $\mid \omega (x)\mid\ll
1$, de manera que
$$
\delta g_{\mu\nu}=\omega(x)\,g_{\mu\nu}\, \,\, \,, \eqno{(C.9a)}
$$
$$
\delta g^{\mu\nu}=-\omega(x)\,g^{\mu\nu}\, \,\, \,, \eqno{(C.9b)}
$$
$$
\delta\phi_A=\frac{W}{2}\,\omega(x)\phi_A\, \,\, \,. \eqno{(C.9c)}
$$

Entonces si $E^A(\phi)=0$ son las ecuaciones de movimiento de los
campos, la variación conforme según (C,9) de la acción (C,8) es
$$
\delta_\omega S=\frac{1}2\,\int
d^Nx\,\sqrt{-g}\,\omega(x)\,(T+W\phi_A\,E^A(\phi)) \, \,\, \,,
\eqno{(C.10)}
$$
donde $T$ es la traza del tensor momento-energía de Belinfante
asociado a los campos $\phi_A$. De esto sigue que, para todo
$\omega(x)$ infinitesimal, la acción (C,8) es invariante conforme
si
$$
T=-W\phi_A\,E^A(\phi) \, \,\, \,, \eqno{(C.11)}
$$
lo cual equivale a decir que $S$ es invariante conforme si el
tensor de Belinfante posee traza nula sobre las ecuaciones de
movimiento de los campos. Como ejemplo de esto, puede mostrarse
que esta condición es satisfecha consistentemente en el caso del
campo escalar real con acción $S(\phi, \zeta(N))$, ya que el
tensor de Belinfante es proporcional a $2-N+4(1-N)\zeta(N)$.

Otro caso que puede verificarse, y que da pie para nuestra
discusión es el correspondiente al caso de la teoría de
Chern-Simons abeliana. La acción de ésta teoría es
$$
S_{CS}=\frac{m}{2}\,\int d^3x\,\epsilon^{\mu\nu\lambda}a_\mu
\partial_\nu a_\lambda\, \,\, \,, \eqno{(C.12)}
$$
cuya invariancia conforme puede verificarse pensando en que el
campo autodual posee peso conforme $W=0$\, u observando que el
tensor de Belinfante de esta teoría es nulo.

Sin embargo, esta situación cambia cuando pensamos en una suerte
de teoría de Chern-Simons abeliana ''masiva'', podemos pasar a un
modelo de tipo autodual de espín 1, o sea
$$
S_{ad}=-\frac{1}{2}\,\int
d^3x\,\sqrt{-g}\,(m\,\varepsilon^{\mu\nu\lambda}a_\mu
\partial_\nu a_\lambda + m^2g^{\mu\nu}a_\mu a_\nu)\, \,\, \,. \eqno{(C.13)}
$$
Para esta teoría, el tensor de Belinfante es
$$
T^{\mu\nu}=m^2(a^\mu a^\nu - \frac{g^{\mu\nu}}{2}\,a^\alpha
a_\alpha)\, \,\, \,, \eqno{(C.14)}
$$
y su traza (en la capa de masas)
$$
T=-\frac{m^2}{2}\,a^\alpha a_\alpha \neq 0\, \,\, \,,
\eqno{(C.15)}
$$
lo cual nos indica que tal teoría no posee invariancia conforme.
Este hecho también podría notarse al verse que el peso conformal
del campo autodual no está bien definido cuando se examinan los
dos términos que constituyen la acción (C,13).

Contrastando con el caso de Chern-Simons puro abeliano, si se
desea mantener un término cuadrático en los campos y buscar la
invariancia conforme, entonces uno podría pensar en un posible
camino a seguir como el de extender el cambio del tipo (C,6) a
$$
m^2\longrightarrow \kappa R \, \,\, \,, \eqno{(C.16)}
$$
$$
m\longrightarrow m_{(R)} \, \,\, \,, \eqno{(C.17)}
$$
donde $\kappa $ y $m_{(R)}$ son un parámetro y una función del
escalar de curvatura, respectivamente. Con esto escribimos la
acción
$$
S(a_\mu,\kappa)=-\frac{1}{2}\,\int
d^3x\,\sqrt{-g}\,(m_{(R)}\,\varepsilon^{\mu\nu\lambda}a_\mu
\partial_\nu a_\lambda + \kappa R\,g^{\mu\nu}a_\mu a_\nu)\, \,\, \,. \eqno{(C.18)}
$$
Entonces, realizando una transformación conforme infinitesimal de
tipo (C,9) con $\phi_A\equiv a_\mu$, la acción (C,18) cambia como
$$
\delta_\omega S(a_\mu,\kappa)=-\frac{1}{2}\,\int
d^3x\,\sqrt{-g}\,\big[(\delta m_{(R)}+W\omega m_{(R)}
)\,\varepsilon^{\mu\nu\lambda}a_\mu
\partial_\nu a_\lambda\,+
$$
$$
+\, \kappa \omega(\Box{}-(\frac{1}2-W)R  )a^\mu a_\mu\big]\, \,\,
\,, \eqno{(C.19)}
$$
y nos dice que la acción (C,18) es invariante conforme, para todo
$a_\mu(x)$ y $\omega(x)$ si
$$
\kappa =0\, \,\, \,, \eqno{(C.20)}
$$
$$
\delta m_{(R)}+W\omega m_{(R)} =0\, \,\, \,. \eqno{(C.21)}
$$
La relación (C.20) indica que la invariancia conforme obliga la
eliminación del término masivo cuadrático en los campos,
reduciendo la acción a una parecida a la de ''Chern-Simons'' pero
con un factor $m_{(R)}$. Mientras tanto, la relación (C,21) nos
indica que en el caso de que el campo $a_\mu$ posea peso conformal
cero, simplemente la función del escalar de Ricci se reduce a
$m_{(R)}=constante$.

Pensando en el criterio de Polchinski, lo anterior se verifica
cuando se  calcula el tensor momento-energía de Belinfante
asociado a $a_\mu$, según la acción (C,18), y cuya traza sobre las
ecuaciones de movimiento es proporcional al parámetro $\kappa$,
indicando que la invariancia conforme demanda la extracción del
término masivo cuadrático en los campos.

Si recordamos particularmente el caso autodual de espín 2 en
dS/AdS, se puede mostrar que una discusión similar a lo anterior,
ocurre entre el término lagrangiano de tipo ''Chern-Simons'',
$m\,\varepsilon ^{\mu \nu \lambda }{h_\mu }^\alpha \nabla _\nu
h_{\lambda \alpha }$ y el de tipo masivo cuadrático, $M^2\,(h_{\mu
\nu }h^{\nu \mu }-h^2 )$ de la acción (\ref{d21}). Esto es así, ya
que la traza del tensor de Belinfante sobre las ecuaciones de
movimiento es proporcional  a $M^2$, lo que obliga nuevamente a
eliminar el término cuadrático para recuperar la invariancia
conformal.

\newpage

\end{document}